\documentclass[11pt]{article}
\usepackage[utf8]{inputenc}
\usepackage[toc,page]{appendix}
\usepackage{a4wide} 
\usepackage{multirow}
\usepackage{hyperref}
\usepackage{amsmath,amsthm,amssymb}
\usepackage{bm}
\usepackage{bbm} 
\usepackage{mathtools}
\usepackage{derivative}
\usepackage{natbib}
\usepackage{parskip} 
\usepackage{array}
\usepackage{tikz}
\usepackage{caption,subcaption}
\usepackage{floatrow}
\newcommand\correspondingauthor{\thanks{Corresponding author, email address: dan.leonte16@imperial.ac.uk.}}

\usepackage{soul}
\usepackage[normalem]{ulem}
\usepackage{enumerate}
\usepackage{float}
\usepackage{adjustbox} 

\theoremstyle{plain}
\newtheorem{theorem}{Theorem}[section]

\newtheorem{lemma}[theorem]{Lemma}

\theoremstyle{definition}
\newtheorem{definition}[theorem]{Definition}

\newtheorem{example}[theorem]{Example}
\theoremstyle{remark}
\newtheorem{remark}[theorem]{Remark}

\newfloatcommand{capbtabbox}{table}[][\FBwidth]
\newcommand{\bt}{\boldsymbol{\theta}}
\newcommand{\x}{\mathbf{x}}

\newcommand{\defeq}{\vcentcolon=}
\newcommand{\eqdef}{=\vcentcolon}

\DeclareMathOperator*{\argmax}{arg\,max}
\DeclareMathOperator*{\argmin}{arg\,min}

\newcommand{\R}{\mathbb{R}}
\newcommand{\Z}{\mathbb{Z}}
\newcommand{\calS}{\mathcal{S}}
\newcommand{\ev}{\mathbb{E}}
\DeclareMathOperator{\Var}{Var}
\DeclareMathOperator{\Corr}{Cor}
\DeclareMathOperator{\Cov}{Cov}

\begin{document}
\title{Likelihood-based inference and forecasting for trawl processes: a stochastic optimization approach}
\author{Dan Leonte\correspondingauthor
\; and  Almut E. D. Veraart\\
	Department of Mathematics, Imperial College London}
\maketitle
%
 \begin{abstract} 
 \noindent We consider trawl processes, which are stationary and infinitely divisible stochastic processes and can describe a wide range of statistical properties, such as heavy tails and long memory. In this paper, we develop the first likelihood-based methodology for the inference of real-valued trawl processes and introduce novel deterministic and probabilistic forecasting methods. Being non-Markovian, with a highly intractable likelihood function, trawl processes require the use of composite likelihood functions to parsimoniously capture their statistical properties. We formulate the composite likelihood estimation as a stochastic optimization problem for which it is feasible to implement iterative gradient descent methods. We derive novel gradient estimators with variances that are reduced by several orders of magnitude. We analyze both the theoretical properties and practical implementation details of these estimators and release a Python library which can be used to fit a large class of trawl processes. In a simulation study, we demonstrate that our estimators outperform the generalized method of moments estimators in terms of both parameter estimation error and out-of-sample forecasting error. Finally, we formalize a stochastic chain rule for our gradient estimators. We apply the new theory to trawl processes and provide a unified likelihood-based methodology for the inference of both real-valued and integer-valued trawl processes.\newline


\end{abstract}

\maketitle
\section{Introduction}
In many real-world settings, traditional statistical models assuming normality and independence may not adequately capture the complex and persistent dynamics of the system of interest. 
For example, time series with long memory can be observed in astronomy, agriculture and geophysics \citep{robinson2003time}. Similarly, time series with skewed and heavy-tailed distributions can be observed in macroeconomics, e.g.~in financial asset returns \citep{bradley2003financial} and in geology, e.g.~in natural phenomena such as earthquakes and floods \citep{Caers1999}. 
Such statistical properties can be modelled directly under the framework of trawl processes, which was independently developed by \cite{ole_brazilian} to model turbulence and \cite{taqqu} to model workloads for network communications. The versatility of trawl processes within the class of stochastic processes comes from their flexibility. Indeed, trawl processes can produce any infinitely divisible marginal distribution, as well as very flexible autocorrelation structures; further, the marginal distribution and autocorrelation structure can be modelled independently of one another. Since their introduction, trawl processes and their extensions have been successfully employed for theoretical study in other settings as well, such as spatio-temporal statistics \citep{nguyen_DG_RG_comparison}, brain imaging \citep{brain_imaging} and finance \citep{barndorff2014integer}. In spite of their early success, trawl processes have only been applied to practical settings in which analytic expressions were available to fit the parameters of the process \citep{noven_phd_thesis,cl_integer_trawl}. 

In this paper, we develop the first likelihood-based methodology for the inference of continuous-time, real-valued trawl processes and demonstrate in a simulation study the superior finite sample properties of our estimator compared to the existing generalized method of moments (GMM) estimator. Trawl processes are particular cases of moving averages, hence stationary and ergodic (see \cite{barndorff2014integer}) and moment-based estimation is consistent \citep{matyas1999}. Nevertheless, previous simulation studies for integer-valued trawl processes \citep{barndorff2014integer} show that empirical moments and GMM estimators can be slow to convergence. Our experiments suggest convergence is even slower for long-memory trawl processes, which are weakly mixing. The next step is maximum likelihood estimation, yet trawl processes are not Markovian and have a highly intractable likelikood function. For integer-valued trawl processes, \cite{cl_integer_trawl} propose the use of the pairwise likelihood (PL) as a composite likelihood function which captures information about both the dependence structure and the marginal distribution. Although this method improves on the finite sample properties of the GMM estimator, its applicability can not be easily extended to real-valued trawl processes. Indeed, in the integer-valued case, the PL function and its gradients are given by finite sums, which can be easily computed on a computer. In contrast, in the real-valued case, the PL function is given by integrals; even in simple cases, the integrands are ill-behaved. \cite{noven_phd_thesis} tries to use numerical  integration to approximate the PL function for trawl processes in a latent variable model, but finds that the likelihood maximization procedure fails to converge due to loss of precision and, further, that other methods aiming to improve accuracy are not computationally feasible. In the Gaussian case, the author finds that the optimization procedure can be performed with the closed-form expression for the PL function, but not with the approximation. Thus, the issue lies with the accurate estimation of the objective function, and not with the PL approach. With this in mind, we propose a novel approach to estimate the PL function and its gradients using Monte Carlo (MC) methods and solve the likelihood maximization problem with simulation-based optimization techniques. The task is then to formulate the PL function of the trawl process and its gradients as expectations, which can then be approximated by simulation inside an iterative, gradient-based optimization scheme. 

Let $\bt$ be the parameters of the trawl process and $\xi(\cdot)$ the PL function. By properties of the trawl processes discussed in the paper, we have that
\begin{equation*}
\xi(\bt) = \int f(z,\bt) \mu(\mathrm{d}z;\bt) = \ev[f(Z,\bt)] ,
\end{equation*}
where $f$ is a smooth function and the law of $Z$ is given by the probability measure $\mu(\mathrm{d}z;\bt)$, which is parameterized by $\bt$. Notice the atypical setting in which the integrator, and not just the integrand, depends on $\bt$. This prevents the use of the usual differentiation under the integral sign to construct an MC estimator for $\nabla_{\bt}\xi(\bt)$. One could estimate $\xi(\bt)$ with samples and plug this estimate in a finite difference approximation, or use the score function (SF) trick \citep{rubinstein1969}, but we find empirically that the resulting estimators are not feasible due to high variance. This is a known issue in sensitivity analysis \citep{rubinstein1992sensitivity} and deep learning \citep{reparam_tutorial}. To derive lower-variance estimators, we leverage the pathwise gradients (PG) and measure-valued gradients (MVG)  methodologies introduced in \cite{Ho1983559} and \cite{pflug2012optimization}, respectively. For real-valued trawl processes, we extensively use PG which aim to remove the dependency of the integrator $\mu$ on $\bt$ and `push' it into $z$ to obtain $\xi(\bt) = \int f \left(z(\bt),\bt\right) \mu(\mathrm{d}z;\bt).$ Then by the chain rule
\begin{align}
\nabla_{\bt}\xi(\bt) &= \int \left(\pdv{f}{z}\left(z(\bt),\bt\right) \nabla_{\bt}z + \nabla_{\bt}f(z,\bt) \right) \mu(\mathrm{d}z) = \ev\left[f(Z,\bt) \nabla_{\bt} Z + \nabla_{\bt}f(Z,\bt)\right] \label{eq:intro_1} \\
&\approx \frac{1}{N}\sum_{i=1}^N\left(f(Z_i,\bt) \nabla_{\bt} Z_i + \nabla_{\bt}f(Z_i,\bt) \right)\label{eq:intro_2},
\end{align}
where the two terms in \eqref{eq:intro_1} account for the dependency of $f$ on $\bt$ through the first and second arguments, $z(\bt)$ and $\bt$, respectively. The gradient $\nabla_{\bt}\xi(\bt)$ can then be estimated using MC samples as in \eqref{eq:intro_2}. We also extend our methodology to deal with integer-valued trawl processes by using the MVG technique, which interprets $\nabla_{\bt}\mu(\cdot;\bt)$ as a signed measure and allows differentiation under the integral sign. Our contributions to the parameter inference of trawl processes are as follows. 

We provide extensive theoretical and practical analyses for our MC estimators and combine the PG and MVG techniques with other variance reduction methods, e.g.~control variates and coupling. We demonstrate in a simulation study that our estimators have their variance reduced by one to two orders of magnitude and that PL inference for real-valued trawl processes is both accurate and computationally efficient. Further, we provide just-in-time, compiled Python implementations using the JAX framework at \cite{Leonte_Ambit_Stochastics_2022} and integrate our methodology within the Autograd automatic differentiation engine. This not only greatly improves code performance, but also makes our implementations easy to adapt to fit the parameters of a wide range of real-valued trawl processes, with minimal changes. It is noteworthy that our implementation eliminates the need to compute the gradients by hand, for each trawl process, as was required in \cite{cl_integer_trawl}. 

Having established that PL is feasible, we derive a novel conditional mean forecasting formula and present the first methodology for the probabilistic forecasting of continuous-time, real-valued trawl processes. We demonstrate in a simulation study that the PL estimator outperforms the GMM estimator in both parameter estimation and out-of-sample forecasting error, regardless of the metric used to evaluate the results: mean squared or absolute error, median absolute error or KL divergence.

The rest of the paper is structured as follows. Section \ref{section:trawl_processes_background} defines L\'evy bases, sets the notation and theoretical framework for trawl processes and discusses the joint structure of both integer-valued and real-valued trawl processes. Section \ref{section:parameter_inference} introduces the GMM and PL methodologies and outlines the challenges associated with PL inference for real-valued trawl processes. We also discuss the formulation of the pairwise density as an MC estimator and illustrate the pathwise gradient (PG) and measure-valued gradient (MVG) methodologies for approximating the gradients with low-variance MC estimators. 
Next, Subsection \ref{subsection:pg} shows that PG can be easily, accurately and efficiently implemented on a computer and Subsection \ref{subsection_linear_control_variateas} explores further variance reduction with control variates, which we use in conjunction with PG. 
Section \ref{section:forecasting} derives a novel conditional mean forecasting formula and presents the first methodology for probabilistic forecasting of continuous-time, real-valued trawl processes. Section \ref{section:simulation_study} demonstrates in a simulation study that the PL estimator outperforms the GMM estimator in both parameter inference and out-of-sample forecasting errors. Additional details on the technical derivations, as well as the practical implementation details and an extended simulation study are available in the supplementary material. 
\section{Trawl processes: Background}\label{section:trawl_processes_background}
We give an overview of L\'evy bases, which can be viewed as non-Gaussian extensions of Gaussian white noise and discuss their elementary properties. Following \cite{ole_brazilian}, we define the trawl process $X_t = L(A_t)$ as the L\'evy basis $L$ evaluated over a
collection of time-indexed sets $A_t$ and discuss its autocorrelation structure, marginal and joint distributions. 

\textbf{Notation and preliminaries}

Let $\mathcal{B}_\text{Leb}(\R^d)$ denote the collection of Borel measurable subsets of $\R^d$ with finite Lebesgue measure. We say that the measure $l$ is finite if $l(\R) < \infty$ and infinite otherwise. By a L\'evy measure $l$ on $\R$ we mean a (possibly infinite) Borel measure with $l({0}) =0$ and $\int_{\R} \min{(1,y^2)} l(\mathrm{d}y) < \infty.$ We write $X\stackrel{d}{=}Y$ if $X$ and $Y$ have the same law. We use the term density for both the probability mass function and probability density function when there is no risk of confusion.   \qed

\textbf{L\'evy bases}

\begin{definition}[L\'evy basis]
\label{def:levy_basis}
A L\'evy basis $L$ is a collection of infinitely-divisible, real-valued random variables $\left\{L(A): A \in \mathcal{B}_\text{Leb}(\R^d)\right\}$ such that for any countable sequence of disjoint sets $A_1,A_2,\ldots \in \mathcal{B}_\text{Leb}(\R^d),$ the random variables $L(A_1),L(A_2),\ldots$ are independent and further, if  $\cup_{j=1}^{\infty} A_{j} \in \mathcal{B}_\text{Leb}(\R^d)$, then $L\left(\cup_{j=1}^{\infty} A_{j}\right)=\sum_{j=1}^{\infty} L\left(A_{j}\right)$ a.s.
\end{definition}
We restrict our attention to homogeneous L\'evy bases, i.e. L\'evy bases for which there exist $\xi \in \R$, $a \in \R_{\ge 0}$ and a L\'evy measure $l$ on $\R$ such that for any $A \in \mathcal{B}_\text{Leb}(\R^d),$ the following holds 
\begin{equation} 
  \ev\left[e^{i t L(A)}\right]
  = \exp{\left[ \left( i t \zeta-\frac{1}{2} t^{2} a+\int_{\mathbb{R}}\left(e^{i t y}-1-i t y \mathbf {1}_{[-1,1]}(y)\right) l(\mathrm{d} y) \right) \mathrm{Leb}(A)\right]}. \label{eq:cumulant}
\end{equation} 
We say a real-valued random variable $L'$ is a L\'evy seed of $L$ if 
\begin{equation*} 
  \ev\left[e^{i t L'}\right]
  = \exp{\left( i t \zeta-\frac{1}{2} t^{2} a+\int_{\mathbb{R}}\left(e^{i t y}-1-i t y \mathbf {1}_{[-1,1]}(y)\right) l(\mathrm{d} y) \right) }.
\end{equation*} 
We can then associate to each L\'evy basis $L$ the L\'evy-Khintchine triplet $(\xi,\,a,\,l)$ of $L'$, which fully determines the distributional properties of $L$. In the above triplet, $\xi$ denotes the drift term, $a$ the variance of the Gaussian component and $l$ the L\'evy measure of the jump part \citep[cf.][p. 37]{ken1999levy}. For a detailed discussion of the homogeneity property of L\'evy bases, see Chapter 5.1 of \cite{ambit_book}. Differentiating \eqref{eq:cumulant} once, respectively twice with respect to $t$, we obtain that $\ev\left[L(A)\right] = \textrm{Leb}(A) \ev\left[L^{'}\right]$ and $\Var{\left(L(A)\right)} = \textrm{Leb}(A) \Var{\left(L^{'}\right)}$. By taking higher derivatives, the moments of $L(A)$ can be expressed in terms of $\textrm{Leb}(A)$ and the moments of $L^{'}$.  Finally, to construct a trawl process, we need to choose the trawl sets. We restrict our attention to monotonic trawls in $d=2$ dimensions, i.e.~trawl processes with trawl sets of the form
\begin{equation*}
    A_t = A + (t,0), \qquad A = \{(s,x) \in \R^2 \colon  s < 0, 0 <  x < \phi(s) \},
\end{equation*}
where $\phi \colon (-\infty,0] \to \mathbb{R}_{\ge 0}$ is a smooth, increasing function. Define the trawl process $X = \left(X\right)_{t \ge 0}$ by the L\'evy basis evaluated over the trawl set $X_t = L(A_t).$  We note that, while the trawl process $X$ is defined to take values in $\R,$ the trawl set is chosen as a subset of $\R^2,$ i.e.~it includes an abstract spatial dimension in addition to the temporal dimension. 

Trawl processes form a rich class within that of stationary, infinitely divisible stochastic processes. An advantage from the statistical modelling perspective is that trawl processes can realize any positive, decreasing and differentiable autocorrelation function and any infinitely divisible marginal distribution, and that the autocorrelation structure and marginal distribution can be chosen independently. We illustrate this flexibility with examples.

\textbf{Correlation structure}

The shape of the trawl set $A$, specified by the trawl
function $\phi$, determines the autocorrelation structure of $X_t$. Trawl processes are stationary and, as shown in Proposition $56$ of \cite{ambit_book}, the following holds
\begin{equation*}
    \rho(h)  \defeq \Corr(X_t,X_{t+h}) = \frac{\mathrm{Leb}\left(A \cap A_h\right)}{\mathrm{Leb}\left(A\right)} = \frac{\int_{-h}^0 \phi(s) \mathrm{d}s}{\int_{-\infty}^0 \phi(s) \mathrm{d}s}, \text{ for }h>0.
\end{equation*}
We present several examples, which interpolate between short and long term memory. 
\begin{example}[Exponential trawl function]\label{exponential_trawl_function_example}If $\phi(t) = e^{\lambda t}$ for $t \le 0$ and $ \lambda >0$, then $\rho(h) = e^{-\lambda h}$ for $h \ge 0$. A more flexible autocorrelation structure is given by the superposition of exponentials $\phi(t) = \sum_{i=1}^J w_j e^{\lambda_j t}$ for $t \le 0$ with $\sum_{j=1}^J w_j= 1$ and $\lambda_j >0$, $ w_j \ge 0$. Then $\int_{-\infty}^0 \phi(t) = \sum_{j=1}^J \frac{w_j}{\lambda_j}$ and $\rho(h) = \sum_{j=1}^J \frac{w_j}{\lambda_j}e^{-\lambda_j h} / \left(\sum_{j=1}^J \frac{w_j}{\lambda_j}\right)$.
\end{example}
To extend beyond finite superpositions of exponentials, consider $\phi(t) = \int_0^\infty e^{\lambda t} \pi(\mathrm{d}\lambda)$ for $t \leq 0$, where $\pi$ is a probability measure on $\R$. Then $\phi$ can be interpreted as a randomized mixture of exponentials, where the exponential rate of decay is chosen according to $\pi$. The previous examples are obtained when $\pi$ is the Dirac delta measure $\delta_\lambda$, respectively $\pi = \sum_{j=1}^J w_j \delta_{\lambda_j}$. The next two examples give sub-exponential and polynomial decay of the autocorrelation function; full derivations can be found in Section S3 of \cite{cl_integer_trawl}.

\begin{example}[Inverse Gaussian trawl function]If $\phi$ follows an Inverse Gaussian distribution, i.e.~$\pi(\mathrm{d}x) =\sqrt{\frac{\lambda}{2 \pi x^3}}  e^{-\frac{\lambda(x-\mu)^2}{2 \mu^2 x}} \mathrm{d}x $, then $\phi(t) = \left(1- \frac{2\mu^2t}{\lambda}\right)^{-1/2} e^{\frac{\lambda}{\mu} \left(1-\sqrt{1-\frac{2 \mu^2 t}{\lambda}}\right) } $ for $t \le 0$ and $\gamma, \, \delta \geq 0$, then $\rho(h) = e^{-\frac{\lambda}{\mu} \left(\sqrt{1+ \frac{2\mu^2h}{\lambda}}-1\right)}$ for $h\geq 0$.
\end{example}
\begin{example}[Gamma trawl function]\label{gamma_trawl_function_example}If $\pi$ follows a $\textrm{Gamma}(1+H,\delta)$ distribution, i.e.~$\pi(\mathrm{d}x) = \frac{\delta^{1+H}}{\Gamma(1+H)}x^{1+H}e^{-\delta x} \mathrm{d}x$  for $x\geq 0$ and $\delta,\, H>0$, then $\phi(t) = \left(1-\frac{t}{\delta}\right)^{-(H+1)}$ for $t \le 0$ and $\rho(h) = \left(1+\frac{h}{\delta}\right)^{-H}$ for $h \geq 0$.
\end{example}
\textbf{Marginal distribution}

Trawl processes can have any infinitely divisible marginal distribution. In the following, we concentrate on distributions for which the density is available analytically and the marginal distribution of $L(A)$ is in the same named family as that of $L^{'}$. These include the integer-valued L\'evy bases from Examples \ref{ex:distr_poisson} - \ref{ex:distr_skellam}, the positive-valued L\'evy bases from Examples \ref{ex:distr_gamma} and \ref{ex:distr_ig} and the real-valued L\'evy bases from Examples \ref{ex:distr_gaussian} and \ref{ex:distr_nig}. The list of parameterizations for the probability distributions used below is available in the Appendix.
\begin{example}[Poisson L\'evy basis]\label{ex:distr_poisson}
Let $L^{'}\sim \text{Poisson}(\nu)$ with $\nu >0$. Then \\ $X_t \sim \text{Poisson}(\nu\mathrm{Leb}\left(A\right)).$ 
\end{example}
\begin{example}[Negative Binomial L\'evy basis]\label{ex:distr_negbin}
Let $L^{'}\sim \textrm{NB}(m,p)$ with $m, p > 0$. Then $X_t \sim \textrm{NB}\left(m \textrm{Leb}(A),p\right)$.
\end{example}
\begin{example}[Skellam L\'evy basis]\label{ex:distr_skellam}
Let $L'\sim \text{Skellam}(\mu_1,\mu_2),$ i.e.~$L^{'}\sim N_1 - N_2$ with $N_1,N_2$ independent and Poisson distributed with intensities $\mu_1,\mu_2 >0$.  Then \newline $X_t  \sim \text{Skellam}(\mu_1 \mathrm{Leb}\left(A\right),\mu_2 \mathrm{Leb}\left(A\right)).$
\end{example}
\begin{example}[Gamma L\'evy basis]\label{ex:distr_gamma}
Let $L^{'} \sim \text{Gamma}(\alpha,\beta)$ with $\alpha,\beta>0$. Then $X_t  \sim \text{Gamma}(\alpha \mathrm{Leb}\left(A\right),\beta).$
\end{example}
\begin{example}[Inverse Gaussian L\'evy basis]\label{ex:distr_ig}
Let $L^{'} \sim \text{IG}(\mu,\lambda)$ with $\mu,\lambda >0$. Then $X_t \sim \textrm{IG}(\mu \mathrm{Leb}(A),\lambda \mathrm{Leb}^2(A))$.
\end{example}
\begin{example}[Gaussian L\'evy basis]\label{ex:distr_gaussian}
Let $L^{'}\sim \mathcal{N}(\mu,\,\sigma^2)$ with $\sigma >0$. Then \\ $X_t \sim \mathcal{N}\left(\mu \mathrm{Leb}\left(A\right), \sigma^2 \mathrm{Leb}\left(A\right)\right).$
\end{example}
An important class of infinitely divisible distributions is that of Normal variance-mean mixtures (see Definition \ref{def:normal_variance_mean_mixture}), which include the following example.
\begin{example}[Normal-inverse Gaussian L\'evy basis]\label{ex:distr_nig}
Let $L^{'} \sim \textrm{NIG}(\alpha,\beta,\delta,\mu)$ with $|\alpha| > |\beta|$. Then $X_t = L(A_t) \sim \textrm{NIG}\left(\alpha,\beta,\delta\mathrm{Leb}\left(A\right),\mu \mathrm{Leb}\left(A\right)\right).$ 
\end{example}
In general, there are ID distributions for which the density is not available in closed form, e.g.~L\'evy $\alpha$-stable distributions, for which the density is estimated by inverting the characteristic function. Even if $L^{'}$ is from a named family of distributions with known density, the marginal distribution of $L(A)$ might not be in the same family, e.g.~if $L^{'}$ has a generalized hyperbolic distribution (see \cite{GIG_not_closed}). 

\textbf{Joint structure}

\begin{figure}[h]
    \centering
    \subfloat[\centering ]{{\includegraphics[height=4.5cm,keepaspectratio]{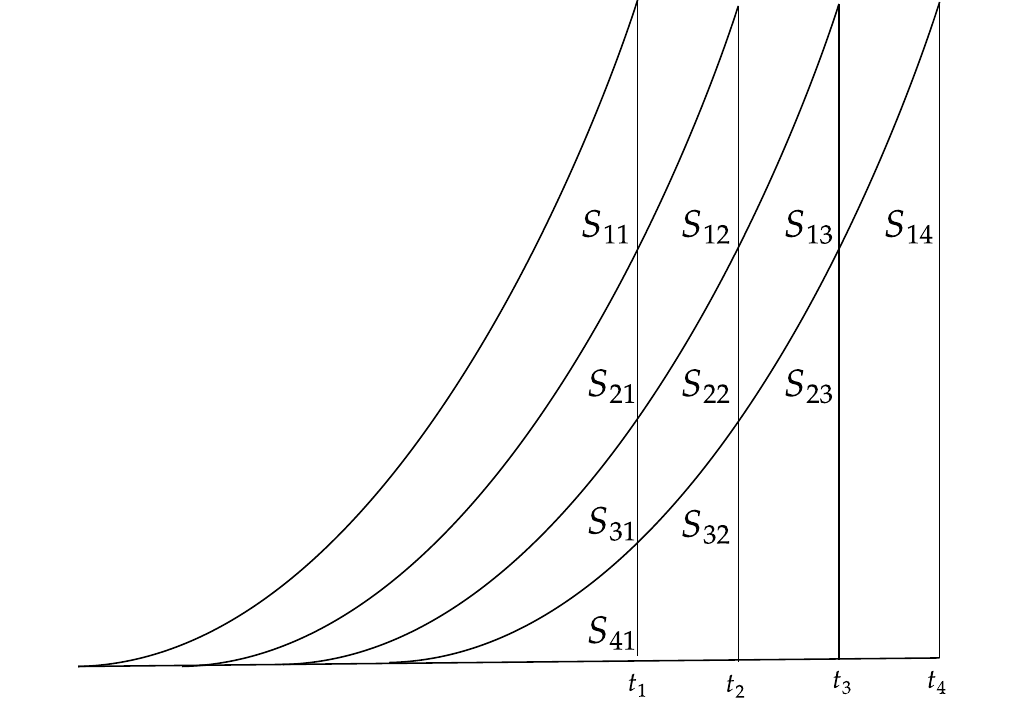} }\label{fig:slice_partition_in_general}}
    \qquad
    \subfloat[\centering ]{{\includegraphics[height=4.5cm,keepaspectratio]{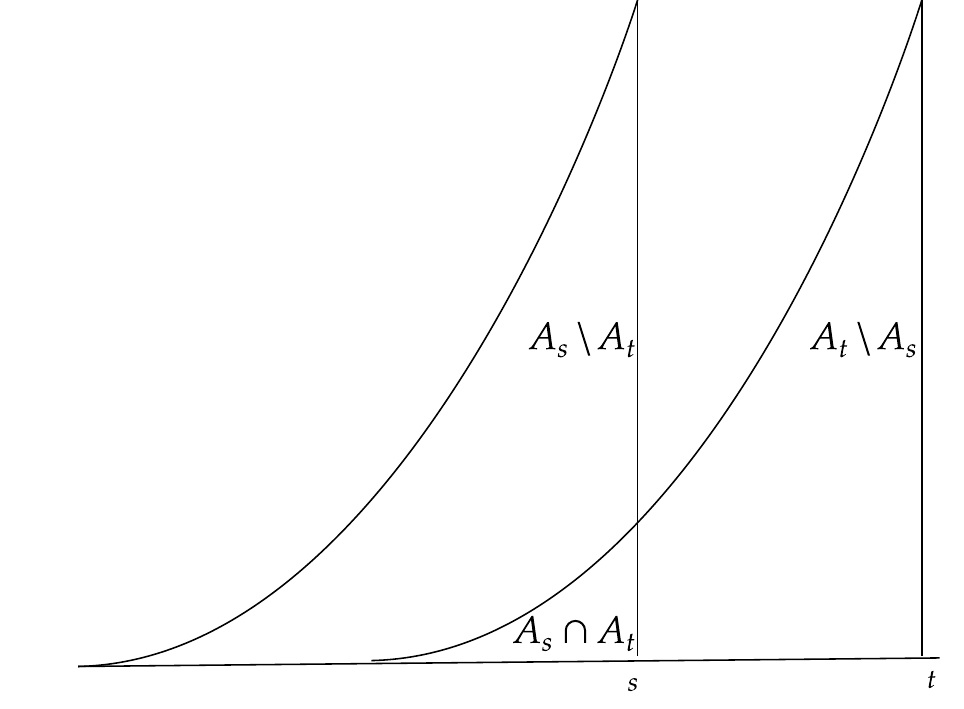} }\label{fig:just_2_trawls}}%
    \caption{a) The slice partition into disjoint sets of the trawl sets $A_{t_1},\ldots,A_{t_n}$ for $n=4$. In general, there are $n(n-1)/2$ sets. b) The slice partition for two sets $A_s$ and $A_t$ only has three sets and the pairwise density $p_{X_s,X_t}(x_s,x_t)$ can be written as an integral in one dimension.}%
\label{fig:slice_partition_and_2_trawls}%
\end{figure}

Trawl processes are in general not Markovian. By taking the slice partition
\begin{equation*}
    S_{ij} = \left(A_j \cap A_{i+j-1}\right) \backslash A_{i+j},\ 1\le i,j \le n, \ i+j \le n+1,
\end{equation*}as considered in \cite{noven_phd_thesis, leonte2022simulation_arxiv} and displayed in Figure \ref{fig:slice_partition_in_general}, we obtain an integral representation of the finite marginal distributions 
\begin{equation}
p_{X_{t_1},\ldots,X_{t_n}}\left(x_{1},\ldots,x_{n}\right) = \int_{\Delta(x_{1},\ldots,x_{n})} \prod_{1 \le i,j \le n \colon i+j \le n+1 } p_{L\left(S_{ij}\right)}(s_{ij}) \, \mathrm{d}s_{11}\mathrm{d}s_{21}\ldots \mathrm{d}s_{1k},\label{eq:joint_structure}
\end{equation}
where the $n(n+1)/2$ variables $s_{ij}$ corresponding to the slices $S_{ij}$ and where \begin{equation*}\Delta(x_{_1},\ldots,x_{_n}) \defeq  \{s_{ij} : 1 \le i, j \le n, \, i+j \le n+1 \text{ and } \sum_{i}s_{ij} = x_j\}.
\end{equation*}
The random variables $L\left(S_{ij}\right)$ have infinitely divisible distributions with L\'evy-Khintchine triplets given by $(\xi \, \mathrm{Leb}(S_{ij}),\, a \, \mathrm{Leb}(S_{ij}),\, l \, \mathrm{Leb}(S_{ij}))$, which only depends on the slices $S_{ij}$ through their areas. Thus the joint distributions are determined by the autocorrelation function of the trawl process and by the L\'evy-Khintchine triplet of $L'$.

Calculating the joint densities is computationally expensive, as it requires evaluating multiple integrals. Thus maximum likelihood estimation is not feasible. By comparison, the expression for the bivariate densities contains only one integral (see Figure \ref{fig:just_2_trawls})
\begin{equation}
    p_{X_s,X_t}(x_s,x_t) = \int p_{L(A_t\cap A_s)}(z) \,  p_{L(A_t\backslash A_s)}(x_t-z) \,  p_{L(A_s\backslash A_t)}(x_s-z) \,  \mathrm{d}z,
    \label{eq:pairwise_likelihood}
\end{equation}
making it a more numerically tractable approximation, which can be used for parameter inference. Note that if the L\'evy seed is discretely supported, the integrals from \eqref{eq:joint_structure} and \eqref{eq:pairwise_likelihood} are replaced by summations. For example, for a Poisson L\'evy basis with $L(A) \sim \text{Poisson}(\nu\mathrm{Leb}\left(A\right))$, $p_{X_s,X_t}(x_s,x_t)$ is given by 
\begin{equation*}
 \sum_{k=0}^{\min{\{x_s,x_t\}}} \mathrm{Poisson}(k; \nu s_{21}) \, \mathrm{Poisson}(x_t-k;\nu s_{12}) \, \mathrm{Poisson}(x_s -k; \nu s_{11}), 
\end{equation*}
where $s_{11},\, s_{21}, \, s_{12}$ are the Lebesgue measures of the slices $S_{11}, \, S_{21},\, S_{12} $; for a Skellam L\'evy basis with $L(A) \sim \mathrm{Skellam}(\nu_1 \mathrm{Leb}(A),\nu_2 \mathrm{Leb}(A))$,  $p_{X_s,X_t}(x_s,x_t)$ is given by
\begin{equation}
     \sum_{k=-\infty}^{\infty} \mathrm{Skellam}(k; \nu_1 s_{21}, \nu_2 s_{21}) \, \mathrm{Skellam}(x_t-k;\nu_1 s_{12}, \nu_2 s_{12}) \, \mathrm{Skellam}(x_s -k; \nu_1 s_{11}, \nu_2 s_{11}).\label{eq:discrete_ivt_skellam}
\end{equation}
The summation is taken over a finite set if $L^{'}$ is supported on the positive integers and over a countable set if $L^{'}$ is supported on the integers.  In the following we compare and contrast with the discrete case, but concentrate on parameter inference for L\'evy bases with continuously supported marginal distributions.
\section{Parameter inference for trawl processes }\label{section:parameter_inference}
We want to perform parameter inference given observations $\x = (x_{\tau},\ldots,x_{n\tau})$ of the trawl process $X$ at times $\tau,\ldots,n\tau$ for $\tau >0$. The parameters to be inferred are the parameters $\bt_{L'}$ which specify the law of the L\'evy seed $L^{'}$, together with the parameters $\bt_{\phi}$ of the trawl function $\phi$. Let $\bt = (\bt_{L^{'}},\bt_\phi)$. Since trawl processes are in general not Markovian, the likelihood function is intractable and maximum likelihood estimation is infeasible. Two alternative methodologies have been explored in the literature: generalised method of moments (GMM) and pairwise likelihood (PL). 

\subsection{Inference by GMM and PL}\label{subsection:gmm_and_pl_intro}
The first approach, used in \cite{barndorff2014integer}, is motivated by the fact that trawl processes are stationary and ergodic (see \cite{barndorff2014integer}), hence moment-based estimation is consistent \citep{matyas1999}. The parameters $\bt_{L^{'}}$ can be estimated from the empirical moments of $x_\tau,\ldots,x_{n\tau}$; the parameters $\bt_{\phi}$ can be estimated by matching the empirical autocorrelation function at lags $1,\ldots,K$
\begin{equation*}
    \argmin_{\bt_{\phi}}\sum_{k=1}^K \left(\rho(h k; \bt_{\phi}) - \tilde{\rho}(k)\right)^2,
\end{equation*}
where $K$ is to be chosen, $\rho(h) = \Corr{(X_t,X_{t+h})}$ is the autocorrelation function and $\tilde{\rho}(k)$ is the empirical autocorrelation function of $X$ at lag $k$ based on observations $x_1,\ldots,x_{n\tau}$. A summary of the GMM applied to trawl processes can be found in Section S3 of \cite{cl_integer_trawl}. Despite the asymptotic properties of the GMM estimator, empirical moments can converge slowly and result in poor finite sample properties of the GMM estimators, especially in the weakly-mixing case of long memory trawl processes. Various simulation studies invesigating the finite sample performance of the GMM estimator for trawl processes have been carried out in \cite{barndorff2014integer,cl_integer_trawl} and \cite{orimar_almut}. 

The second approach, studied in \cite{cl_integer_trawl} for positive, integer-valued trawl processes, proposes the use of the pairwise likelihood as a composite likelihood function which captures information about both the dependence structure and marginal distribution. This method has better finite sample properties, but comes with an increased computational cost. To emphasize the dependency of the pairwise densities on the parameters $\bt$ of the trawl process $X$, we write $p_{X_s,X_t}(x_s,x_t;\bt)$ for $p_{X_s,X_t}(x_s,x_t)$; when there is no risk of confusion, we abridge this to $p(x_s,x_t;\bt)$. Define the pairwise likelihood function at lag $k$ by
\begin{equation*}
    {PL}^{(k)}(\bt;\x) = \prod_{i=1}^{n-k} p(x_{i\tau},x_{(i+k)\tau};\bt), 
\end{equation*}
and the pairwise likelihood function by
\begin{equation*}
    \mathcal{L}(\bt) \defeq \prod_{k=1}^K {PL}^{(k)}(\bt;\x) =  \prod_{k=1}^K \prod_{i=1}^{n-k} p(x_{i\tau},x_{(i+k)\tau};\bt), 
\end{equation*}
where $K$ is the number of lags to be included. The PL estimator is then given by
\begin{equation*}
    \hat{\bt}^{PL} \defeq \argmax_{\bt} \mathcal{L}(\bt).
\end{equation*}
In practice, for numerical stability on finite precision machines, we work with the log-likelihood $\log{\mathcal{L}(\bt)} = \sum_{k=1}^K \sum_{i=1}^{n-k} \log{p(x_{i\tau},x_{(i+k)\tau};\bt)}$ and its gradients. The problem is then to estimate $\log{p(x_s,x_t;\bt)}$ and $\nabla_{\bt} \log{ p(x_s,x_t;\bt)}= \frac{\nabla_{\bt}p(x_s,x_t;\bt)}{p(x_s,x_t;\bt)}$, which in turn reduces to estimating $p(x_s,x_t;\bt)$ and $\nabla_{\bt} p(x_s,x_t;\bt)$. The difficulty in applying the pairwise-likelihood (PL) methodology to real-valued trawl processes lies in accurately and efficiently estimating these two quantities, which are to be used in an iterative, gradient-based optimisation scheme. 

 Unlike in the case of the positive, integer-valued trawl processes studied in \cite{cl_integer_trawl}, the pairwise densities $p(x_s,x_t;\bt)$ are given by the integral in \eqref{eq:pairwise_likelihood} and not finite sums, hence are not generally available analytically. Further, the integrand can be ill-behaved, even for common distributions (see Example \ref{ex:pl_as_exp_for_gamma_levy_basis} with $0 < \alpha_1 < 1$), rendering off-the-shelf numerical integration inefficient. \cite{noven_phd_thesis} attempts to use numerical integration in the Fourier space to approximate the pairwise densities for latent trawl processes, as the Fourier transform of the trawl process is often known analytically, but finds that the likelihood maximization procedure does not converge. Other approaches to improve the accuracy of the PL approximation are investigated, amongst which lowering the error tolerance level of the numerical integration and using spline bases functions. These methods prove to be too computationally expensive. We have also failed to accurately approximate the likelihood function with the quadrature methods available in Scipy, the scientific computing library of Python. \cite{noven_phd_thesis} further analyzes the Gaussian case, for which pairwise densities are available in closed form and notices significant improvements in the convergence properties. We conclude that the lack of convergence in the likelihood optimization procedure is due to errors in the numerical integration and not the loss of efficiency from replacing the likelihood with the pairwise likelihood function. 
\subsection{Adapting inference by PL to real-valued trawls using Monte Carlo methods}\label{subsection:adapting_inference_by_pl}
We propose a novel approach and show that the pairwise likelihood function and its gradient can be accurately estimated with Monte Carlo (MC) samples. We discuss the formulation of the pairwise density as an MC estimator and provide a bias-variance analysis. Subsequently, we exploit the structure of this MC estimator to derive low variance gradient estimators in Subsection \ref{subsection:trawl_inference_as_a_stoch_optim_problem}.

By taking an expectation over $Z = L(A_t \cap A_s)$, the pairwise density $p(x_s,x_t;\bt)$ can be expressed in the following general form
\begin{align}
   p(x_s,x_t;\bt) &= \int p_{L(A_t\cap A_s)}(z) \,  p_{L(A_t\backslash A_s)}(x_t-z) \,  p_{L(A_s\backslash A_t)}(x_s-z) \mathrm{d}z \notag \\   &=  \ev[p_{L(A_t\backslash A_s)}(x_t-Z;\bt) \,  p_{L(A_s\backslash A_t)}(x_s-Z;\bt)] = \ev[f(Z,\bt)], \label{eq:pairwise_density_no_q}
\end{align}
where the function $f(z,\bt) = p_{L(A_t\backslash A_s)}(x_t-z;\bt) \,  p_{L(A_s\backslash A_t)}(x_s-z;\bt)$ depends implicitly on $x_s$ and $x_t$. Similar expressions are obtained if instead we take expectations over $L(A_t\backslash A_s)$ or $L(A_s \backslash A_t)$. For some L\'evy bases, $f$ contains terms that cancel out, allowing for simplification. We discuss one such example. Recall that $\bt_{\phi}$ parameterizes the trawl function $\phi$ of the trawl process $X$.
\begin{example}[Gamma L\'evy basis]\label{ex:pl_as_exp_for_gamma_levy_basis}
Let $x_s,x_t \in \R,\, L^{'} \sim \mathrm{Gamma}(\alpha,\beta)$ and $\rho(\cdot;\bt_{\phi})$ be the autocorrelation function parameterized by $\bt_{\phi}$.  Let $l_1 = \min{(x_s,x_t)}$ and $l_2 = \max{(x_s,x_t)}$ and further define $\alpha_0 = \alpha \rho(h;\bt_{\phi})$ and $\alpha_1 = \alpha (1 - \rho(h;\bt_{\phi}))$, where $h =  t-s$. With $\bt = (\alpha,\beta,\bt_{\phi})$, we have that
\begin{align}
    p(x_s,x_t;\bt) =& \int_0^{l_1} p_{L(A_t\cap A_s)}(z;\bt) \ p_{L(A_t \backslash A_s) }(x_t -z ;\bt)  \ p_{L(A_s \backslash A_t)}( x_s -z ;\bt) \mathrm{d}z \\ = &\int_0^{l_1} \text{Gamma}(z;\alpha_0,\beta) \,  \text{Gamma}(x_t-z;\alpha_1,\beta) \, \text{Gamma}(x_s-z;\alpha_1,\beta) \mathrm{d}z \notag \\
    =& \frac{\beta^{(\alpha+ \alpha_1)}  l_1^{\alpha-1}}{e^{\beta(l_1+l_2)}   \Gamma(\alpha)  \Gamma(\alpha_1)}\int_0^{l_1} \left[\frac{\Gamma(\alpha)}{l_1^{\alpha-1} \Gamma(\alpha_0)  \Gamma(\alpha_1)} z^{\alpha_0-1}  (l_1-z)^{\alpha_1-1}\right] \left(l_2-z\right)^{\alpha_1-1} e^{\beta z} \mathrm{d}z\notag \\ 
    =& \frac{\beta^{(\alpha+ \alpha_1)}  l_1^{\alpha-1}}{e^{\beta(l_1+l_2)}   \Gamma(\alpha)  \Gamma(\alpha_1)} \ev_{Z \sim \text{Beta}(\alpha_0,\alpha_1)}\left[ \left(l_2-l_1 Z\right)^{\alpha_1-1} e^{\beta l_1 Z}\right],
   \label{eq:when_beta_is_large}
\end{align}
where $Z \sim \text{Beta}(\alpha_0,\alpha_1)$ and $f(z,\bt) = \frac{\beta^{(\alpha+ \alpha_1)}  l_1^{\alpha-1}}{e^{\beta(l_1+l_2)}   \Gamma(\alpha)  \Gamma(\alpha_1)} \left(l_2-l_1 z\right)^{\alpha_1-1} e^{\beta l_1 z}$. \end{example}

The key aspect is that the density of $Z$, call it $q(\cdot;\bt)$, depends on $\bt$, hence interchanging differentiation and integration in \eqref{eq:pairwise_density_no_q} does not immediately produce an MC estimator for the gradient $\nabla_{\bt} p(x_s,x_t;\bt)$. More precisely, we have that $\nabla_{\bt} p(x_s,x_t;\bt)$ is equal to
\begin{align*}
\nabla_{\bt} \ev\left[f(Z,\bt)\right] =& \nabla_{\bt} \int f(z,\bt) q(z;\bt) \mathrm{d}z = \int  \left(\nabla_{\bt} q(z;\bt) f(z,\bt) + q(z;\bt) \nabla_{\bt}f(z,\bt) \right)    \mathrm{d}z
 \\ =& \int  f(z,\bt) \nabla_{\bt} q(z;\bt) \mathrm{d}z + \ev_{q(z;\bt)}\left[ \nabla_{\bt}f(z,\bt)\right].%
       \label{eq:gradient_of_likelihood}
\end{align*}
Nevertheless, having written $p(x_s,x_t;\bt)$ as an expectation, we can also estimate $\nabla_{\bt}p(x_s,x_t;\bt)$ with MC samples, e.g.~by using an MC approximation of $p(x_s,x_t;\bt)$ inside a finite difference approximation. Unfortunately, we find empirically that the gradient estimators constructed by finite differences have large bias and variances, making gradient descent optimization routines diverge. We note that increasing the number of samples $N$ used in the estimation of each pairwise density is computationally expensive, as the MC procedure is repeated for each of the pairs $\{(x_{i\tau},x_{(i+k)\tau})\}_{i=1}^{n-k}$ and for each $1 \le k \le K$. This is impractical when either $n$ or $K$ is large, i.e.~when a long path $x_{\tau},\ldots,x_{n\tau}$ is available, or when we observe a long memory trawl process, for which a large $K$ must be chosen to capture the slowly decaying autocorrelation. These are important limitations, as we seek a parameter inference method which scales well with more data and which can be applied to various settings, such as long memory processes.  To understand the issue, we do a bias-variance analysis for our estimators.   

Assume that $(U_i)_{i=1}^N$ and $(V_i)_{i=1}^N$ are vectors with iid entries of consistent and unbiased estimators for $\log{ p(x_s,x_t;\bt)}$, respectively $\nabla_{\bt} \log{ p(x_s,x_t;\bt)}$ and that $U_i >0$. We do not require vectors $(U_i)_{i=1}^N$ and $(V_i)_{i=1}^N$ to be independent. 
Let $\overline{U} = \sum_{i=1}^N U_i/N$ and $\overline{V}= \sum_{i=1}^N V_i/N$. Then the log and ratio estimators 
\begin{align*}
     \log{p(x_s,x_t;\bt)} \approx & \log{\left(\overline{U}\right)}\\
\nabla_{\bt}\log{p(x_s,x_t;\bt)} \approx  & \overline{V} / \overline{U}
\end{align*} are consistent, yet 
only asymptotically unbiased, with skewed distributions for which closed-form densities are not available. 
Fix $x_s$ and $x_t$. The bias and variance for $\log{(\overline{U})}$ are given by
\begin{equation}
    -\frac{1}{N}\frac{\Var{(U_1)}}{2\,  p(x_s,x_t;\bt)^2} + O\left(\frac{1}{N^2}\right) \text{ and }   \frac{1}{N} \frac{\Var{(U_1)}}{{p(x_s,x_t;\bt)}^2} + O\left(\frac{1}{N^2}\right), \text{ respectively};\label{eq:bias_var_expression_for_log_u}
\end{equation} similarly, the bias and variance for $ \overline{V}/\overline{U}$ are given by
\begin{align*} &\frac{1}{N} \frac{\nabla_{\bt}\log{{p(x_s,x_t;\bt)}}\Var{(U_1)}-  \Cov{(U_1,V_1)}}{{p(x_s,x_t;\bt)}^2}   + O\left(\frac{1}{N^2}\right) \text{ and } \\
&\frac{1}{N} \frac{\Var{\left(\nabla_{\bt} \log{p(x_s,x_t;\bt)}U_1 - V_1\right)}}{{p(x_s,x_t;\bt)}^2} + O\left(\frac{1}{N^2}\right),
\end{align*} 
respectively (see the Appendix 
for non-asymptotic, probabilistic bounds on the estimation error). 

It is tempting to conjecture that the loss of accuracy from the previous paragraph is mainly due to $\overline{U}$, as we take the log of $\overline{U}$ and also divide by $\overline{U}$, which is numerically unstable for triplets $(x_s,x_t, \bt)$ with $p(x_s,x_t;\bt)$ close to $0$. Nevertheless, we find empirically that when constructed via finite differences, the estimator $V_1$ has much higher variance than $U_1$ or $\nabla_{\bt} \log{p(x_s,x_t;\bt)}U_1$, hence the dominant terms in the bias and variance of the ratio estimator $\overline{V}/\overline{U}$ are
\begin{equation}
        \frac{1}{N} \frac{\Cov{(U_1,V_1)}}{p(x_s,x_t;\bt)^2} \text{ and } \frac{1}{N}\frac{\Var{(V_1)}}{p(x_s,x_t;\bt)^2}, \label{eq:bias_var_expression_for_v_over_u}
\end{equation}
respectively. We derive some insights which motivate the techniques employed later in the paper. Firstly, by the empirical observation above and for large $N$, we have that
\begin{equation*}
    \Var{\left(\overline{V}/\overline{U}\right)} \approx \frac{1}{N}\frac{\Var{(V_1)}}{p(x_s,x_t;\bt)^2} > \frac{1}{N}\frac{\Var{(U_1)}}{p(x_s,x_t;\bt)^2} \approx \Var{\left(\log{\overline{U}}\right)},
\end{equation*}
and the most important task is to employ a new methodology to estimate $\nabla_{\bt}p(x_s,x_t;\bt)$. Secondly, the dominant terms in the biases and variances of the log and ratio estimators are proportional to $\Var{(U_1)},\, \Cov(U_1,V_1)$ and $\Var{(V_1)}$, hence reducing the variances of $U_1$ and $V_1$ also reduces the dominant terms in the bias and variances of the log and ratio estimators by the same factor. We address the high variance of $U_1$ and $V_1$ separately and start with the latter. We use techniques from sensitivity analysis to derive better estimators of $\nabla_{\bt} p(x_s,x_t;\bt)$.
\subsection{Parameter inference as a simulation-based stochastic optimization problem}
\label{subsection:trawl_inference_as_a_stoch_optim_problem}
%
%
In the following, $f \colon \R \times \R^d \to \R$ is a $C^1$ function  and $q\left(\cdot;\bt\right)$ a density parameterized by a vector of parameters $\bt \in \R^d$ such that $\nabla_{\bt} q(\cdot;\bt)$ exists and is continuous at each $\bt$. We write $f(z,\bt)$ and $q(z;\bt)$ to emphasize $f$ as a function of two variables and $q$ as a density in $z$, parameterized by $\bt$. Throughout this paper, the random variable $Z$ has density $q(\cdot;\bt)$ and we write $\ev_{q(z;\bt)}\left[f(z,\bt)\right]$ and $\ev[f(Z,\bt)]$ interchangeably; we use the former when the dependency of $q$ on $\bt$ needs to be explicitly stated and the latter otherwise. If $z$ also depends on $\bt$, we write the total derivative (TD) of $f$ with respect to $\bt$ as  $\nabla_{\bt}^{\textrm{TD}}f = \pdv{f}{z} \nabla_{\bt}z + \nabla_{\bt}f$. Finally, for a family of distributions parameterized by $\bt,$ we denote by $\mathcal{D}(\cdot;\bt)$ the density of the distribution $\mathcal{D}$ with parameters $\bt$. For example, $\mathcal{N}(z;\mu,\sigma^2)$ is the density of the normal distribution with mean $\mu$ and variance $\sigma^2$, evaluated at $z$, where $\bt = (\mu,\sigma^2)$. 

As seen before, the pairwise density $ p(x_s,x_t;\bt)$ can be expressed in the following general form
\begin{equation}
    \ev[f(Z,\bt)] =  \ev_{q(z;\bt)}[f(z,\bt)], \label{eq:general_form}
\end{equation}
where $Z = L (A_t \cap A_s)$ has density $q(\cdot;\bt)$ and $f(z,\bt) = p_{L(A_t\backslash A_s)}\left(x_t - z\right) \, p_{L(A_s\backslash A_t)}\left(x_s -z\right)$. Assuming there are no obvious cancellations, such as the ones in Example \ref{ex:distr_gamma}, there are two cases: if $L^{'}$ is supported on the real line, the distribution $q$ under which we take expectations in \eqref{eq:general_form} is that of $L(A)$ for some set $A$; if $L^{'}$ is supported on the positive real line, $f(\cdot;\bt)$ is supported on $[0,\min{(x_s,x_t)}]$ and thus $q$ is the truncation to $[0,\min{(x_s,x_t)}]$ of the distribution of $L(A)$ for some set $A$. 

We next study the estimation of $\nabla_{\bt}\ev[f(z,\bt)]$ by MC methods for the class of distributions $q$ specified above. By interchanging integration and differentiation,  we obtain
\begin{align}
       \nabla_{\bt} \ev_{q(z;\bt)}\left[f(z,\bt)\right]=& \int \nabla_{\bt}\left(q(z;\bt) f(z,\bt) \right) \mathrm{d}z = \int  \left(\nabla_{\bt} q(z;\bt) f(z,\bt) + q(z;\bt) \nabla_{\bt}f(z,\bt) \right)    \mathrm{d}z\notag\\
       =& \int  f(z,\bt) \nabla_{\bt} q(z;\bt) \mathrm{d}z + \ev_{q(z;\bt)}\left[ \nabla_{\bt}f(z,\bt)\right]  .%
       \label{eq:gradient_of_likelihood}
\end{align}
The two terms correspond to the dependency on $\bt$ of the sampling measure $q$ and function $f$, respectively. There are at least three different methodologies for estimating the first term from the above equation with samples: the score function (SF), measure-valued gradients (MVG) and the pathwise gradients (PG), where we follow the terminology from \cite{reparam_tutorial}. We first present the three methods and illustrate each in the Gaussian case $q(z;\bt) = \mathcal{N}(z;\mu,\sigma^2)$, where $\bt = (\mu,\sigma) \in \R^2$, as in this case the necessary formulae are available analytically. We then provide an extensive analysis of the properties and practical implementations of these methodologies.

\textbf{Score function (SF)} The first method is the most general one, only requiring access to the SF
\begin{equation*}
\nabla_{\bt} \ev_{q(z;\bt)}\left[f(z,\bt)\right] = \ev_{q(z;\bt)}\left[ f(z,\bt) \nabla_{\bt} \log{q(z;\bt)} + \nabla_{\bt}f(z,\bt)  \right].
\end{equation*}
In the Gaussian case $q(z;\bt) = \mathcal{N}(z;\mu,\sigma^2)$, 
 the gradient is given by
 \begin{equation}
  \ev_{q(z;\bt)}\left[f(z,\bt) \begin{pmatrix} (z-\mu)/\sigma^2 \\ (z-\mu)^2/(4\sigma^3)- 1/(2\sigma)
    \end{pmatrix}\right] +    \ev_{q(z;\bt)}\left[\nabla_{\bt} f(z,\bt)\right]. \label{eq:SF_estimator_gaussian_case}
\end{equation}
\textbf{Measure-valued gradients (MVG)} Alternatively, the MVG method uses the decomposition of the signed measure induced by the unnormalized density $\left(\nabla_{\bt} q(\cdot;\bt)\right)_i$ into $c^{+}_i q^{+}_i(\cdot; \bt) - c^{-}_i q^{-}_i(\cdot;\bt)$, where $c^{+}_i, \, c^{-}_i$ are positive constants, $q^{+}_i(\cdot;\bt), \, q^{-}_i(\cdot;\bt)$ are 
probability measures parameterized by $\bt$ for $i=1,\ldots,d$. In shorthand, we have 
\begin{equation*}
\nabla_{\bt}q(\cdot;\bt) = \bm{c}^{+} \bm{q}^{+}(\cdot;\bt) -  \bm{c}^{-} \bm{q}^{-}(\cdot;\bt),
\end{equation*}
where $\bm{c}^{+}$ and $\bm{c}^{-}$ are $d$-dimensional vectors with positive entries, $\bm{q}^{+}(\cdot;\bt)$ and $\bm{q}^{-}(\cdot;\bt)$ are $d$-dimensional vectors of probability measures and the products are componentwise. We then have that
\begin{equation*}
    \nabla_{\bt} \ev_{q(z;\bt)}\left[f(z,\bt)\right] =   \bm{c}^{+} \ev_{\bm{q}^{+}(\bm{z}^{+};\bt)}\left[f(\bm{z}^{+},\bt)\right] -\bm{c}^{-} \ev_{\bm{q}^{-}(\bm{z}^{-};\bt)}\left[f(\bm{z}^{-},\bt)\right] + \ev_{q(z;\bt)}\left[\nabla_{\bt}f(z;\bt)\right].
\end{equation*}
In the Gaussian case $q(z;\bt) = \mathcal{N}(z;\mu,\sigma^2)$, by \eqref{eq:gradient_of_likelihood}, the partial derivative with respect to $\sigma$ is given by
\begin{equation*}
\pdv{}{\sigma} \ev_{\mathcal{N}(z;\mu,\sigma)}\left[ f(z,\bt)\right] =  
\int f(z,\mu,\sigma) \pdv{}{\sigma}\mathcal{N}(z;\mu,\sigma^2) \mathrm{d}z + \ev_{\mathcal{N}(z;\mu,\sigma)}\left[ \pdv{f}{\sigma}(z,\mu,\sigma) \right],
\end{equation*}
where the first term can be written as 
\begin{align}
&\int f(z,\mu,\sigma) \mathcal{N}(z;\mu,\sigma^2) \left( \frac{(z-\mu)^2}{\sigma^3}- \frac{1}{\sigma}\right) \mathrm{d}z \notag\\
= & \frac{1}{\sigma}\int f(z,\mu,\sigma) \left( \frac{(z-\mu)^2}{\sigma^2}\mathcal{N}(z;\mu,\sigma^2)\right) \mathrm{d}z - \frac{1}{\sigma} \int f(z,\mu,\sigma)  \mathcal{N}(z;\mu,\sigma^2)  \mathrm{d}z \notag \\
=&\frac{1}{\sigma} \ev_{\mathcal{M}(z;\mu,\sigma^2)}\left[f(z)\right] - \frac{1}{\sigma} \ev_{\mathcal{N}(z;\mu,\sigma^2)}\left[f(z)\right] \label{eq:MVG_estimator_gaussian_case},
\end{align}
where $\mathcal{M}(z;\mu,\sigma^2) = \frac{(z-\mu)^2}{\sigma^2}\mathcal{N}(z;\mu,\sigma^2)$ is the density of a doubled-sided Maxwell distribution (see the Appendix). 
Then the constants in the MVG formulation are $\frac{1}{\sigma}$ and the two sampling measures are double-sided Maxwell and Gaussian. We leave to the reader the similar and more tedious derivation for the partial derivative with respect to $\mu$.

Note that MVG can be applied even when $\bt \mapsto q(\cdot;\bt)$ is not differentiable. Then $\nabla_{\bt}q(\cdot;\bt)$ is understood as a weak derivative and a decomposition into positive and negative parts exists by the Hahn-Jordan theorem. Multiple decompositions exists even in simple cases (see \cite{pflug2012optimization} for an example and for a rigorous treatment of calculus with weak derivatives). As with the SF, the MVG does not require a smooth $f$. 

\textbf{Pathwise gradients (PG)} By comparison, the PG method requires $f$ to be differentiable with respect to $z$. The idea is to replace $\ev_{q(z;\bt)}\left[ f(z,\bt) \nabla_{\bt} \log{q(z;\bt)}\right]$ with  $\ev_{z(x;\bt)}\left[\pdv{f}{z}(z,\bt) \nabla_{\bt}z \right]$, by removing the dependency of $q$ on $\bt$ and 'pushing' it into $f$, where $\nabla_{\bt}z$ is a pathwise vector-valued gradient to be defined. \cite{rubinstein1992sensitivity} calls this the push-in method. Then
\begin{equation*}\nabla_{\bt} \ev_{q(z;\bt)}\left[f(z,\bt)\right] = \ev_{q(z;\bt)}\left[ \pdv{f}{z}(z,\bt) \nabla_{\bt} z + \nabla_{\bt}f(z,\bt)  \right].
\end{equation*}
In the Gaussian case $q(z;\bt) = \mathcal{N}(z;\mu,\sigma^2)$, we take advantage of the location-scale property to compute the pathwise gradient $\nabla_{\bt}z$. Define $\varepsilon = \frac{z-\mu}{\sigma}$. By the chain rule and by taking the total derivative with respect to $\bt =(\mu,\sigma^2)$, we obtain that
\begin{align*}\nabla_{\bt} \ev_{\mathcal{N}(z;\mu,\sigma^2)}\left[f\left(z,\bt \right)\right] =& \nabla_{\bt} \ev_{\mathcal{N}( \varepsilon;0,1)}  \left[  f(\mu + \sigma \varepsilon,\bt) \right]= \ev_{\mathcal{N}( \varepsilon;0,1)}  \left[ \nabla_{\bt}^{\textrm{TD}} f(\mu+\sigma\varepsilon,\bt) \right]  \\   
=&   \ev_{\mathcal{N}(\varepsilon;0,1)} \left[\pdv{f}{z}\left(\mu+\sigma\varepsilon,\bt\right) \nabla_{\bt}\left(\mu+\sigma\varepsilon\right) \right] + \ev_{\mathcal{N}(\varepsilon;0,1)}  \left[\nabla_{\bt} f(\mu+\sigma\varepsilon,\bt)\right] \\
=& \ev_{\mathcal{N}(\varepsilon;0,1)} \left[\pdv{f}{z}\left(\mu+\sigma\varepsilon,\bt\right) \begin{pmatrix} 1 \\ \varepsilon
\end{pmatrix} \right] + \ev_{\mathcal{N}(\varepsilon;0,1)}  \left[\nabla_{\bt} f(\mu+\sigma\varepsilon,\bt)\right].
\end{align*} 
Finally, by the change of variable formula,
\begin{equation}
\nabla_{\bt} \ev_{\mathcal{N}(z;\mu,\sigma^2)}\left[f\left(z,\bt \right)\right] = \ev_{q(z;\bt)}\left[\pdv{f}{z}(z,\bt) \begin{pmatrix} 1 \\ (z-\mu)/\sigma
\end{pmatrix}\right] + \ev_{q(z;\bt)}\left[\nabla_{\bt} f(z,\bt)\right],\label{eq:PG_estimator_gaussian_case}
\end{equation}
and the pathwise gradient $\nabla_{\bt}z$ is equal to $\left(1,\frac{z-\mu}{\sigma}\right)$. By considering the deterministic and differentiable mapping
\begin{align*}
  \mathcal{S}\colon  \R \times \R^d &\longrightarrow \R \qquad (d=2)\\
    (z;\bt) = (z;(\mu,\sigma)) &\longmapsto  \frac{z-\mu}{\sigma}, 
\end{align*} 
we were able to transform a sample $z$ from $\mathcal{N}(\mu,\sigma^2)$ into a sample $\varepsilon$ from $\mathcal{N}(0,1)$, which does not depend on $\bt$. Let $\mathcal{S}^{-1}$ be the inverse of $\mathcal{S}$ with respect to its first argument. Then $\varepsilon= \mathcal{S}(z,\bt)$ implies $z = \mathcal{S}^{-1}(\varepsilon;\bt)$ and $\nabla_{\bt}z = \nabla_{\bt}\mathcal{S}^{-1}(\epsilon;\bt)$. In general, the trick lies in determining a base distribution which does not depend on $\bt$, which can be differentiably transformed into the required distribution and for which the pathwise gradient $\nabla_{\bt}z$ can be computed efficiently. 

\textbf{Comparison of gradient estimation methodologies}

According to \cite{kleijnen1996optimization}, the use of the SF to account for the gradient of the sampling measure with respect to the design parameters $\bt$ was pioneered independently by different researchers in the late 1960s, amongst which we mention \cite{miller1967}, \cite{mikhailov} and \cite{rubinstein1969}. Nevertheless, we determine in a simulation study that, in the context of inference for trawl processes, the estimator for $\ev_{q(z;\bt)}\left[f(z,\bt) \nabla_{\bt} \log{q(z;\bt)}\right]$ has high variance, just as the finite difference estimator, rendering gradient-based optimization practically unfeasible. 
Although there is no universal ranking of the three estimators, as shown in the simulation studies from \cite{pflug2012optimization,FU2006575,reparam_tutorial}, MVG and PG tend to perform better. We compare the latter two methods based on the range of distributions for which they can be used, degree of variance reduction reported in the literature, compatibility with other variance reduction methods and computational cost and ease of implementation.

There are multiple difficulties in the practical implementation of MVG. Firstly, this method requires knowledge of the decomposition of $\nabla_{\bt}q$ into $\bm{q}^{+}$ and $\bm{q}^{-}$, limiting its applicability in the real-valued case to few L\'evy seeds. The variance of the resulting estimator depends on the chosen decomposition and finding the optimal one is usually not possible. One approach is to use the Hahn-Jordan decomposition when available, for which $\bm{q}^{+}$ and $\bm{q}^{-}$ have disjoint supports, yet this is not optimal in general \citep[cf.][Examples 4.21 and 4.28]{pflug2012optimization}. 
Finally, note that we require twice as many samples to estimate the gradients and that although MVG can provide low variance gradients, a case-by-case implementation is required.

These difficulties can be circumvented in the real-valued case with the PG method, which exploits the differentiability of $f$ in $z$ to remove the dependency of $q$ on $\bt$. The PG estimator has the simplest form (see \eqref{eq:SF_estimator_gaussian_case}, \eqref{eq:MVG_estimator_gaussian_case} and \eqref{eq:PG_estimator_gaussian_case}) and performs significantly better than the SF in a variety of stochastic optimization tasks, such as training variational autoencoders \citep{kingma2013auto} and Bayesian logistic regression \citep{fan2015fast}. In the following section, we show that PG can be efficiently implemented on a computer for a large class of real-valued distributions and that the method can be incorporated in Automatic Differentiation (AD) engines, thus avoiding tedious differentiation by hand. Further, we show that PG can be used in conjunction with control variates, and that even using a Taylor polynomial of degree $1$ as control variate removes the bias and provides significant variance reduction. Based on the above, we select PG as our candidate methodology for estimating the gradients of the pairwise densities of real-valued trawl processes and demonstrate the major improvement over the SF methodology in a simulation study. Nevertheless, PG are generally not available for discretely supported distributions, such as those of integer-valued trawl processes. We address this case in the supplementary material and develop the theory for hybrid gradient estimators, which combines PG and MVG. We formalize a chain rule for stochastic transformations for which at least one of PG and MVG are available and develop a unified composite likelihood inference for both integer-valued and real-valued trawl processes in Section \ref{supplementary_section:measure_valued_grad}.


\section{Variance reduction methods}\label{section:variance_reduction_methods}
In the following, we formally define, then efficiently and accurately compute the pathwise gradients $\nabla_{\bt}z$, which are to be used inside an iterative, gradient-based optimization scheme. We then discuss the use of control variates for the estimation of both the pairwise density and its gradient, thus combining two variance reduction methods for the estimation of the gradient. Finally, we discuss the computer implementation and describe the limitations of PG and control variates in the real-valued case. A simulation study demonstrating the effectiveness of our methods is presented in Subsection \ref{subsection:variance_reduction_sim_study}.

 We use the $\ev_{q(z;\bt)}[f(z,\bt)]$ and $\ev[f(Z,\bt)]$ interchangeably (recall the notation at the beginning of Subsection \ref{subsection:trawl_inference_as_a_stoch_optim_problem}). In the former notation, the dependency of $q$ on $\bt$ is explicit, whereas in the latter the dependency is implicit.
\subsection{Pathwise gradients}
\label{subsection:pg}
The PG methodology was initially introduced under the name `infinitesimal perturbation analysis' by \cite{Ho1983559} for the optimization of discrete queuing models and later expanded upon by \cite{pflug2012optimization} and \cite{glasserman}. Recently, it has become more widely used in the deep learning community, under different names:  reparameterization trick in \cite{kingma2013auto}, stochastic back-propagation rule in \cite{rezende2014stochastic} and implicit reparameterization gradients in \cite{implicit_figurnov}. As noted before, by interchanging differentiation and integration, we obtain that
\begin{equation*}
   \nabla_{\bt} \ev_{q(z;\bt)}\left[f(z,\bt)\right]=  \int \nabla_{\bt}\left( f(z,\bt) q(z;\bt) \right)\mathrm{d}z =   \int f(z,\bt) \nabla_{\bt} q(z;\bt) \mathrm{d}z + \ev_{q(z;\bt)}\left[\nabla_{\bt}f(z,\bt)\right].  
\end{equation*}
The term $\int f(z,\bt) \nabla_{\bt} q(z;\bt) \mathrm{d}z $ only accounts for the dependency of the sampling measure $q$ on $\bt$, and not that of $f$ on $\bt$. Thus, to deal with this term, it is enough to obtain low variance gradients for $\nabla_{\bt}\ev_{q(z;\bt)}\left[g(z)\right]= \int g(z) \nabla_{\bt} q(z;\bt)\mathrm{d}z$, where $g$ is solely a function of $z$. In the following, we define the pathwise gradient $\nabla_{\bt}z$ by defining $z$ as a deterministic, differentiable function of the parameters $\bt$ and of a sample $\varepsilon$ from the uniform distribution on $[0,1]$, which does not depend on $\bt$. We first compute the pathwise gradients for the class of distributions $q$ with tractable probability density and cumulative distribution functions and then extend to the class of certain transformations of tractable distributions. 

To begin with, let $F(\cdot;\bt)$ be the cumulative distribution function corresponding to $q(\cdot;\bt)$ and $\mathcal{U}(\cdot;0,1)$ the density of the uniform distribution on $[0,1]$. By interchanging differentiation and integration and by the chain rule, we obtain that
\begin{align}
\nabla_{\bt}\ev_{q(z;\bt)}\left[g(z)\right]&= \nabla_{\bt} \ev_{ \mathcal{U}(\varepsilon;0,1)}\left[ g \circ F^{-1}\left(\varepsilon;\bt\right)\right] = \ev_{ \mathcal{U}(\varepsilon;0,1)}\left[g'\left(F^{-1}\left(\varepsilon;\bt\right)\right) \nabla_{\bt} F^{-1}(\varepsilon;\bt)\right]\notag \\
&= \ev_{q(z;\bt)}\left[g'(z) \nabla_{\bt}F^{-1}\left(F(z;\bt);\bt\right)\right] = \ev_{q(z;\bt)}\left[g'(z)\nabla_{\bt}z\right]
,\label{eq:implicit_grad}
\end{align}
where $\nabla_{\bt}z \defeq \nabla_{\bt}F^{-1}\left(F(z;\bt);\bt\right) = \nabla_{\bt} F^{-1}(\varepsilon;\bt)$. Note that the samples from $q(\cdot;\bt)$ do not have to be generated via inversion. Although initially coupled through the quantile function $F^{-1}(\cdot;\bt)$ in the left-hand side of \eqref{eq:implicit_grad}, the sampling and differentiation are decoupled in the right-hand side of the same equation, i.e.~in $\ev_{q(z;\bt)}\left[g'(z)\nabla_{\bt}z\right]$. We can use the same pathwise gradient regardless of the sampling method. Next, we explain how to accurately and efficiently compute the pathwise gradient for distributions with tractable density $q(\cdot;\bt)$  and cumulative distribution function $F(\cdot;\bt)$.

The quantile function $F^{-1}$ is often not available in closed form and is calculated by root-finding methods. \cite{Knowles2015StochasticGV} uses this inside a finite difference quotient to approximate $\nabla_{\bt}F^{-1}$. \cite{pflug2012optimization} shows in Chapter 3.2.3 that $\nabla_{\bt}F^{-1}(\varepsilon;\bt) = -\frac{\nabla_{\bt} F(\varepsilon;\bt)}{q(z;\bt)},$ which only requires knowledge of the density and of the gradient of the cumulative distribution function. When the latter is not available, \cite{Pathwise_Derivatives_Beyond_the_Reparameterization_Trick_Jankowiak} uses closed-formed expressions such as Taylor expansions, Lugannani-Rice saddlepoint expansions and rational polynomial approximations. Concurrently, \cite{implicit_figurnov} applies forward-mode automatic differentiation to the numerical procedure which approximates the cumulative distribution function, therefore extending the above method to any distribution $q$ with numerically tractable cumulative distribution function. Both \cite{Pathwise_Derivatives_Beyond_the_Reparameterization_Trick_Jankowiak} and \cite{implicit_figurnov} improve significantly on the computational time and accuracy of \cite{Knowles2015StochasticGV}. We work with the last two methodologies when inferring the parameters of the trawl processes, as their implementations are already available in Jax and TensorFlow. To derive the formula from \cite{pflug2012optimization}, we remind the reader of the notation $\nabla_{\bt}z = \nabla_{\bt} F^{-1}(\varepsilon;\bt)$. By keeping track of the dependency of $z = F^{-1}(\varepsilon;\bt)$ on $\bt$ when taking the gradient of the equation $\varepsilon = F(z;\bt)$ with respect to $\bt$, we obtain that
\begin{equation*}
    \nabla_{\bt} \varepsilon = 0 = \nabla_{\bt}^{\mathrm{TD}} F(z;\bt)= q(z;\bt) \nabla_{\bt}z  + \nabla_{\bt}F(z;\bt)\Rightarrow \nabla_{\bt}z    = -\frac{\nabla_{\bt} F(z;\bt)}{q(z;\bt)}.
\end{equation*}
Finally, our estimator is 
\begin{equation*}
\nabla_{\bt} \ev_{q(z;\bt)}\left[g(z)\right] = \ev_{q(z;\bt)}\left[g'(z)\nabla_{\bt}z\right],
\end{equation*}
where $\nabla_{\bt}z    = -\frac{\nabla_{\bt} F(z;\bt)}{q(z;\bt)}$. The variance properties of this estimator are analyzed in \cite{fan2015fast} and \cite{gal} in the Gaussian case, while \cite{glasserman} and \cite{CUI2022199} extend the study to other distributions. In particular, Chapter $7.2$ of \cite{glasserman} provides a bound on the variance of the estimator in terms of the Lipschitz constant of $g$ and \cite{CUI2022199} provides sufficient conditions under which the pathwise estimator has a lower variance, although these conditions are hard to check in practice. 
Despite the limited theoretical analysis, the PG method has already been successfully employed in a variety of stochastic optimisation tasks such as estimation of the greeks in finance \citep[Chapter~7.4]{glasserman} and policy learning in reinforcement learning \citep{williams}. An extensive list of applications can be found in \cite{reparam_tutorial}.
\begin{remark}\label{remark:no_loss_of_gen_when_using_g} The PG method is applicable even if $g(z)$ is replaced with $f(z,\bt)$, which also depends on $\bt$. Indeed, we have that
\begin{align*}
     \nabla_{\bt} \ev_{q(z;\bt)}\left[f(z,\bt)\right] =& \nabla_{\bt} \ev_{\mathcal{U}(\varepsilon;0,1)} \left[f\left(F^{-1}(\varepsilon;\bt)),\bt\right)\right] = \ev_{\mathcal{U}(\varepsilon;0,1)} \left[\nabla_{\bt}^{\mathrm{TD}}f\left(F^{-1}(\varepsilon;\bt)),\bt\right)\right] \\
    =&  \ev_{ \mathcal{U}(\varepsilon;0,1)} \left[\pdv{f}{z}\left(F^{-1}(\varepsilon;\bt)),\bt\right) \nabla_{\bt}F^{-1}(\varepsilon;\bt) \right] + \ev_{\mathcal{U}(\varepsilon;0,1)} \left[\nabla_{\bt}f\left(F^{-1}(\varepsilon;\bt)),\bt\right)\right]  \\  =& \ev_{q(z;\bt)}\left[\pdv{f}{z}(z,\bt) \nabla_{\bt}z\right] + \ev_{q(z;\bt)}\left[\nabla_{\bt}f(z,\bt)\right] ,
\end{align*}
where $\nabla_{\bt}z = \nabla_{\bt}F^{-1}(\varepsilon;\bt) = \frac{\nabla_{\bt}F(z;\bt)}{q(z;\bt)}$.
\end{remark}
We discuss some particular cases which are relevant to the parameter inference of trawl process: Examples \ref{ex:gaussian}-\ref{ex:gamma} give pathwise gradients for random variables with numerically tractable density and cumulative distribution functions; Examples \ref{ex:beta} and \ref{ex:truncated} extend the pathwise gradient to deterministic mappings of such random variables; finally, Example \ref{ex:nig} extends the above theory to certain stochastic transformations of tractable distributions through a chain rule for the pathwise gradients of conditional samples.

\begin{example}[Gaussian distribution]\label{ex:gaussian}
If $q(z;\bt)= \mathcal{N}(z;\mu,\sigma^2)$, $\nabla_{\bt}z = 
-\frac{\nabla_{\bt}F(z;\bt)}{q(z;\bt)} = (1,\frac{z-\mu}{\sigma})$, which agrees with the pathwise gradient given by the location-scale transformation from \eqref{eq:PG_estimator_gaussian_case}.
\end{example}
\begin{example}[Inverse Gaussian distribution]
If $q(z;\bt) = \textrm{IG}(z;\mu,\delta)$, then $F(z;\bt) = \newline \Phi{\left(\sqrt{\lambda }\left({\frac {\sqrt{x}}{\mu }}- \frac{1}{\sqrt{x}}\right)\right)} +\exp \left({\frac {2\lambda }{\mu }}\right)\Phi{\left(-{\sqrt{\lambda}}\left({\frac {\sqrt{x}}{\mu }}+\frac{1}{\sqrt{x}}\right)\right)}$
, where $\Phi$ is the cdf of $\mathcal{N}(0,1)$. Then both $\nabla_{\bt}F$ and $q$ are available in closed form. 
\end{example}
\begin{example}[Gamma distribution]\label{ex:gamma}
If $q(z;\bt) = \textrm{Gamma}(z;\alpha,\beta)$, then $F(z;\bt)$ is the regularized Gamma function. Since $\textrm{Gamma}(\alpha,\beta) = \frac{1}{\beta}\textrm{Gamma}(\alpha,1)$, $\nabla_{\beta}z$ can be obtained as for the scale parameter in the Gaussian case. For $\nabla_{\alpha}z$, $q$ is available analytically and $\nabla_{\alpha}F$ can be obtained as discussed above, as in \cite{implicit_figurnov} or \cite{Pathwise_Derivatives_Beyond_the_Reparameterization_Trick_Jankowiak}.
\end{example}
\begin{example}[Beta distribution]\label{ex:beta}
If $X_1\sim \textrm{Gamma}(\alpha,1)$ and $X_2 \sim \textrm{Gamma}(\beta,1)$ are independent, then $X = X_1 / (X_1+X_2) \sim \textrm{Beta}(\alpha,\beta)$. Hence gradients for the Beta distribution can be obtained from these of the Gamma distribution, by the usual product and chain rules from calculus.
\end{example}
Note that if $L^{'}$ is supported on the positive-real line, such as in the two previous examples, the sampling measure $q$ is actually the truncation of the distribution of $L(A)$ for some set $A$ to the interval $[0,b]$, where $b = \min{(x_s,x_t)}$. This can easily be handled, as long as $F$ is available numerically.
\begin{example}[Truncated distributions]\label{ex:truncated} Consider a distribution supported on the positive real line for which the density $q$, cdf $F$ and gradient $\nabla_{\bt}F$ are available and let $\tilde{q},\, \tilde{F}$ be the corresponding restrictions to $[0,b].$ Then
\begin{equation*}
-\frac{\nabla_{\bt}\tilde{F}(z;\bt)}{\tilde{q}(z;\bt)} = \left( \frac{F(z;\bt)}{q(z;\bt)} \frac{\nabla_{\bt}F(b;\bt)}{F(b;\bt)} - \frac{\nabla_{\bt}F(z;\bt)}{q(z;\bt)}\right)\mathbbm{1}_{0 < z < b }.
\end{equation*}
\end{example}
Based on the above building blocks, we can develop calculus rules to determine $\nabla_{\bt} z$ for more general distributions, even without numerically tractable expressions for the density and cdf. The key property to generalizing the method of pathwise gradients is the existence of smooth, invertible functions such as the quantile or the location-scale transformations, which sequentially remove the dependency of $q$ on $\bt$.
\begin{definition}[\textbf{Standardization functions}]\label{def:standardization_function}
We say $\mathcal{S} \colon \R \times \R^d \to \R$ is a standardization function for the density $q(\cdot;\bt)$ of $Z$ if the law of $\mathcal{S}(Z;\bt)  \eqdef \mathcal{E}$ has density $\cal{T}$ which does not depend on $\bt$, if $\calS(\cdot;\bt)$ is invertible with respect to its first argument and further if both $\calS$ and $\mathcal{S}^{-1}$ are $C^1$ in both the argument $z$ and parameter $\bt$, where $\mathcal{S}^{-1}$ is the inverse with respect to the first argument. We then have 
    \begin{align}
    \nabla_{\bt} \ev_{q(z;\bt)}\left[g(z)\right] =& \nabla_{\bt} \ev_{\cal{T}(\varepsilon)}\left[g \circ \calS^{-1}(\varepsilon;\bt)\right] = \ev_{\cal{T}(\varepsilon)}\left[g'\circ\calS^{-1}(\varepsilon)\nabla_{\bt}\calS^{-1}(\varepsilon;\bt)\right] \label{eq:interchange_diff_and_exp_stand_function}\\ = & \ev_{q(z;\bt)}\left[g'(z) \nabla_{\bt}\calS^{-1}(\calS(z;\bt),\bt)\right].\notag
    \end{align} 
\end{definition}
\begin{lemma}[\textbf{Chain rule for pathwise gradients}]\label{lemma:chain_rule_pathwise_grads_with_cal_S} Let $\mathcal{S}_1,\ldots,\mathcal{S}_n  \colon \R \times \R^d$ be invertible with respect to the first argument and $C^1$ as above and $\varepsilon,\varepsilon_1,\ldots,\varepsilon_n$ be random random variables with densities $\mathcal{T}, \, \mathcal{T}_1,\ldots,\mathcal{T}_n$ which do not depend on $\bt$. Further let $Z$ be a random variable with density $q(\cdot;\bt)$ and define $Y_n = Z$, $Y_{n-1} = \mathcal{S}_n(\varepsilon_n;Y_n), \ldots, Y_1 = \mathcal{S}_2(\varepsilon_2;Y_2)$ and $\varepsilon = \mathcal{S}_1(\varepsilon_1;\bt)$. Then $S = S_n \circ \ldots \circ S_1$ is a standardization function and 
\begin{equation}
\nabla_{\bt} \ev_{q(z;\bt)}[g(z)] = \ev_{q(z,y_{n-1},\ldots,y_1;\bt)}\left[f(z) \ \nabla_{y_{n-1}}z \ \nabla_{y_{n-2}}y_{n-1} \ldots \nabla_{y_1}y_2\ \nabla_{\bt}y_1\right],  \label{eq:chain_rule_pathwise_grads_with_cal_S}
\end{equation}
where $z=y_n,\, q(z,y_{n-1},\ldots,y_1;\bt)$ is the joint density of $(Z,Y_{n-1},\ldots,Y_1)$ and $\nabla_{y_i}y_{i+1} = \nabla_{y_{i+1}}\mathcal{S}_{i+1}(\varepsilon_{i+1};y_{n+1})$, for $1 \le i \le n-1$.
\end{lemma}
Note that the cumulative distribution function is a standardization function for any continuous random variable, hence it satisfies Definition \ref{def:standardization_function} and generalizes \eqref{eq:implicit_grad}. Further, Lemma \ref{lemma:chain_rule_pathwise_grads_with_cal_S} can be extended to allow for standardization functions $S_i$ which depend explicitly on $\bt$,i.e.~$\mathcal{S}_i\left(\varepsilon_i;\left(Y_2,\bt\right)\right)$ for $1 \le i \le n$ by adding some extra terms to \eqref{eq:chain_rule_pathwise_grads_with_cal_S}. The applicability of the result is immediate: if $Z$ can be generated by sequential conditional sampling and each of these samples has numerically tractable pathwise gradients, then the pathwise gradient for $Z$ is also numerically tractable. 
\begin{example}[Normal-inverse Gaussian distribution]    \label{ex:nig}
If $Y \sim \textrm{IG}(1/\gamma,1)$, then $Z|Y \sim \mathcal{N}(\mu + \beta y, \sigma^2 y^2)$ has a $\textrm{NIG}(\mu,\sigma,\alpha,\beta)$ law, where $\gamma = \sqrt{\alpha^2-\beta^2}$. Hence gradients for the $\textrm{NIG}$ distribution can be obtained from these of the Inverse Gaussian and Gaussian distributions, by the chain rule.
\end{example}
In general, we can reparameterize any distribution from the class of normal variance-mean mixtures, by the chain rule, as long as we have pathwise gradients for the mixing distribution (see Definition \ref{def:normal_variance_mean_mixture}). 
\begin{remark}[Beyond the quantile function]
Finding standardization functions other than the ones above is not trivial. To interchange differentiation and expectations in \eqref{eq:interchange_diff_and_exp_stand_function}, we require two critical assumptions: $\calS$ and $\calS^{-1}$ are differentiable and further $\cal{T}$ does not depend on $\bt$. Relaxing these assumptions allows us to move past the obvious use of the quantile functions, or of a sequence of conditional quantile functions, as standardization functions. We mention two such extensions. \cite{naesseth2017reparameterization} proposes reparameterization gradients through acceptance-rejection sampling algorithms, in which $\calS$ is not smooth. \cite{generalized_reparam_gradient} proposes generalized reparameterization gradients, in which $\cal{T}$ depends weakly on $\bt$, i.e.~the first moment of $\cal{T}(\cdot;\bt)$ does not depend on $\bt$. These methods extend the applicability of pathwise gradients to an even wider class of distributions.
\end{remark}
In the real-valued case, the limitations of the composite likelihood approach stem from the lack of a numerically tractable density for the marginal distribution of the trawl process (for example if $L^{'}$ has a generalized inverse Gaussian distribution), and more generally from $\pdv{f}{z}$ or $\nabla_{\bt}f$ being difficult to compute, and not from the lack of pathwise gradients. In the discrete case, there are distributions for which $\nabla_{\bt}z$ does not exist (e.g.~Poisson). For such cases, we use MVG and, more generally, hybrid estimators which combine MVG and PG (see Section \ref{supplementary_section:measure_valued_grad}). Further, we extend the stochastic chain rule from Lemma \ref{lemma:chain_rule_pathwise_grads_with_cal_S} to the setting where each conditional sample $Y_i|Y_{i-1}$ has either MVG or PG, providing a unified approach to gradient estimation.

\subsection{Linear control variates}\label{subsection_linear_control_variateas}
We present the general theory of linear control variates and apply it to derive low-variance MC estimators for the pairwise density $p(x_t,x_s;\bt) = \ev_{q(z;\bt)}\left[f(z,\bt)\right] = \ev\left[f(Z,\bt)\right]$, where $Z$ has density $q(\cdot;\bt)$ and $f$ depends implicitly on the values of $x_s$ and $x_t$, which are fixed.  
We extend the methodology 
to the estimation of the gradients and discuss how the optimal constant should be chosen. We illustrate the effectiveness of the method in a simulation study in the next section.

The idea of linear control variates is to construct a function $\tilde{f}$ such that $p(x_s,x_t;\bt) = \ev\left[f(Z,\bt)\right] = \ev\left[\tilde{f}(Z,\bt)\right]$ and $\Var{\left(\tilde{f}\left(Z,\bt\right)\right)} < \Var{\left(f\left(Z,\bt\right)\right)}$. A common approach is to rely on a function $h(z,\bt)$ for which $\ev\left[h(Z,\bt)\right]$ is known analytically. Assuming we have such an $h$, define
\begin{equation*}
\tilde{f}\left(z,\bt\right) = f(z,\bt) - \gamma \left(h(z,\bt) - \ev\left[h(Z,\bt)\right]\right),
\end{equation*}
where $\gamma$ is a constant to be determined. Then $\ev\left[f(Z,\bt)\right] = \ev\left[\tilde{f}(Z,\bt)\right]$, the quadratic function
\begin{equation*}
    \Var{\left(\tilde{f}(Z,\bt)\right)} = \gamma^2  \Var{\left( h(Z,\bt)\right)} - 2 \gamma \Cov{\left(f(Z,\bt),h(Z,\bt)\right)} + \Var{(f(Z,\bt))},
\end{equation*}
is minimized for $\gamma^{*} = \frac{\Cov{\left(f(Z,\bt),h(Z,\bt)\right)}}{\Var{\left(h(Z,\bt)\right)}}$ and the variance reduction when $\gamma^{*}$ is used is given by $\frac{\Var{\left(\tilde{f}(Z,\bt)\right)}}{\Var{\left(f(Z,\bt)\right)}} = 1 - \Corr{(f(Z,\bt),h(Z,\bt))}^2$. In general, $\Cov{\left(f(Z,\bt),h(Z,\bt)\right)}$ and $\Var{\left(h(Z,\bt)\right)}$ are not known analytically and have to be estimated from samples, in a pilot study. 

We specify $h$ as a Taylor polynomial and carry out the variance analysis in this case. Fix a positive integer $m$ and $z_0 \in \R$. In the following, assume that $f$ can be continuously differentiated $m$ times in $z$ and once in $\bt$ and that the moments $M_{l}(\bt) \defeq \ev\left[(Z-z_0)^l\right]$ and their gradients $\nabla_{\bt}M(\bt)$ exist and are known analytically for all $l \in \{1,\ldots,m\}$. Let
\begin{equation*}
    T^m_{f}(z,\bt) = \sum_{l=1}^m \pdv[order=l]{f}{z}(z_0,\bt)\frac{(z-z_0)^l}{l!},
\end{equation*}
and set $h = T^m_{f}$. We omit $m$ when there is no risk of confusion. Then \begin{equation*}
p(x_s,x_t;\bt) = \ev\left[f(Z,\bt) -\gamma_0 T_{f}(Z,\bt)\right] + \gamma_0 \sum_{l=1}^m \pdv[order=l]{f}{z}(z_0,\bt)\frac{M_l(\bt)}{l!}.
\end{equation*}
with optimal 
\begin{equation}
    \gamma^{*}_0 = \frac{\Cov{(f(Z,\bt),T_f(Z,\bt))}}{\Var{\left(T_f(Z,\bt)\right)}}.\label{eq:optimal_gamma_0}
\end{equation}
Intuitively, for fixed $\bt$ and large $m$, the graphs of the functions $f(\cdot,\bt)$ and $T_{f}(\cdot,\bt)$ are similar, hence the random variables $f(Z,\bt)$ and $T_{f}(Z,\bt)$ are correlated and $\Var{(\tilde{f}(Z,\bt))} < \Var{(f(z,\bt))}$. This argument is formal when $q$ is compactly supported, for which $\lVert f(\cdot,\bt)-T_{f}(\cdot,\bt) \rVert_{\infty} \to 0$ as $m \to \infty$. The caveat is that a very large $m$ might be needed if the partial derivatives $\pdv[order=l]{f}{z}(z_0,\bt)$ are large  (for example, take $\beta$ large in \eqref{eq:when_beta_is_large}), if $d$ is large or $q$ is not compactly supported. 
We turn our attention to the estimation of the gradients. We have
\begin{equation}\label{eq:gradient_pre_tricks}
    \nabla_{\bt} p(x_s,x_t;\theta) = \nabla_{\bt} \ev\left[f(Z,\bt)\right] = \ev\left[\nabla_{\bt} f(Z,\bt)\right] + \ev\left[\pdv{f}{z}(Z,\bt)\nabla_{\bt}Z(\bt)\right].
\end{equation}
In the following, $T_{\nabla_{\bt} f}$ is the vector of univariate Taylor polynomials associated to the partial derivatives of $f$ with respect to $\bt$. The first term from \eqref{eq:gradient_pre_tricks} can be dealt with as before
\begin{equation*}
\ev\left[\nabla_{\bt} f(Z,\bt)\right] = \ev\left[\nabla_{\bt} f(Z,\bt) - \bm{\gamma}_1 T_{\nabla_{\bt} f}\left(Z,\bt\right)\right] + \bm{\gamma}_1 \sum_{l=1}^m  \pdv[order=l]{\left(\nabla_{\bt} f\right)}{z}(z_0,\bt) \frac{ M_l(\bt)}{l!}. 
\end{equation*}
Noting that partial derivatives commute and $T_{\nabla_{\bt}f} = \nabla_{\bt} T_f$, the optimal vector is 
\begin{equation}
\bm{\gamma}_1^{*} = \frac{\Cov{\left(\nabla_{\bt}f(Z,\bt), T_{\nabla_{\bt}f}(Z,\bt) \right)}}{\Var{( T_{\nabla_{\bt}f}(Z,\bt) )}} = \frac{\Cov{\left(\nabla_{\bt}f(Z,\bt),\nabla_{\bt}T_f(Z,\bt)\right)}}{\Var{\left(\nabla_{\bt}T_f(Z,\bt)\right)}}.\label{eq:optimal_gamma_1}
\end{equation}
Note that $\Var{\left(\nabla_{\bt} T_f\right)}$ can be computed explicitly if we have analytic expressions for the first $2m$ moments $M_1(\bt),\ldots,M_{2m}(\bt)$. Combining control variates with PG, the second term in \eqref{eq:gradient_pre_tricks} becomes
\begin{equation*}
  \ev\left[\pdv{f}{z}(Z,\bt) \nabla_{\bt}Z\right]=     \ev\left[\left(\pdv{f}{z}(Z,\bt) - \bm{\gamma}_2 T_{\pdv{f}{z}}(Z,\bt)\right) \nabla_{\bt}Z\right] + \bm{\gamma}_2 
   \ev\left[T_{\pdv{f}{z}}(Z,\bt) \nabla_{\bt}Z\right],
\end{equation*}
with optimal vector
\begin{equation}
    \bm{\gamma}_2^{*} = \frac{\Cov{\left(\pdv{f}{z}(Z,\bt)\nabla_{\bt}Z(\bt),T_{\pdv{f}{z}}(Z,\bt)\nabla_{\bt}Z(\bt)\right)}}{\Var{\left(T_{\pdv{f}{z}}(Z,\bt) \nabla_{\bt}Z(\bt)\right)}},\label{eq:optimal_gamma_2}
\end{equation}
and where 
\begin{align}&\ev\left[T_{\pdv{f}{z}}(Z,\bt) \nabla_{\bt}Z(\bt)\right]  = 
\nabla_{\bt} \ev\left[T_f(Z,\bt)\right] -\ev\left[\nabla_{\bt}T_f(Z,\bt)\right] \label{eq:compute_moment_expression_for_grad} \\
  =&\nabla_{\bt} \left( \sum_{l=1}^m  \pdv[order=l]{f}{z}(z_0,\bt) \frac{ M_l(\bt)}{l!}\right) - \sum_{l=1}^m  \pdv[order=l]{\left(\nabla_{\bt} f\right)}{z}(z_0,\bt) \frac{ M_l(\bt)}{l!} =  \sum_{l=1}^m  \pdv[order=l]{f}{z}(z_0,\bt) \frac{ \nabla_{\bt}M_l(\bt)}{l!}. \notag
\end{align}
Combining the control variate estimators for $p(x_s,x_t;\bt)$ and $\nabla_{\bt}p(x_s,x_t;\bt)$, we have that 
\begin{align*} 
&\nabla_{\bt} \log{p(x_s,x_t;\bt)} = \frac{\nabla_{\bt}p(x_s,x_t;\bt)}{p(x_s,x_t;\bt)} = \frac{\ev\left[\nabla_{\bt} f(Z,\bt)\right] + \ev\left[\pdv{f}{z}(Z,\bt)\nabla_{\bt}Z(\bt) \right]}{\ev\left[f(Z,\bt)\right]}\notag\\ =&\frac{\ev\left[\nabla_{\bt} f(Z,\bt) - \hat{\bm{\gamma}}_1 T_{\nabla_{\bt} f}\left(Z,\bt\right)\right] + \hat{\bm{\gamma}}_1 c_1(\bt)  + \ev\left[\left(\pdv{f}{z}(Z,\bt) - \hat{\bm{\gamma}}_2 T_{\pdv{f}{z}}(Z,\bt)\right) \nabla_{\bt}Z(\bt)\right] + \hat{\bm{\gamma}}_2
  c_2(\bt) }{\ev\left[f(Z,\bt) -\hat{\gamma}_0 T_{f}(Z,\bt)\right] + \hat{\gamma}_0 c_0(\bt)},
\end{align*}
where \begin{align*}
    c_0(\bt) &= \ev\left[T_{ f}\left(Z,\bt\right)\right] = \sum_{l=1}^m \pdv[order=l]{f}{z}(z_0,\bt)\frac{M_l(\bt)}{l!}, \\
    c_1(\bt) &=\ev\left[T_{\nabla_{\bt} f}\left(Z,\bt\right)\right] =  \sum_{l=1}^m  \pdv[order=l]{\left(\nabla_{\bt} f\right)}{z}(z_0,\bt) \frac{ M_l(\bt)}{l!},\\
    c_2(\bt) &=  \ev\left[T_{\pdv{f}{z}}(Z,\bt) \nabla_{\bt}Z(\bt)\right] = \sum_{l=1}^m  \pdv[order=l]{f}{z}(z_0,\bt) \frac{ \nabla_{\bt}M_l(\bt)}{l!}, 
\end{align*} 
can be computed analytically if the first $m$ moments $M_1(\bt),\ldots,M_m(\bt)$ are available.
\begin{remark}
    In \eqref{eq:gradient_pre_tricks} we applied control variates to the PG estimators, yet the same control variate methodology applies to the SF and MVG gradient estimators. For example, with the SF estimator, we have
\begin{equation*}
    \nabla_{\bt} p(x_s,x_t;\theta) = \nabla_{\bt} \ev\left[f(Z,\bt)\right] = \ev\left[\nabla_{\bt} f(Z,\bt)\right] + \ev\left[f(Z,\bt)\nabla_{\bt}\log{q(Z;\bt)}\right],
\end{equation*}
where the second term is equal to
\begin{equation*}
  \ev\left[\left(f(Z,\bt) - \bm{\gamma}_{\text{SF}}\ T_{f}(Z,\bt)\right)  \nabla_{\bt}\log{q(Z;\bt)}\right] + \bm{\gamma}_{\text{SF}} \
   \ev\left[T_{f}(Z,\bt) \nabla_{\bt}\log{q(Z;\bt)}\right],
\end{equation*}
with optimal vector
\begin{equation*}
    \bm{\gamma}_{\text{SF}}^{*} = \frac{\Cov{\left(f(Z,\bt)\nabla_{\bt}\log{q(Z;\bt)},T_{f}(Z,\bt)\nabla_{\bt}\log{q(Z;\bt)}\right)}}{\Var{\left(T_{f}(Z,\bt) \nabla_{\bt}\log{q(Z;\bt)}\right)}},
\end{equation*}
and where 
\begin{equation*}\ev\left[T_{f}(Z,\bt) \nabla_{\bt}\log{q(Z;\bt)}\right]  = 
\nabla_{\bt} \ev\left[T_f(Z,\bt)\right] -\ev\left[\nabla_{\bt}T_f(Z,\bt)\right]  
   =  \sum_{l=1}^m  \pdv[order=l]{f}{z}(z_0,\bt) \frac{ \nabla_{\bt}M_l(\bt)}{l!}
\end{equation*}
follows as in \eqref{eq:compute_moment_expression_for_grad}.
\end{remark}

\textbf{Theoretical considerations in the implementation of control variates}

We discuss two key theoretical aspects: the biases incurred by using estimates instead of the optimal constants $\gamma_0^{*}, \, \bm{\gamma}_1^{*}$ and $\bm{\gamma}_2^{*}$ and the availability of the moments $M_l(\bt)$ in closed form.

The optimal constants $\gamma_0^{*}, \, \bm{\gamma}_1^{*}$ and $\bm{\gamma}_2^{*}$ can be approximated from samples by estimating the corresponding numerators and denominators from Equations \eqref{eq:optimal_gamma_0}, \eqref{eq:optimal_gamma_1} and \eqref{eq:optimal_gamma_2}. The ratio estimators $\hat{\gamma}_0, \, \hat{\gamma}_1$ and $\hat{\gamma}_2$ are consistent and biased; if the same number of samples $N$ is used for both the numerator and denominator, the bias and variance are of order $O(1/N)$ (see Subsection \ref{supplementary_subsection:asymptotic_log_ratio_estimators} of the Supplementary material). We notice significant variance reductions even for small values of $N$, e.g. in the range $100-500$. This is not the only biased quantity in control variates. To this end, let $\hat{\bm{\gamma}}_1 = \hat{\bm{\gamma}}_1(Z_1,\ldots,Z_N)$ be the optimal constant as estimated from $Z_1,\ldots, Z_N \stackrel{\textrm{iid}}\sim q$. Then the estimator \begin{equation*}\frac{1}{N}\sum_{i=1}^N \left(f(Z_i,\bt) - \hat{\gamma}_1(Z_1,\ldots,Z_N) h(Z_i,\bt) \right) + \hat{\bm{\gamma}}_1(Z_i,\ldots,Z_N) \ev\left[h(Z,\bt)\right]
\end{equation*}
is biased, i.e.~its expectation is not $\ev[f(Z,\bt)]$. \cite{glasserman} concludes in Chapter 4.1.3 that the bias is typically small and that the cost of estimating the optimal coefficient $\gamma$ in a pilot study is unattractive. The same comments apply to the bias induced by the estimation of $\hat{\bm{\gamma}}_1$ and $\hat{\bm{\gamma}}_2$.

Using the Taylor polynomial  $T_f$ as a control variate requires knowledge of analytic expressions for the moments $M_l(\bt) = \ev[(Z-z_0)^l]$, which is equivalent to knowing the non-central moments $\ev[Z^l]$. This is often the case for the distributions encountered in the composite likelihood optimization of real-valued trawl processes. If $L^{'}$ is supported on $\R$ and its characteristic function is available analytically (e.g.~Gaussian, Normal-inverse Gaussian, Variance-gamma), the distribution $q$ is from the same family as that of $L^{'},$ hence the characteristic function and moments of all orders for $q$ are known analytically. However if $L^{'}$ is supported on $\R^{+}$, we are not aware of analytic expression for the moments of $q$, apart from particular cases in which terms cancel in the integral expression of $p(x_s,x_t;\bt)$ (see Example \ref{ex:pl_as_exp_for_gamma_levy_basis}). If there are no such cancellations, $q$ is the truncation of an infinitely divisible distribution and moments are not immediately available. 

\section{Simulation study}\label{section:simulation_study}
We start by demonstrating the effectiveness of our variance reduction methodologies in a simulation study in Subsection \ref{subsection:variance_reduction_sim_study}. Having established that the PL function and its gradients can be efficiently and accurately approximated with MC samples, we demonstrate that the PL estimator significantly outperforms the GMM estimator in terms of parameter estimation error in Subsection \ref{subsection:parameter_inference_results}. 
\subsection{Variance reduction}
\label{subsection:variance_reduction_sim_study}
We conduct our experiments in the setting of the trawl process with Gamma L\'evy seed from Example \ref{ex:pl_as_exp_for_gamma_levy_basis}. We remind the reader of the notation. Let $X = (X_t)_{t \ge 0}$ be a trawl process with $L^{'} \sim \mathrm{Gamma}(\alpha,\beta)$ and recall that $\bt_{\phi}$ parameterizes the trawl function $\phi$ of $X$.  Let $\rho(\cdot;\bt_{\phi})$ be the autocorrelation of $X$, let $x_s,x_t \in \R$ and $l_1 = \min{(x_s,x_t)},\, l_2 = \max{(x_s,x_t)}$. Further define $\alpha_0 = \alpha \rho(h;\bt_{\phi})$ and $\alpha_1 = \alpha (1 - \rho(h;\bt_{\phi}))$, where $h =  t-s$. With $\bt = (\alpha,\beta,\bt_{\phi})$, we have that
\begin{align}
 p(x_s,x_t;\bt) &= \frac{\beta^{(\alpha+ \alpha_1)}  l_1^{\alpha-1}}{e^{\beta(l_1+l_2)}   \Gamma(\alpha)  \Gamma(\alpha_1)} \ev_{Z \sim \text{Beta}(\alpha_0,\alpha_1)}\left[ \left(l_2-l_1 Z\right)^{\alpha_1-1} e^{\beta l_1 Z}\right] \label{eq:for_grad_to_ref}\\&=
  \frac{\beta^{(\alpha+ \alpha_1)}  l_1^{\alpha-1}}{e^{\beta(l_1+l_2)}   \Gamma(\alpha)  \Gamma(\alpha_1)} \ev_{q(z;\bt)}\left[f(z,\bt)\right],\notag
\end{align}
where $Z$ has density $q =  \text{Beta}(\alpha_0,\alpha_1)$ and $f(z,\bt) = \frac{\beta^{(\alpha+ \alpha_1)}  l_1^{\alpha-1}}{e^{\beta(l_1+l_2)}   \Gamma(\alpha)  \Gamma(\alpha_1)} \left(l_2-l_1 z\right)^{\alpha_1-1} e^{\beta l_1 z}$ depends implicitly on $l_1$ and $l_2$. For most of the Subsection \ref{subsection:variance_reduction_sim_study}, we perform our analysis on the realization $\x = (x_\tau,\ldots,x_{n\tau})$ displayed in Figure \ref{fig:trawl_path_realization} of the trawl process $X$ with autocorrelation function $\rho(h;\lambda) = e^{-\lambda |h|}$ and with simulation parameters $\tau = 1, \, n= 150$ and $\bar{\bt} = (\alpha,\beta,\lambda)= (3,0.5,0.35)$. The next step is to choose a value for $\bt$ which is representative of the values encountered in the iterations of the composite likelihood optimization procedure. We consider the pairwise densities $p\left(x_i,x_{i+1};\bt = \hat{\bt}\right)$ at lag $k=1$ and their gradients $\nabla_{\bt} p\left(x_i,x_{i+1};\bt = \hat{\bt}\right)$ for $1 \le i \le 149$, 
where $\hat{\bt}$ is estimated by GMM from $\x$. The two main reasons for studying the gradient estimators at $\bt = \hat{\bt}$ are as follows. Firstly, the composite likelihood procedure requires a starting point, which we take to be $\hat{\bt}$ in our experiments. Secondly, $\hat{\bt} \approx \bar{\bt}$ when $n$ is large, which ensures that $\hat{\bt}$ is in the parameter region of interest and further motivates our choice. Next, we show in a simulation study that the PG methodology provides lower variance estimators than the SF one for the gradient of the pairwise densities. Afterwards, we perform a similar simulation study for the control variate methodology.  
\subsubsection{Pathwise gradients simulation results}
\label{subsubsection:simulation_study_pg}
Let $l^i_1 = \min{(x_i,x_{i+1})}$ and $l^i_2 = \max{(x_i,x_{i+1})}$ for $1 \le i \le 149$. From \eqref{eq:for_grad_to_ref}, we see that for trawl processes with Gamma marginal distribution, the PG and SF methodologies give the following expressions for the gradient
\begin{align*}
 &\nabla_{\bt} \ev_{q(z;\bt)}\left[ \left(l^i_2-l^i_1 z\right)^{\alpha_1-1} e^{\beta l^i_1 z}\right] \\
 &= \ev_{ q(z;\bt)}\left[\pdv{}{z}\left( \left(l^i_2-l^i_1 z\right)^{\alpha_1-1} e^{\beta l_1 z}\right)\nabla_{\bt}z  + \nabla_{\bt}\left(\left(l^i_2-l^i_1 z\right)^{\alpha_1-1} e^{\beta l^i_1 z}\right)\right]\\
 &=\ev_{ q(z;\bt)}\left[ \left(l^i_2-l^i_1 z\right)^{\alpha_1-1} e^{\beta l_1 z} \nabla_{\bt}\log{q(z;\bt)} + \nabla_{\bt}\left({\left(l^i_2-l^i_1 z\right)}^{\alpha_1-1} e^{\beta l^i_1 z}\right) \right],
\end{align*}
for $1 \le i \le 149$. Considering only the first term in the above equations allows us to isolate the part of the gradient which accounts for the dependency of the sampling measure $\text{Beta}(\alpha_0,\alpha_1)$ on $\bt$. Note that $\alpha_0$ and $\alpha_1$ do not depend on $\beta$ and we report just the partial derivatives with respect to $\alpha$ and $\lambda$ of the first term in the expression for the pairwise density. Let us settle the notation. With $\square \in \{\alpha, \lambda\}$, define 
\begin{align*}
 f^{\text{PG}}_{\square}(z,x_i,x_{i+1};\bt) = &  \pdv{}{z} \left(\left(l^i_2-l^i_1 z\right)^{\alpha_1-1} e^{\beta l^i_1 z}\right)\pdv{}{\square}z ,   \\
 f^{\text{SF}}_{\square}(z,x_i,x_{i+1};\bt) = &  \left(l^i_2-l^i_1 z\right)^{\alpha_1-1} e^{\beta l^i_1 z}\pdv{}{\square}\log{q(z;\bt)}.
\end{align*}
Further define the gradients $p_{\square}(x_i,x_{i+1};\bt)$, the gradient estimators $f^{\text{PG}}_{\square}(Z,x_i,x_{i+1};\bt)$ and $f^{\text{SF}}_{\square}(Z,x_i,x_{i+1};\bt)$, their standard deviations $\mathrm{sd}^{\text{PG}}_{\square}(x_i,x_{i+1};\bt)$ and $\mathrm{sd}^{\text{SF}}_{\square}(x_i,x_{i+1};\bt)$ and ratio of standard deviations $r_{\square}(x_i,x_{i+1};\bt)$
\begin{align}
    p_{\square}(x_i,x_{i+1};\bt) &\defeq \ev\left[f^{\text{PG}}_{\square}(Z,x_i,x_{i+1};\bt)\right] = \ev\left[f^{\text{SF}}_{\square}(Z,x_i,x_{i+1};\bt)\right],\notag  \\
    \mathrm{sd}^{\text{PG}}_{\square}(x_i,x_{i+1};\bt) &\defeq \sqrt{\Var{\left[ f^{\text{PG}}_{\square}\left(Z,x_i,x_{i+1};\bt\right)\right]}}, \ \   
    \mathrm{sd}^{\text{SF}}_{\square}(x_i,x_{i+1};\bt) \defeq \sqrt{\Var{\left[f^{\text{SF}}_{\square}\left(Z,x_i,x_{i+1};\bt\right)\right]}},\notag\\
    r_{\square}(x_i,x_{i+1};\bt) &\defeq \frac{\mathrm{sd}^{\text{PG}}_{\square}(x_i,x_{i+1};\bt)}{\mathrm{sd}^{\text{SF}}_{\square}(x_i,x_{i+1};\bt)}\label{eq:r_ratio_st_dev_PG_SF},
\end{align}
where the expectation is taken over $Z$ with density $q(\cdot;\bt) = \text{Beta}(\cdot;\alpha_0,\alpha_1)$ and where we make the dependency of $f$ on the pairs $(x_i,x_{i+1})$ is explicit.  

\begin{figure}[h!]
 \centering
\subfloat[Trawl path realization]{\label{fig:trawl_path_realization}\includegraphics[width=.33\linewidth]{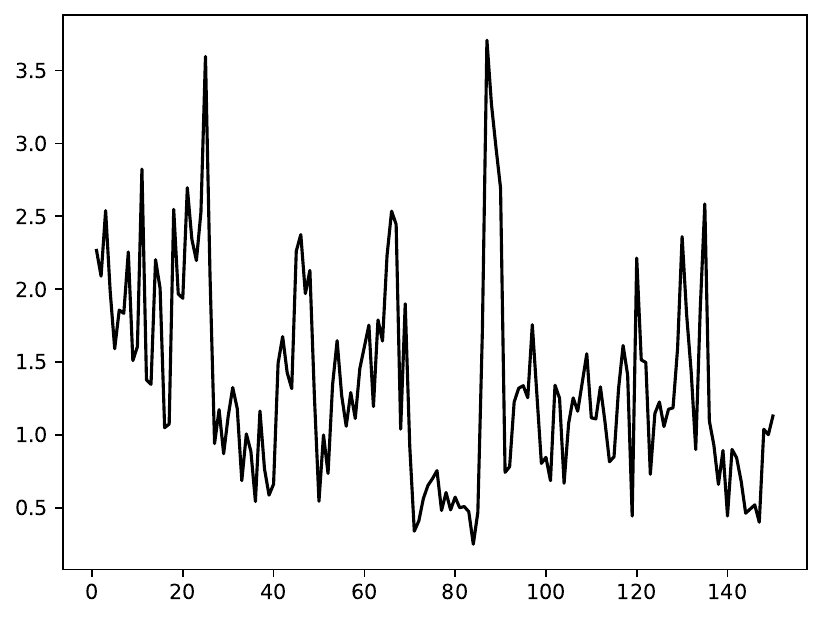}}\hfill
\subfloat[$p_{\alpha}(x_i,x_{i+1};\hat{\bt}),\, 1 \le i \le 149$]{\label{fig:true_grad_alpha}\includegraphics[width=.33\linewidth]{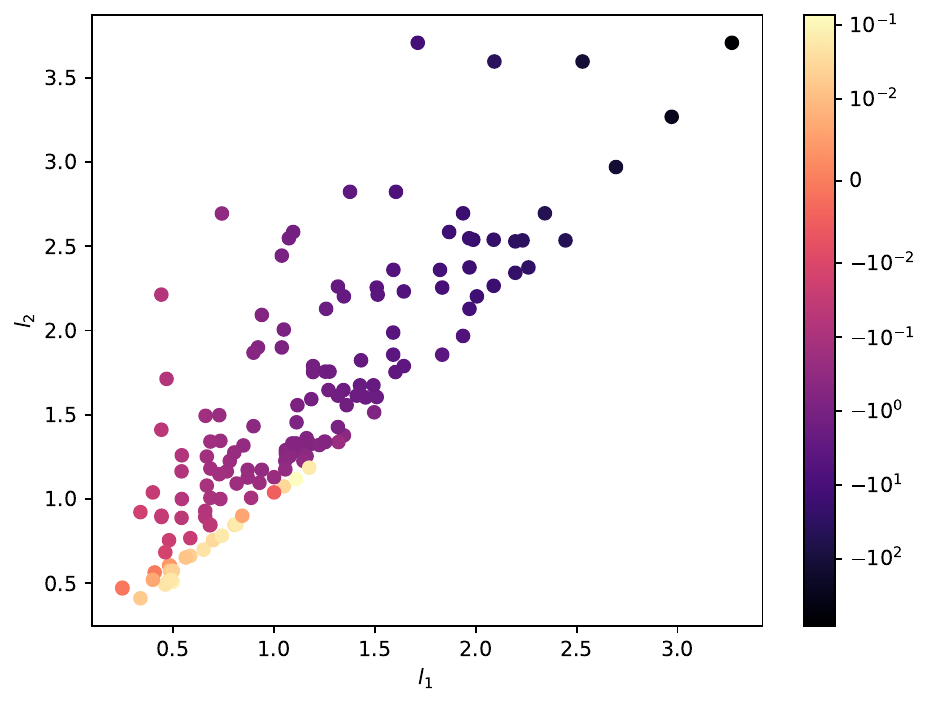}}\hfill
\subfloat[$p_{\lambda}(x_i,x_{i+1};\hat{\bt}),\, 1 \le i \le 149$]{\label{fig:true_grad_lambda}\includegraphics[width=.33\linewidth]{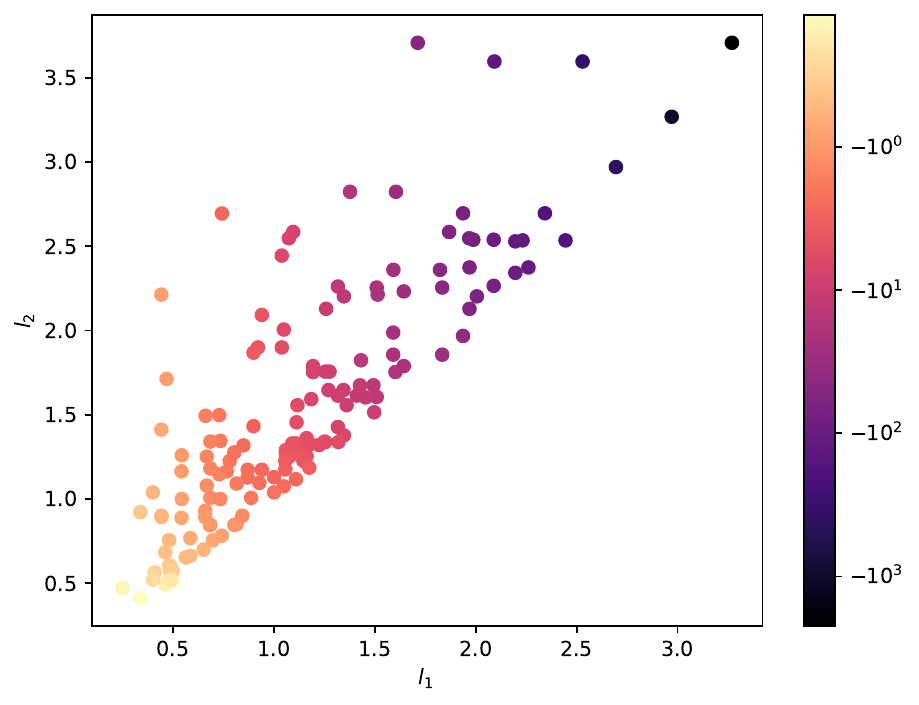}}\par 
\subfloat[$\mathrm{sd}^{\text{SF}}_{\alpha}(x_i,x_{i+1};\hat{\bt}),\, 1 \le i \le 149$]{\label{fig:std_grad_alpha_SF}\includegraphics[width=.33\linewidth]{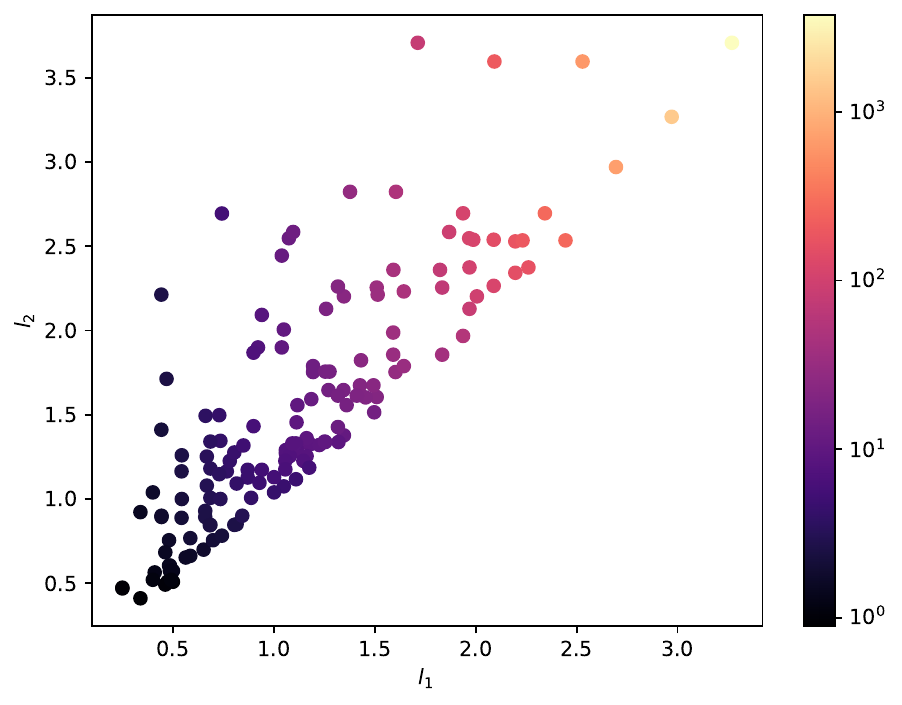}}\hfill
\subfloat[$r_{\alpha}(x_i,x_{i+1};\hat{\bt}),\, 1 \le i \le 149$]{\label{fig:std_reduction_grad_alpha}\includegraphics[width=.33\linewidth]{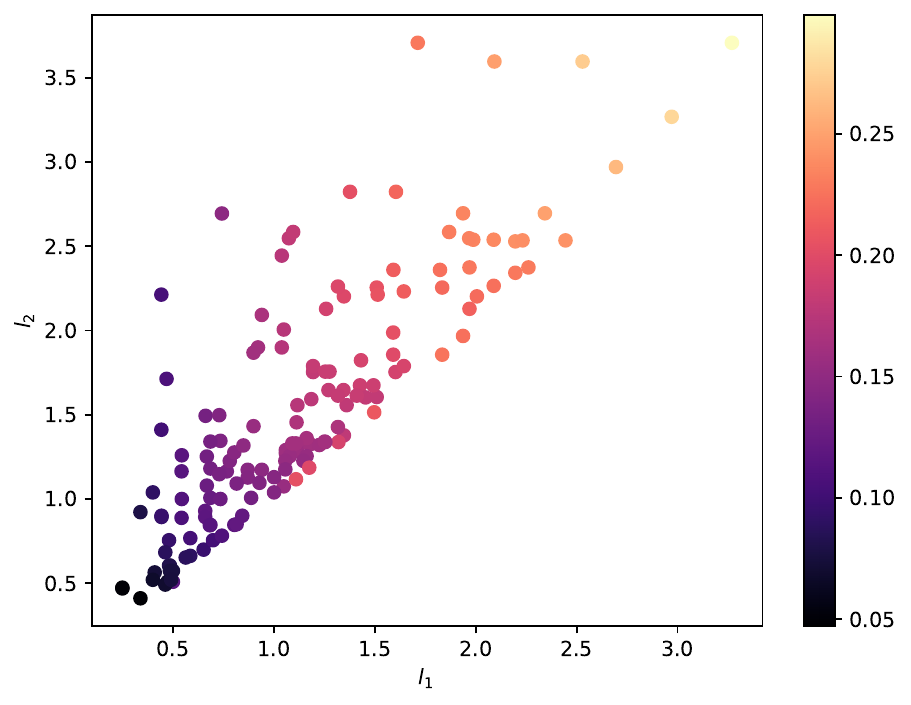}}\hfill
\subfloat[Simulation study displaying the variability of the empirical quantiles of $\left\{r_{\alpha}\left(x_i,x_{i+1};\hat{\bt}\right)\right\}_{i=1}^{749}$ across $1000$ trawl paths.]{\label{fig:sim_study_st_red_grad_alpha}\includegraphics[width=.33\linewidth]{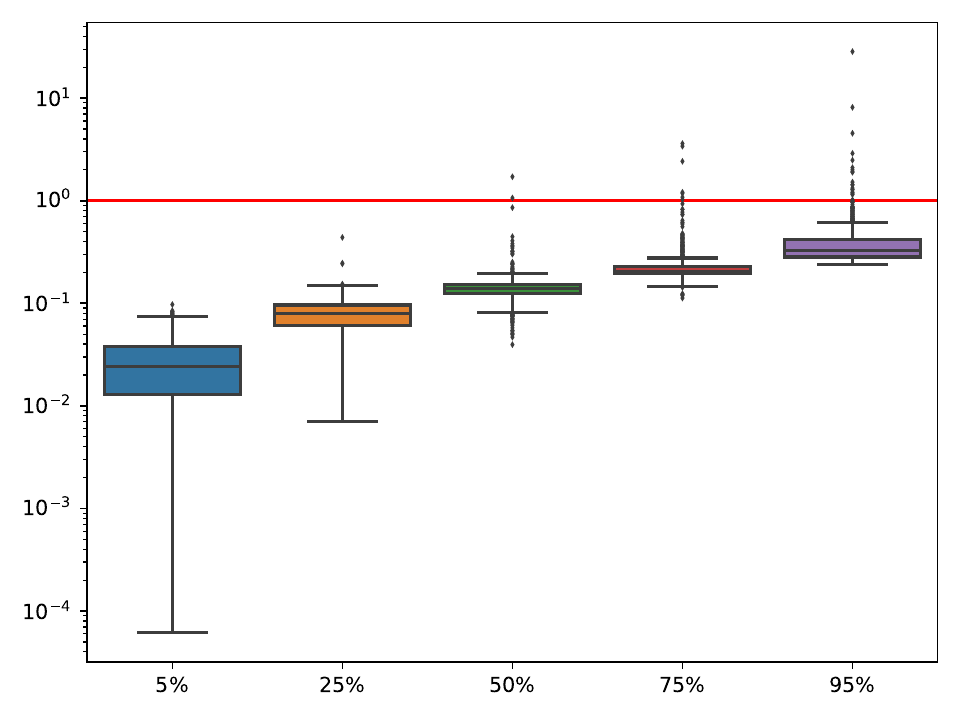}}\par 
\subfloat[$\mathrm{sd}^{\text{SF}}_{\lambda}(x_i,x_{i+1};\hat{\bt}),\, 1 \le i \le 149$]{\label{fig:std_grad_lambda_SF}\includegraphics[width=.33\linewidth]{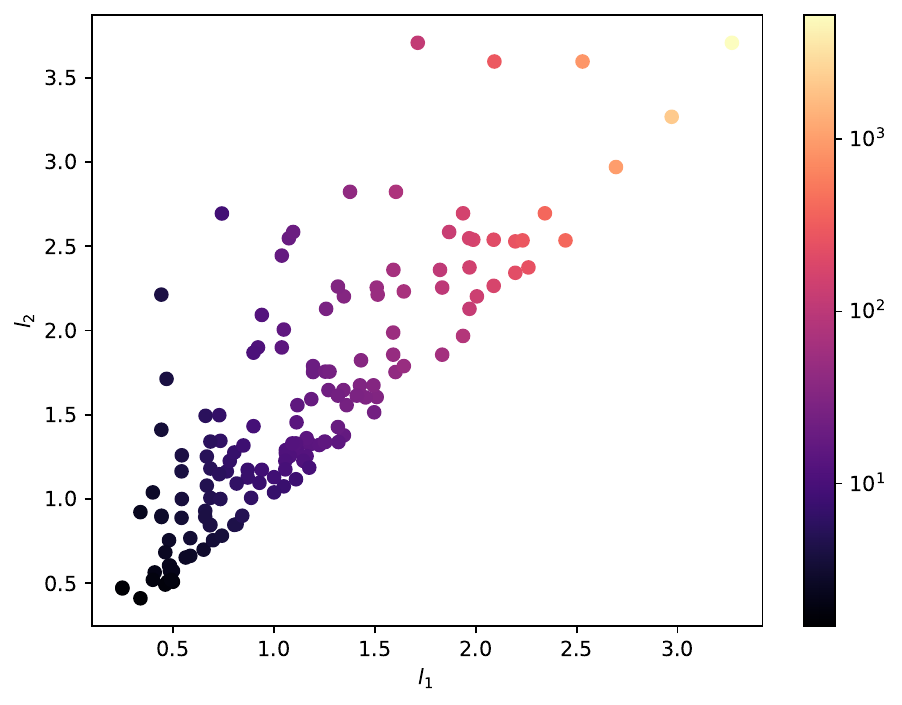}}\hfill
\subfloat[$r_{\lambda}(x_i,x_{i+1};\hat{\bt}),\, 1 \le i \le 149$]{\label{fig:std_reduction_grad_lambda}\includegraphics[width=.33\linewidth]{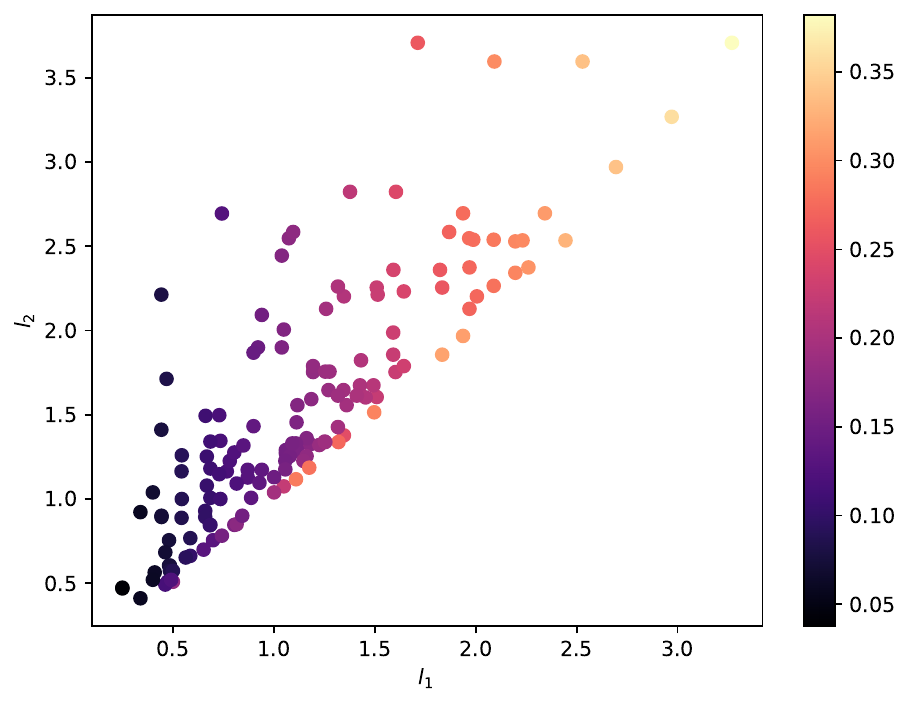}}\hfill
\subfloat[Simulation study displaying the variability of the empirical quantiles of $\left\{r_{\lambda}\left(x_i,x_{i+1};\hat{\bt}\right)\right\}_{i=1}^{749}$ across $1000$ trawl paths.]{\label{fig:sim_study_st_red_grad_lambda}\includegraphics[width=.33\linewidth]{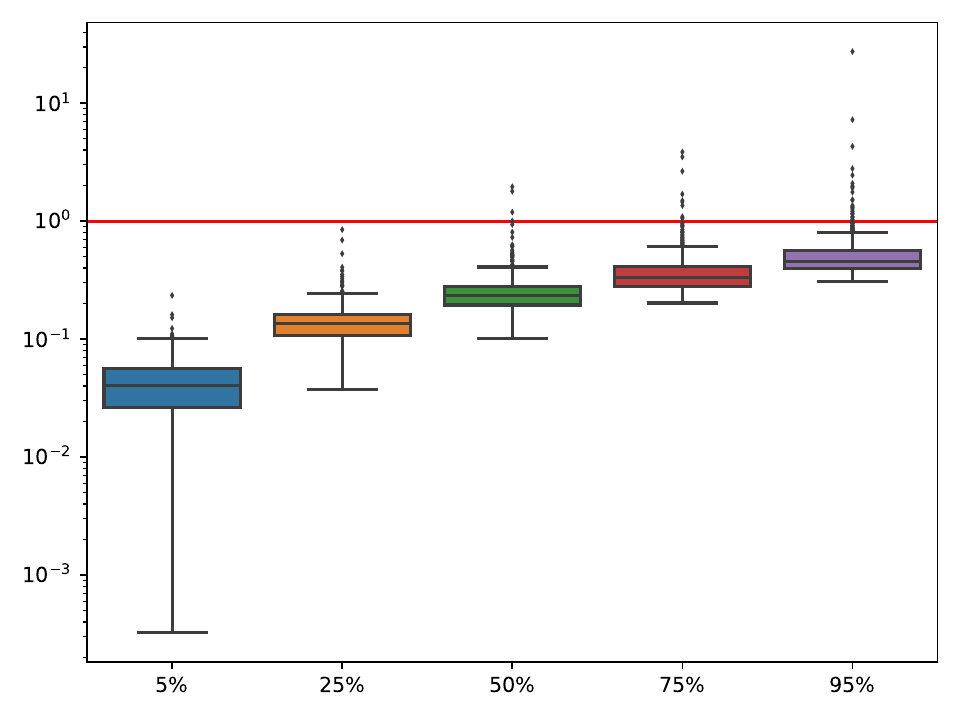}}
\caption{Illustration of the variance reduction properties of the PG methodology for the estimation of the gradients of the pairwise densities.}
\label{fig:grads_whole_fig}
\end{figure}

In the following, we display the values for the gradients of the pairwise densities, as well as the standard deviations of the corresponding MC estimators. We represent the pair $(x_i,x_{i+1})$ as the pair $(\min{(x_i,x_{i+1})},\max{(x_i,x_{i+1})}) = (l_1^i,l_2^i)$ and the corresponding quantity of interest through a color scheme, with a linear or log-scale color bar. Since the trawl process is stationary, we have that $p(x_i,x_{i+1};\bt) = p(x_{i+1},x_i;\bt)$ and $\nabla_{\bt} p(x_i,x_{i+1};\bt) = \nabla_{\bt}p(x_{i+1},x_i;\bt)$ for any value of $\bt$. Thus sorting the pairs $(x_i,x_{i+1})$ allows for an easier visual representation in the upper-triangular corner of the figures while making no theoretical difference in the gradient analysis. Figures \ref{fig:true_grad_alpha} and  \ref{fig:true_grad_lambda} display the values of the partial derivatives $p_{\alpha}(x_i,x_{i+1};\hat{\bt})$ and $p_{\lambda}(x_i,x_{i+1};\hat{\bt})$ for $1 \le i \le 149$. Figures
 \ref{fig:std_grad_alpha_SF} and \ref{fig:std_grad_lambda_SF} show the standard deviations of the SF estimators $\mathrm{sd}^{\text{SF}}_{\alpha}(x_i,x_{i+1};\hat{\bt})$ and $\mathrm{sd}^{\text{SF}}_{\lambda}(x_i,x_{i+1};\hat{\bt})$ for $1\le i \le 149.$ Note that the standard deviation of the SF estimators can even be two orders of magnitude higher than the corresponding values of the gradients, making accurate gradient calculations computationally expensive. Figures \ref{fig:std_reduction_grad_alpha} and \ref{fig:std_reduction_grad_lambda} show the ratios of the standard deviations of the PG and SF estimators $ r_{\alpha}(x_i,x_{i+1};\hat{\bt})$ and $ r_{\lambda}(x_i,x_{i+1};\hat{\bt})$, as defined in \eqref{eq:r_ratio_st_dev_PG_SF}; a value below $1$ favors the PG estimator over the SF one. Notice that the PG estimator always performs better and that it has a standard deviation which is reduced by a factor between $3$ and $20$ as compared to the standard deviation of the SF estimator; this improvement comes without an increase in the computation time. Finally, we show in Figures \ref{fig:sim_study_st_red_grad_alpha} and \ref{fig:sim_study_st_red_grad_lambda} that this improvement is maintained across a wide range of parameter regimes. To this end, we draw $1000$ random samples $\bar{\bt} = (\alpha, \beta, \lambda)$ from the priors $\alpha, \beta \sim \textrm{Gamma}(6,4),\, \lambda \sim \textrm{Gamma}(4,4)$. For each replication, we simulate a realisation $\x = (x_{\tau},\ldots,x_{n\tau})$ of the trawl process with parameters $\bar{\bt}$ and $\tau = 0.5, \, n=750$. From each $\x$, we compute the ratios of the standard deviations of the gradient estimators $\left\{r_{\alpha}\left(x_i,x_{i+1};\hat{\bt}\right)\right\}_{i=1}^{749}$ and $\left\{r_{\lambda}\left(x_i,x_{i+1};\hat{\bt}\right)\right\}_{i=1}^{749}$; we summarize the distributions of the two ratios above by computing the empirical $5\%,25\%,50\%,75\%$ and $95\%$ quantiles.
We then display the variability of these empirical quantiles across the $1000$ simulations with boxplots in Figures \ref{fig:sim_study_st_red_grad_alpha} and \ref{fig:sim_study_st_red_grad_lambda} respectively. This allows us to study the variability in the distribution of the ratio of the standard deviations. For example, Figure \ref{fig:sim_study_st_red_grad_alpha} shows that in the vast majority of the $1000$ simulations, the median  of the ratio of standard deviations is about $10^{-1}$, as displayed in the boxplot for the $50\%$ quantile. This gives a variance reduction by a factor of $10^2$. We also note that the PG estimator increases the variance of the estimator only in very few simulations, as can be seen from the boxplot outliers. We conclude that PG is an effective variance reduction method.

\subsubsection{Control variate simulation results}
\label{subsubsection:simulation_study_linear_control_variates}
\begin{figure}[p]
\centering
\subfloat[$p(x_i,x_{i+1};\hat{\bt}),\, 1 \le i \le 149$]{\label{l_cv_a}\includegraphics[width=.33\linewidth]{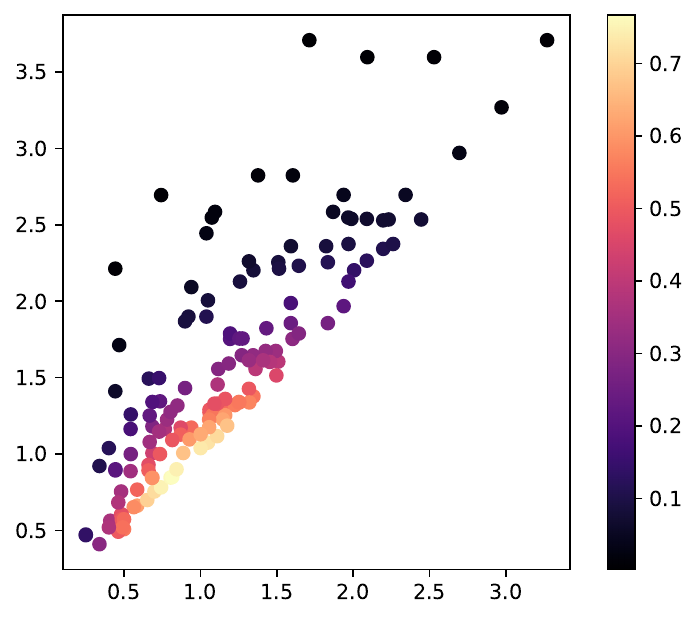}}\hfill
\subfloat[$\mathrm{sd}(x_i,x_{i+1};\hat{\bt}),\, 1 \le i \le 149$]{\label{l_cv_b}\includegraphics[width=.33\linewidth]{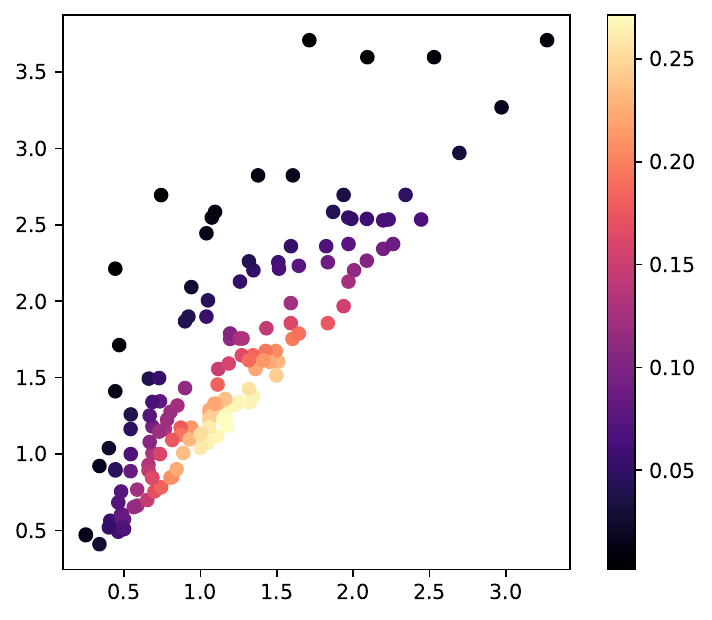}}\hfill
\subfloat[kernel density estimates for the estimators of $\log{\mathcal{L}(\hat{\bt})}$]{\label{l_cv_c}\includegraphics[width=.33\linewidth]{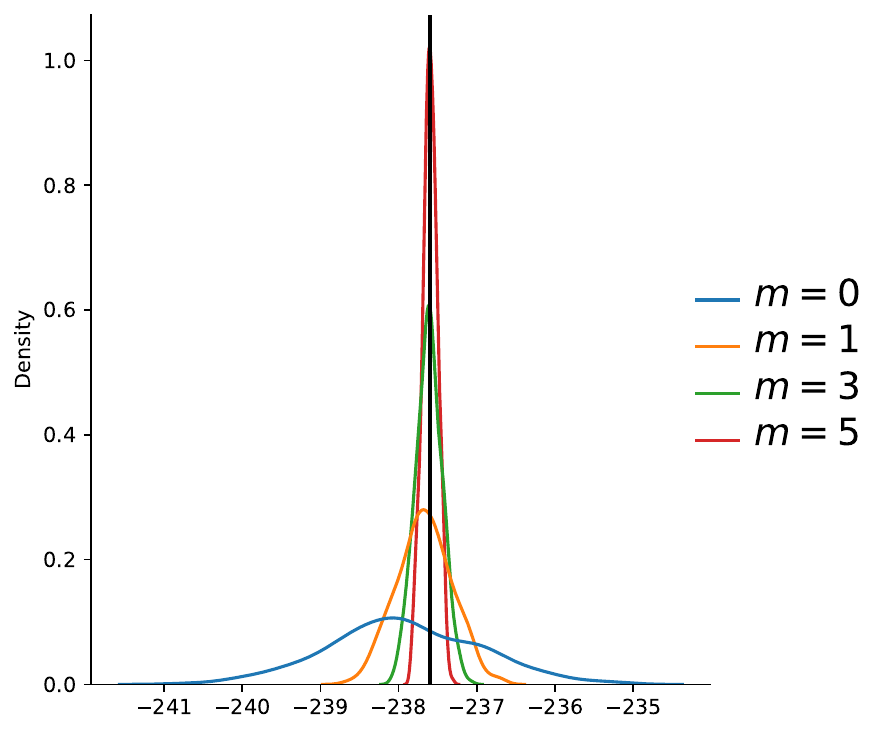}}\par 
\subfloat[$r^1(x_i,x_{i+1};\bt), \, 1 \le i \le 149$]{\label{l_cv_1}\includegraphics[width=.33\linewidth]{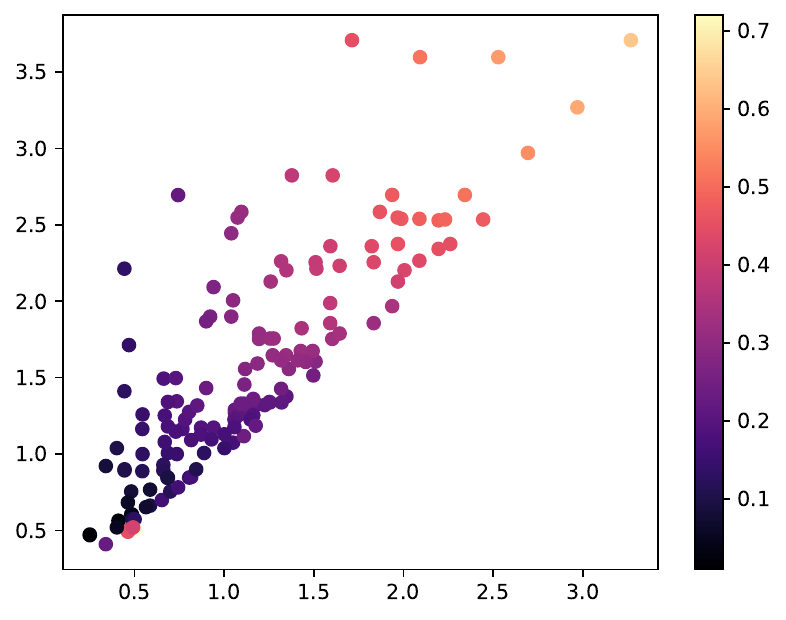}}\hfill
\subfloat[$r^3(x_i,x_{i+1};\bt), \, 1 \le i \le 149$]{\label{l_cv_3}\includegraphics[width=.33\linewidth]{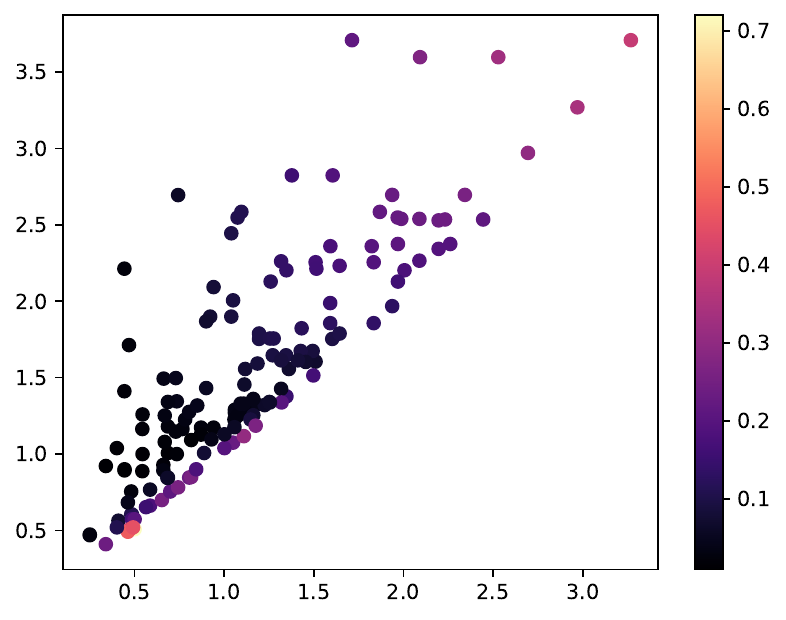}}\hfill
\subfloat[$r^5(x_i,x_{i+1};\bt), \, 1 \le i \le 149$]{\label{l_cv_5}\includegraphics[width=.33\linewidth]{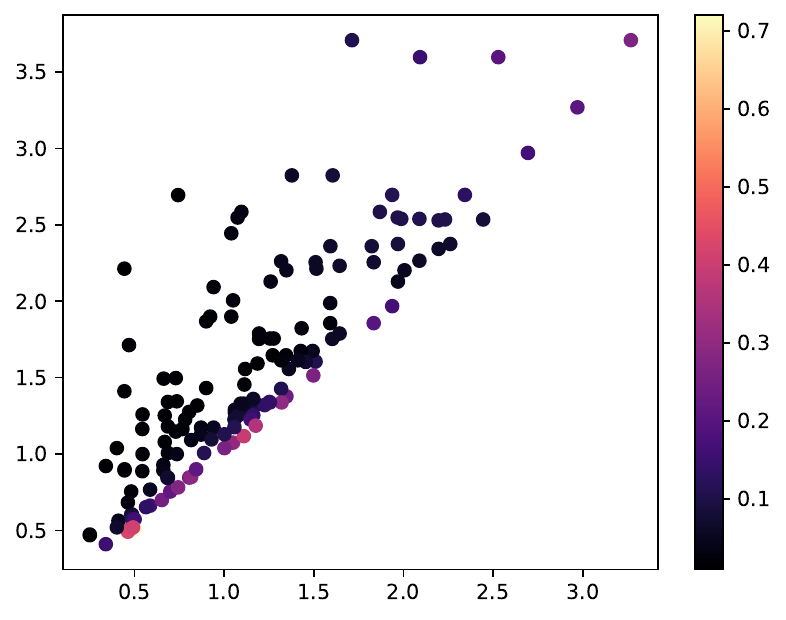}}\par
\subfloat[Boxplot summary of the distribution of the $5\%,25\%,50\%,75\%$ and $95\%$ quantiles of $\left\{r^m\left(x_i,x_{i+1};\hat{\bt}\right)\right\}_{i=1}^{n-1}$ across $1000$ simulated trawl processes for $ 1 \le m \le 5$.]{\label{adffsd}\includegraphics[width=0.7\linewidth]{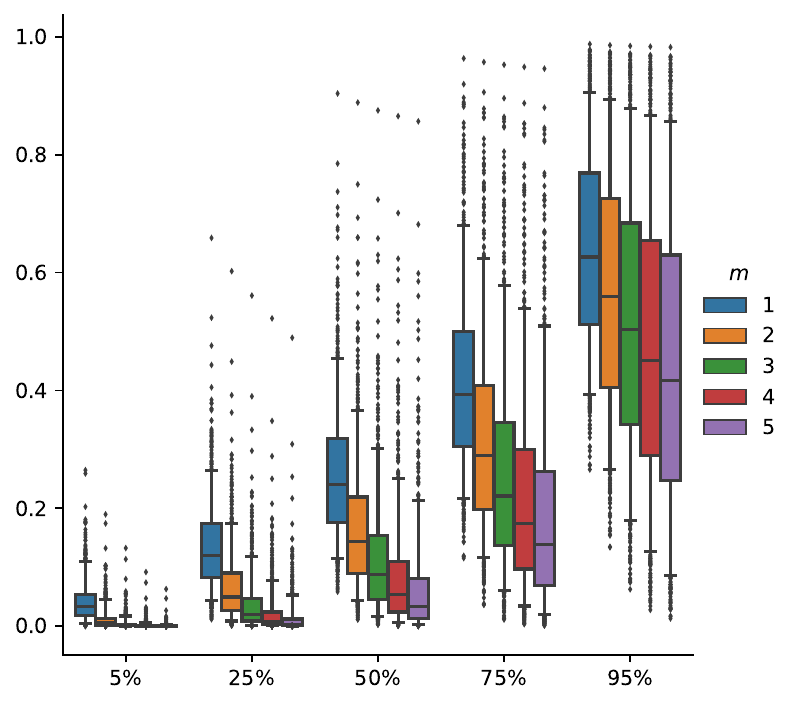}}
\caption{Illustration of the variance reduction properties of the control variates methodology applied to the estimation of the pairwise densities and log pairwise likelihood.}
\label{fig:likelihood_cv_whole_fig}
\end{figure}

We show empirically 
that the control variates methodology can be employed to significantly reduce the variance of the estimators for both the pairwise densities and their gradients and that only a Taylor polynomial of a low degree is required. This also leads to a large reduction in the bias of the estimators for the log pairwise likelihood $\log{\mathcal{L}}$ and its gradient $\nabla_{\bt}\log{\mathcal{L}}$. We display the quantities of interest via a color bar, as in Figure \ref{fig:grads_whole_fig}, first using the trawl path from Figure \ref{fig:trawl_path_realization} for Figures \ref{l_cv_a}-\ref{l_cv_5} and then with a simulation study in Figure \ref{adffsd}. We define the standard deviation of the pairwise density estimators $\mathrm{sd}(x_i,x_{i+1};\bt)$, the standard deviation of the same estimator with a Taylor polynomial of degree $m$ as control variate $\mathrm{sd}^m(x_i,x_{i+1};\bt)$ and the corresponding standard deviation ratio $r^m(x_i,x_{i+1};\bt)$ as follows
\begin{align}
\mathrm{sd}(x_i,x_{i+1};\bt) &\defeq \sqrt{\Var{\left[ f\left(Z,x_i,x_{i+1};\bt\right)\right]}},\notag \\
    \mathrm{sd}^m(x_i,x_{i+1};\bt) &\defeq \sqrt{\Var{\left( f\left(Z,x_i,x_{i+1};\bt\right)- \hat{\gamma}_1(x_i,x_{i+1}) T^m_f\left(Z,x_i,x_{i+1};\bt\right)\right)}},\label{eq:ambiguity}\\ 
    r^m(x_i,x_{i+1};\bt) &\defeq \frac{\mathrm{sd}^m(x_i,x_{i+1};\bt)}{\mathrm{sd}(x_i,x_{i+1};\bt)}\notag,
\end{align}
where $f$ is from Example \ref{ex:pl_as_exp_for_gamma_levy_basis} and where we make explicit the dependency of the optimal constant $\hat{\bm{\gamma}}_1$ on $(x_i,x_{i+1})$. The definition from \eqref{eq:ambiguity} is ambiguous without specifying how $\hat{\bm{\gamma}}_1$ is estimated. When employing this notation, we use the same samples to estimate the pairwise densities and the optimal constant and we specify how many samples we use. 

Figures \ref{l_cv_a} and \ref{l_cv_b} show the pairwise densities $p(x_i,x_{i+1};\hat{\bt})$ and the standard deviations $\mathrm{sd}(x_i,x_{i+1};\hat{\bt})$ of the corresponding estimators for $1 \le i \le 149$, where $\hat{\bt}$ is the GMM estimator. Figure \ref{l_cv_c} shows kernel density estimates of the estimators for $\log{\mathcal{L}(\bt)} = \sum_{i=1}^{149} \log{p(x_i,x_{i+1};\hat{\bt})}$ with no control variate, i.e.~$m=0$, and with Taylor polynomials of degrees $m=1,\, 3,\,5$ applied as control variates to each pair $(x_i,x_{i+1})$. To produce the plot, we use $25$ independent samples for each pair, from which we estimate both the pairwise density and $\hat{\bm{\gamma}}_1$. The kernel density estimate is based on $10^3$ replications, in each of which we use $149 \cdot 25$ independent samples $Z$ from $q(\cdot;\bt)$, $25$ for each pair. The true value of $\log{\mathcal{L}(\bt)}$, as estimated from $10^5$ samples with $m=5$, is displayed through a vertical line. We note substantial bias and standard deviation reductions as we increase $m$, even when we only use $25$ samples to calibrate $\hat{\bm{\gamma}}_1$. This further shows that in our setting, the performance of control variates is not sensitive to the miscalibration of the optimal constants. Finally, we show that the improvements are significant for a wide range of parameter regimes. We proceed as for Figures \ref{fig:sim_study_st_red_grad_alpha} and \ref{fig:sim_study_st_red_grad_lambda}. To this end, we draw $1000$ random samples $\bar{\bt} = (\alpha, \beta, \lambda)$ from the prior $\alpha, \beta \sim \textrm{Gamma}(6,4),\, \lambda \sim \textrm{Gamma}(4,4)$. For each replication, we simulate a realisation $\x = (x_{\tau},\ldots,x_{n\tau})$ of the trawl process processes with parameters $\bar{\bt}$ and $\tau = 0.5, \, n=750$. From each $\x$ we compute the empirical $5\%,25\%,50\%,75\%$ and $95\%$ quantiles of $\left\{r^m\left(x_i,x_{i+1};\hat{\bt}\right)\right\}_{i=1}^{749}$; we then display the variability of these empirical quantiles across the $1000$ simulations with a boxplot in Figure \ref{adffsd}. We notice excellent variance reduction properties. In particular, even for $m=1$, for over $95\%$ of the simulated paths, the median of the standard deviation ratio, i.e.~the median of $\left\{r^1(x_i,x_{i+1};\hat{\bt})\right\}_{i=1}^{749}$ is below $0.3$. For $m=3$, in over $95\%$ of the simulated paths, the median of the standard deviation ratio drops below $0.1$, reducing the number of samples required to achieve the same performance as the $m=0$ estimator  by a factor of $100$. We note that higher $m$ gives a higher variance reduction. 

We turn our attention to the combined variance reduction properties of control variates and pathwise gradients for the estimation of the gradients and show that the improvements over the SF methodology are maintained even for long-memory trawl processes. To this end, let $X$ be a trawl process with $L^{'} \sim \mathrm{Gamma}(\alpha,\beta)$ and autocorrelation function $\rho(h;\bt_{\phi}) = \left(1+\frac{h}{\delta}\right)^{-H}$, for $h \geq 0$, where $\bt_{\phi} = (H,\delta)$ and $H,\, \delta >0$. Let $\x = (x_\tau,\ldots,x_{n\tau})$ be a realisation of the trawl process $X$ described above for $\tau = 0.5, \, n= 1500$ and simulation parameters $\bar{\bt} = (\alpha,\beta,H,\delta)= (6,1.75,1.25,1)$. Consider lags $K = (1,3,5,10,15,20)$, let $\hat{\theta}$ be the GMM estimators for $\bt$ and further
\begin{equation*}
\log{\mathcal{L}(\hat{\bt})} = \sum_{k \in K}\sum_{i=1}^{1500-k} \log{p(x_i,x_{i+1};\hat{\bt})}.
\end{equation*}
To obtain an estimator for $\nabla_{\bt} \log{\mathcal{L}(\bt)}$, we use $N = 750$ samples for each pair $(x_i,x_{i+k})$, where $k \in K$ and $1 \le i \le 1500-k$; we use the same $750$ samples to estimate both $p(x_i,x_{i+k};\hat{\bt})$ and $\nabla_{\bt} p(x_i,x_{i+k};\hat{\bt})$. We repeat this computation $1000$ times while keeping $\x$ fixed to estimate the bias and variance of our SF and PG gradient estimators with Taylor polynomials of degree $m$ as control variates and show the results in Table \ref{table:variance_reduction_sim_study_main_body}. Note again that $q$ does not depend on $\beta$, hence the SF and PG results are the same for this variable. We first analyze the case $m=0$, which corresponds to not using control variates; as discussed in Subsection \ref{subsection_linear_control_variateas}, the moments of $q$ are not always available in closed form. We see that the PG significantly outperforms SF for all parameters; for $H$ and $\delta$, the bias and standard deviation are reduced by more than $10$, respectively $7$ times. With control variates, the higher the $m$, the better the bias and standard deviation properties and we see the same pattern as for $m=0$. The PG methodology improves significantly over the SF one for $H$ and $\delta$; regarding the gradients with respect to $\alpha$ for $m\ge 1$, PG has a slightly higher bias, although the standard deviation is significantly smaller than for the SF. We conclude that PG together with $m=1$ or $m=2$ removes the bias and decreases the variance enough to carry a gradient descent optimization procedure. We provide further tables with more evaluation metrics, e.g.~mean absolute error (MAE), root-mean-square error (rMSE) and median absolute error (medAE) in Table \ref{table:extended_sim_study_table} of Subsection \ref{supplementary_section:extendend_simulation_studies}. For example, we see that the PG estimator has a much smaller MAE than the SF one. Thus the lower bias of SF for the partial derivative with respect to $\alpha$ comes from larger errors that cancel out, as opposed to smaller errors that do not cancel for PG.
\begin{remark}
    In Table \ref{table:variance_reduction_sim_study_main_body} we display the results only for one chosen value $\bt = (6,1.75,1.25,1)$ and do not provide simulation studies as in Figures \ref{fig:sim_study_st_red_grad_alpha}, \ref{fig:sim_study_st_red_grad_lambda} and \ref{adffsd}. The SF estimator becomes numerically unstable and returns infinite or non-numeric values when at least one of the parameters of $q = \textrm{Beta}(\alpha_0,\alpha_1)$ is close to $0$, where $\alpha_0 = \alpha \rho(k \tau;\bt),\, \alpha_1 = \alpha (1- \rho(k\tau;\bt)),\, k$ is the lag used and $\tau$ is the spacing between trawl sets. This issue often occurs for small $\alpha$ or for small or large $k$, i.e.~when the autocorrelation $\rho (k\tau;\bt) \approx 0$ or $1$. Thus comparing the two methods in a simulation study is difficult, although it is worth noting that the PG is stable for a wide range of parameter regimes.
\end{remark}
\begin{table}[h!]
\begin{adjustbox}{width=\textwidth}
\begin{tabular}{@{\extracolsep{6pt}}llrrrrrrrrrr@{}}
   &   & \multicolumn{2}{c}{$m=0$} & \multicolumn{2}{c}{$m=1$} & \multicolumn{2}{c}{$m=2$} & \multicolumn{2}{c}{$m=3$}   \\ \cline{3-4} \cline{5-6}  \cline{7-8}  \cline{9-10} 

   &   &  bias & st. dev. &  bias & st. dev. &  bias & st. dev. &  bias & st. dev. \\
\hline
SF & $\alpha$ &  4.19 &     5.25 &  \textbf{0.27} &     1.92 &  \textbf{0.07} &     1.43 & \textbf{0.00} &     1.20 \\
   & $\beta$ & -3.59 &     3.16 & -0.58 &     1.80 & -0.20 &     1.44 & -0.04 &     1.15  \\
   & $H$ &  7.88 &     9.11 &  0.61 &     2.11 &  0.22 &     1.44 &  0.10 &     1.14  \\
   & $\delta$ & -5.07 &     5.33 & -0.39 &     1.44 & -0.14 &     1.02 & -0.06 &     0.81  \\
      \hline   
PG & $\alpha$ &  3.03 &     2.79 &  \textbf{0.44} &     1.43 &  \textbf{0.15} &     1.14 &  \textbf{0.03} &     0.94 \\
   & $\beta$ & -3.59 &     3.16 & -0.58 &     1.80 & -0.20 &     1.44 & -0.04 &     1.15  \\
   & $H$ &  0.70 &     1.25 &  0.02 &     1.18 &  0.00 &     0.84 & -0.03 &     0.67  \\
   & $\delta$ & -0.39 &     0.75 & 0.00 &     0.82 & 0.00 &     0.63 &  0.01 &     0.50  \\
\hline 
\end{tabular}
\end{adjustbox}
\caption{Bias and standard deviation of the SF and PG gradient estimators for each value $m$ of the degree of the Taylor polynomial used as control variate; $m=0$ corresponds to not using control variates. For $m\ge4$ the estimated biases are too small to be measured reliably and we only report results for $m \le 3$. We note that PG improve significantly over the bias and standard deviation of the SF estimators. We present the three values where SF performs better in bold. The true values of the gradient rounded to the nearest integer are $(-77,  17,  70, -56)$. }
\label{table:variance_reduction_sim_study_main_body}
\end{table}

\textbf{Practical considerations in the implementation of control variates}

Our application of the control variates methodology to composite likelihood inference for trawl processes differs from the approach typically found in the modern stochastic optimization literature. Rather than utilizing reverse-mode automatic differentiation (AD) to compute the first-order derivatives of a mapping from high to low dimensional space, as commonly done in machine learning and deep learning, we require efficient computation of higher-order derivatives of a mapping from low to high dimensional, which can only be done using forward-mode AD. In spite of these computational challenges, we demonstrate that efficient implementations are possible and that our methodology is feasible. Additionally, we provide an overview of techniques that can be employed to accelerate the current numerical implementation. As before, let $\x = (x_{\tau},\ldots,x_{n\tau})$ be the discretely observed path of the trawl process $X$ at times $\tau,\ldots,n\tau$, $\bt \in \R^d$ the parameters which specify the distribution and autocorrelation structure of $X$, $m$ the degree of the Taylor polynomial used as control variate and $\{1,\ldots,K\}$  the lags used in the PL estimator. Then there are ${N_0} \defeq (n-1) + \ldots, (n-K) = Kn - k(K+1)/2$ pairwise densities to approximate. Assume we use sets of $N$ i.i.d.~samples $\left\{z^{(i)}_1\right\}_{i=1}^N,\ldots,\left\{z^{(i)}_{{N_0}}\right\}_{i=1}^{N}$ to estimate each of the ${N_0}$ pairwise densities. Finally, let $f_1,\ldots,f_{{N_0}}$ be the functions used to estimate each of the ${N_0}$ pairwise densities.

Firstly, the estimation of the optimal constants $\bm{\gamma}^{*}_1$ and $\bm{\gamma}^{*}_2$ turns out to involve the differentiation of a mapping from low to high dimensional space. For example, to estimate
\begin{equation*}
    \bm{\gamma}^{*}_1 =  \frac{\Cov{\left(\nabla_{\bt}f(Z,\bt),\nabla_{\bt}T_f(Z,\bt)\right)}}{\Var{\left(\nabla_{\bt}T_f(Z,\bt)\right)}},
\end{equation*}
we need to compute the gradient with respect to $\bt$ of the following two mappings 
\begin{align*}
    \R^d &\longrightarrow \R^{N \times {N_0}},\\
    \bt &\longmapsto \left(f_j\left(z^{(i)}_j\right)\right)_{ij}, \\ 
    \bt &\longmapsto \left(T_{f_j}\left(z^{(i)}_j\right)\right)_{ij},
\end{align*}
where $1 \le i \le N ,\, 1 \le j \le {N_0}$. In this case, forward-mode and reverse-mode AD have computational complexity $O(d)$ and $O(N N_0)$, respectively. In general, $d$ is orders of magnitude lower than $N N_0$; in our experiments, forward-mode AD works well, whereas reverse-mode AD fails due to the high complexity of the algorithm. The speed-up boils down to parenthesizing the Jacobian multiplications in the order which requires matrix-vector rather than matrix-matrix multiplications, as in the adjoint methods in design (see Chapter 8.7 of \cite{strang2007computational}). This step can be accelerated even further. Indeed, the optimal constants can be accurately estimated with a fraction of the samples used in the PL procedure; moreover, the optimal constants can be reused in consecutive iterations of gradient descent for an even lower computational cost. 

Secondly, the control variates methodology heavily relies on the efficient computation of the higher order partial derivatives $\pdv[order=l]{f}{z} \in \R$, for $1 \le l \le m$. Although atypical for modern frameworks such as Tensorflow and Torch, this has been studied extensively in previous works \citep{Karczmarczuk,Pearlmutter_Siskind}. In general, there is no formula for the complexity of calculating the first $m$ derivatives of $f$ in terms of the complexity of calculating $f$, as this depends explicitly on the exact operations which put together define $f$. It is thus difficult to define an effective sample size for the control variate methodology when taking into account the extra computational time.  Although the simulation studies from Figure \ref{fig:likelihood_cv_whole_fig} and Table \ref{table:variance_reduction_sim_study_main_body} display significant improvements with only a negligible increase in computational time for $m \le 2$, using $m \ge 4$ proved computationally expensive and the simulation study from Subsection \ref{subsection:parameter_inference_results} was carried out with $m=2$. We argue $m =1$ or $2$ are generally good hyperparameter choices
, as they already provide low-variance estimators which are essentially unbiased.

\begin{remark}
It is not clear if the difficulty we face for large $m$ is due to theoretical reasons or sub-optimal implementations. For example, when applying the chain rule to compute the $n^{\text{th}}$ derivative of a composition of functions $f \circ g $, one can group terms according to Faà di Bruno's formula to avoid recomputing terms, similarly to the product rule for higher order derivatives. Our implementation in JAX does recompute terms, although new approaches to implement Faà di Bruno's formula in JAX are now available \citep{jax_jet}. We suspect this makes up most of the computational time. A task for further research is to see if removing the inefficiencies described above does indeed significantly lower the computational time, which in turn would allow to use little to no MC samples and approximate the pairwise densities by high-degree Taylor polynomials when the moments of $q$ are available analytically. 
\end{remark}

\subsection{Parameter inference results} 
\label{subsection:parameter_inference_results}
\begin{figure}[h!]
 \centering
\subfloat[$(\alpha, \beta, \lambda) = (3,0.75,0.1)$]{\label{fig:estimation_rmse_a}\includegraphics[width=.33\linewidth]{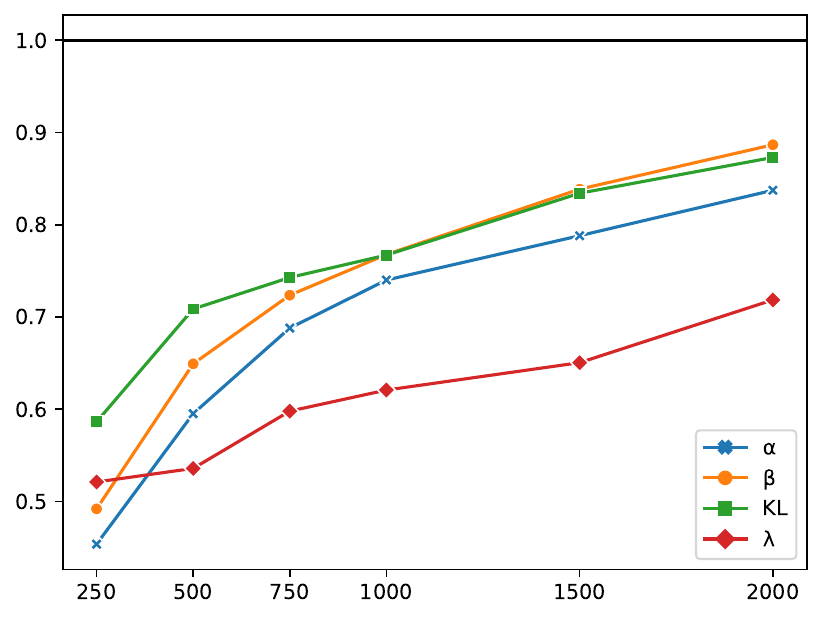}}\hfill
\subfloat[$(\alpha, \beta, \lambda) = (3,0.75,0.25)$]{\label{fig:estimation_rmse_b}\includegraphics[width=.33\linewidth]{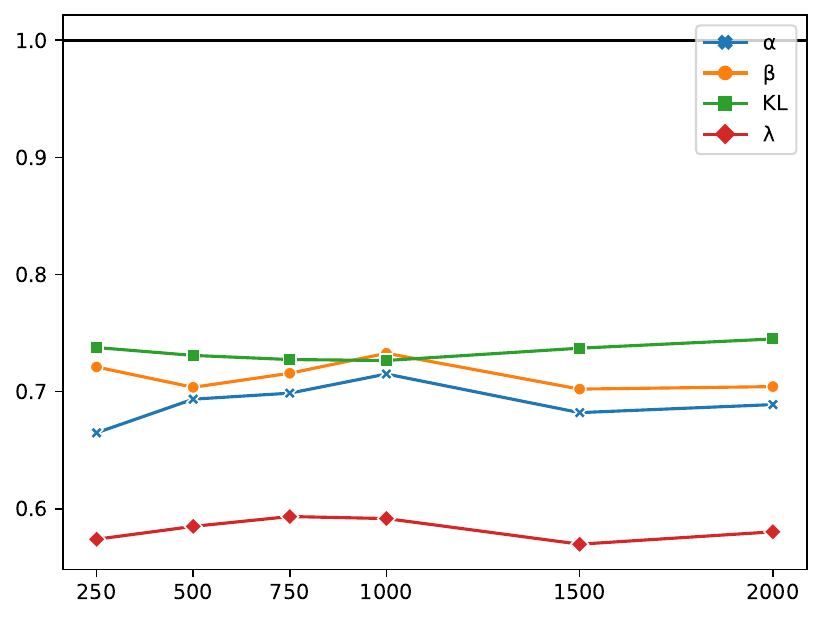}}\hfill
\subfloat[$(\alpha, \beta, \lambda) = (3,0.75,0.4)$]{\label{fig:estimation_rmse_c}\includegraphics[width=.33\linewidth]{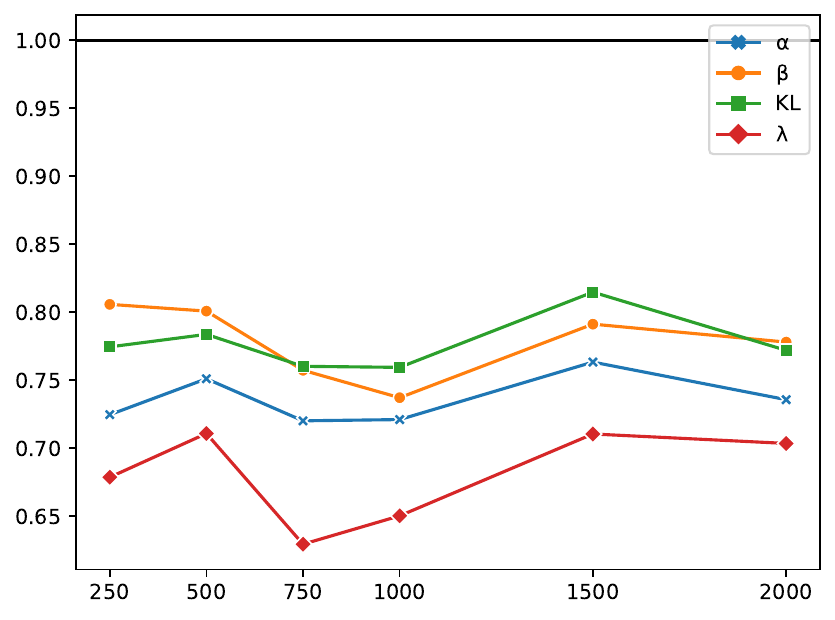}}\par 
\subfloat[$(\alpha, \beta, H, \delta) = (3,0.75,0.5,1)$]{\label{fig:estimation_rmse_d}\includegraphics[width=.33\linewidth]{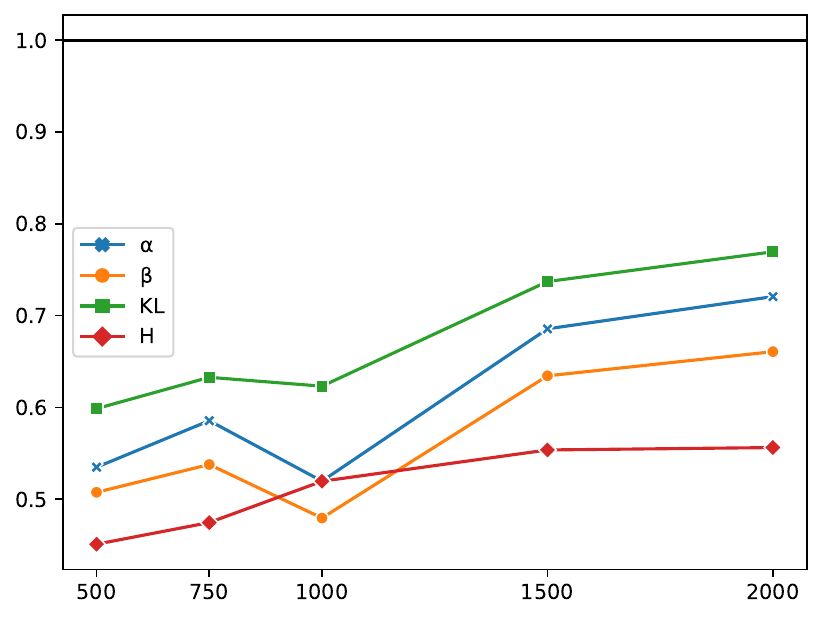}}\hfill
\subfloat[$(\alpha, \beta, H, \delta) = (3,0.75,1.5,1)$]{\label{fig:estimation_rmse_e}\includegraphics[width=.33\linewidth]{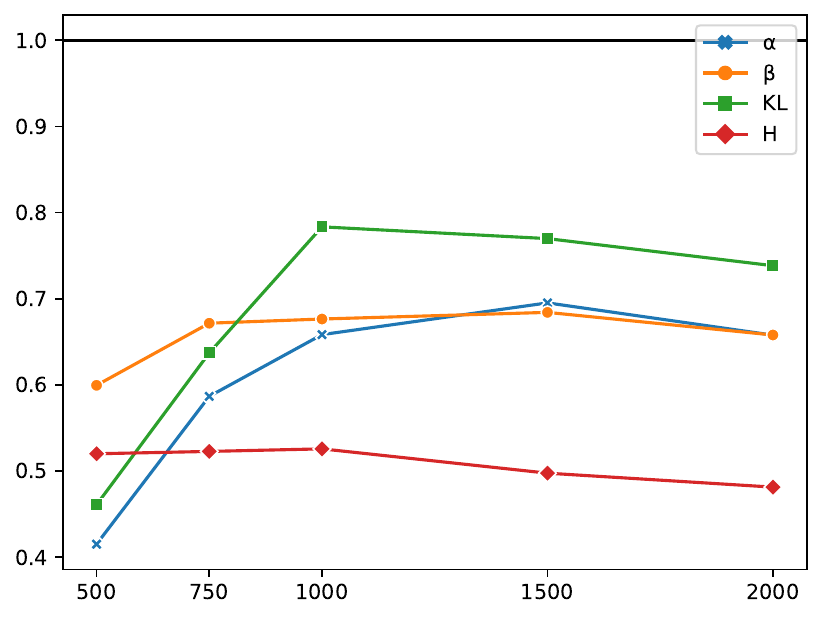}}\hfill
\subfloat[$(\alpha, \beta, H, \delta) = (3,0.75,2.5,1)$]{\label{fig:estimation_rmse_f}\includegraphics[width=.33\linewidth]{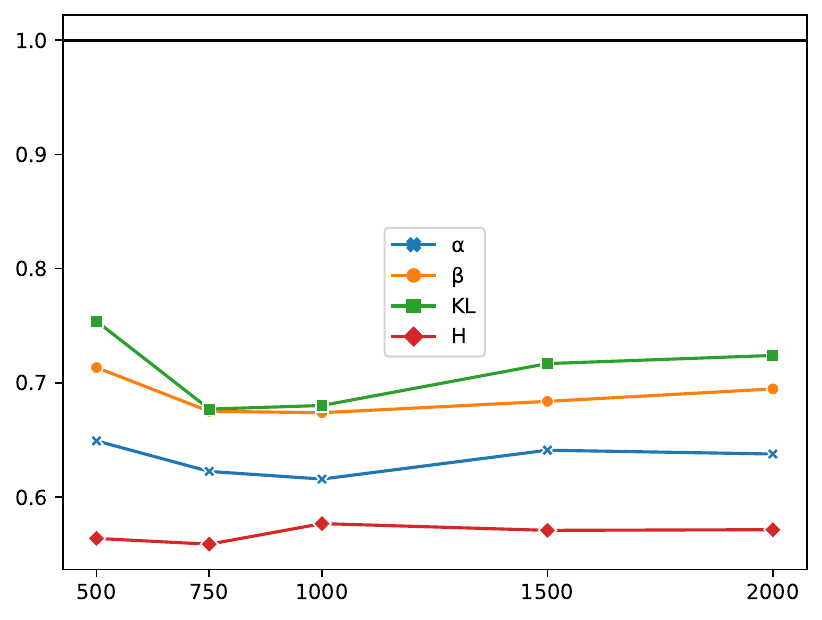}}\par 
\subfloat[$(\alpha, \beta, H, \delta) = (4,3,0.5,0.75)$]{\label{fig:estimation_rmse_g}\includegraphics[width=.33\linewidth]{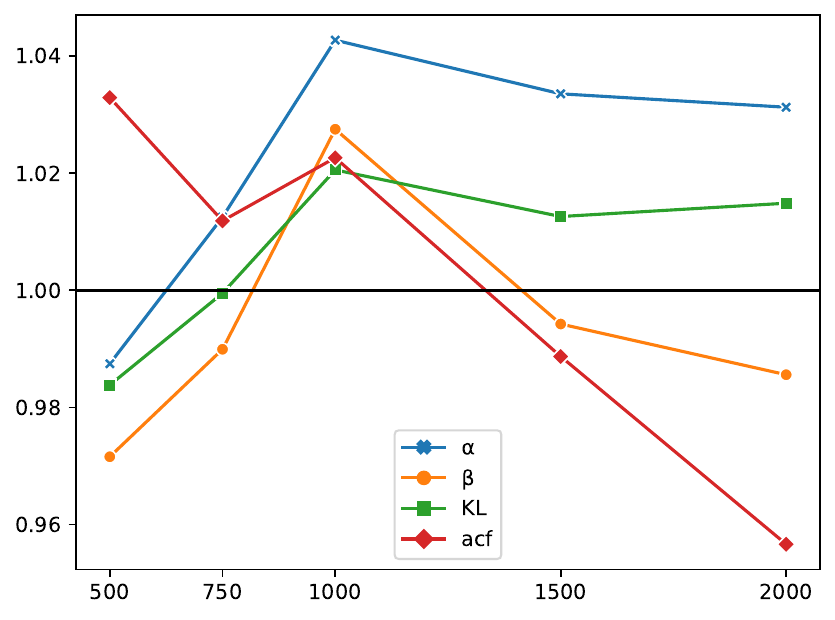}}\hfill
\subfloat[$(\alpha, \beta, H, \delta) = (4,3,1,1)$]{\label{fig:estimation_rmse_h}\includegraphics[width=.33\linewidth]{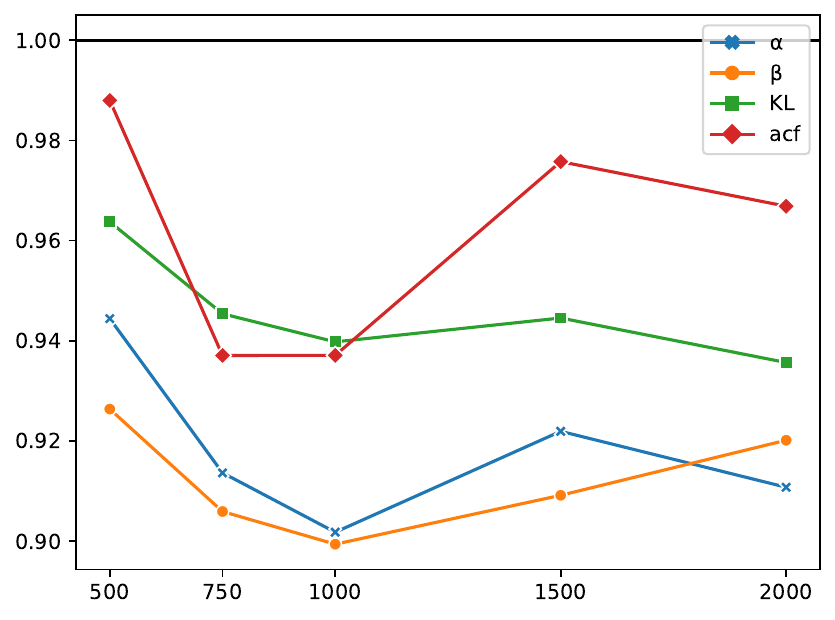}}\hfill
\subfloat[$(\alpha, \beta, H, \delta) = (4,3,2,3) $]{\label{fig:estimation_rmse_i}\includegraphics[width=.33\linewidth]{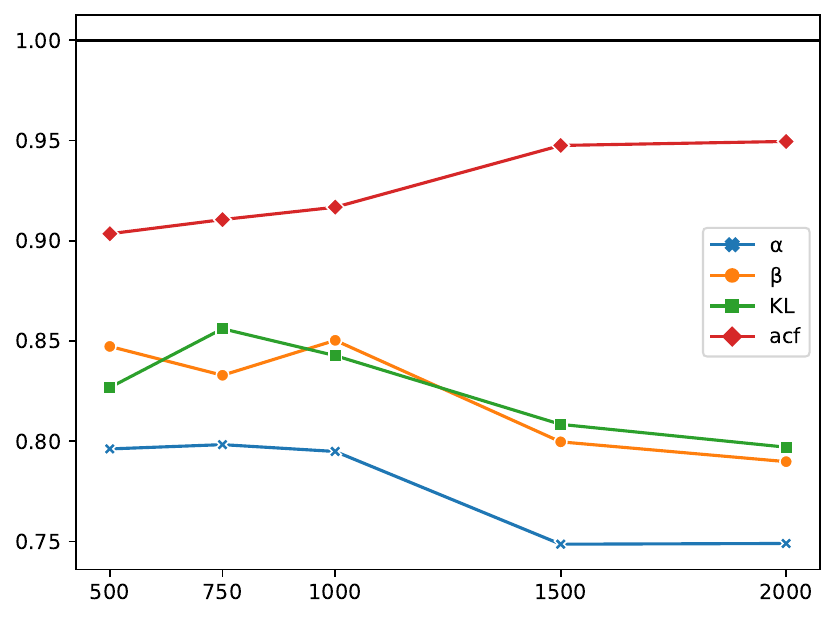}}
\caption{Ratio between estimation errors (rMSE, mean KL divergence and mean weighted $L^2$ distance) of the PL and GMM estimators for the parameters of the trawl process $X$ with Gamma L\'evy seed and trawl function parameterized by $\bt_{\phi}$, based on $100$ simulations. In this figure, we consider the exponential and Gamma trawl functions and $\bt_{\phi}$ is $\lambda$, $H$ or $(H,\delta)$. We display the results as a function of the length of the discretely observed path $(x_1,\ldots,x_n)$ to which we apply PL and GMM estimation, where $n \in \{250,500,750,1000,1500,2000\}$. A result below $1$ favours PL over GMM. We note that the PL performs significantly better.} 
\label{fig:inference_results_gamma_levy_seed_rMSE}
\end{figure}

We demonstrate in a simulation study that the PL estimator achieves a lower estimation error than the GMM estimator outside of the Gaussian case. We note that in the Gaussian case, the two estimators are almost identical. To this end, we conduct a simulation study and consider the trawl processes $X$ with $X_t\sim \textrm{Gamma}(\alpha,\beta)$ and multiple trawl functions. We set $(\alpha,\beta) = (3,0.75) $ for Figures \ref{fig:estimation_rmse_a}-\ref{fig:estimation_rmse_f} and $(\alpha,\beta) = (4,3)$ for Figures \ref{fig:estimation_rmse_g}-\ref{fig:estimation_rmse_i} and use the following parametric forms for the trawl function $\phi \colon (-\infty,0] \to \mathbb{R}_{\ge 0}$
\begin{itemize}
    \item[]Figures \ref{fig:estimation_rmse_a}-\ref{fig:estimation_rmse_c}: the exponential trawl function $\phi(t) = e^{\lambda t}$ from Example \ref{exponential_trawl_function_example} with $\lambda \in \{0.1,0.25,0.4\}$;
     \item[]Figures \ref{fig:estimation_rmse_d}-\ref{fig:estimation_rmse_f}: the Gamma trawl function $\phi(t) = \left(1-t\right)^{-(H+1)}$ from Example \ref{gamma_trawl_function_example} with $H \in \{0.5,1.5,2.5\}$ and where we fixed $\delta=1$;
    \item[]Figures \ref{fig:estimation_rmse_g}-\ref{fig:estimation_rmse_i}: the same Gamma trawl function $\phi(t) = \left(1-\frac{t}{\delta}\right)^{-(H+1)}$ with $(H,\delta) \in \{(0.5, 0.75), (1, 1), (2, 3) \}$.
\end{itemize} 
For \ref{fig:estimation_rmse_a}-\ref{fig:estimation_rmse_c} we infer $\bt = (\alpha,\beta,\lambda)$, for \ref{fig:estimation_rmse_d}-\ref{fig:estimation_rmse_f} we infer $\bt = (\alpha,\beta,H)$ and for \ref{fig:estimation_rmse_g}-\ref{fig:estimation_rmse_i} we infer $\bt = (\alpha,\beta,H,\delta)$. We compute the PL and GMM estimators from the discretely observed paths $(x_1,\ldots,x_n)$ of the trawl processes described above, where $n \in \{250, 500, 750, 1000, 1500, 2000\}$. It remains to compare the two estimators. We remind the reader that the distributional properties of the trawl process are fully determined by its marginal distribution and autocorrelation function, which we can compare to assess the relative performance of the two estimators. For the marginal distribution, we compare the root-mean-square estimation error (rMSE) of $\alpha$ and $\beta$, as well as the mean Kullback–Leibler divergence (KL) between the marginal distributions of the trawl process, as inferred by PL and GMM. For the autocorrelation structure, comparing the estimation error of each parameter is only meaningful for trawl functions with a single parameter, e.g.~$\lambda$ in \ref{fig:estimation_rmse_a}-\ref{fig:estimation_rmse_c} and $H$ in Figures \ref{fig:estimation_rmse_d}-\ref{fig:estimation_rmse_f}. For the two-parameter trawl function from Figures \ref{fig:estimation_rmse_g}-\ref{fig:estimation_rmse_i}, we can have very different values of $(H,\delta)$ giving similar shapes for the autocorrelation functions and comparing the per parameter estimation error is not meaningful. Instead, if $\phi_{\tiny{\text{PL}}}$ and $\phi_{\tiny{\text{GMM}}}$ are the infered autocorrelation functions, we compare the weighted $L^2$ distance given by 
\begin{equation*}
    \left(\int \frac{1}{1+kt^2}\left(\phi_{\tiny{\text{PL}}}(t) - \phi_{\tiny{\text{GMM}}}(t)\right)^2\mathrm{d}t\right)^{1/2},
\end{equation*}
where we set $k=0.01$. We require a weighting function as $\phi_{\tiny{\text{PL}}}$ and $\phi_{\tiny{\text{GMM}}}$ are not necessarily square-integrable, but note that the results are robust to varying $k$. We display the ratio of these quantities (rMSE, mean KL divergence and mean weighted $L^2$ distance) in Figure \ref{fig:inference_results_gamma_levy_seed_rMSE}; a ratio below $1$ favors the PL estimator and a ratio above $1$ favors the GMM estimator. Apart from Figure \ref{fig:estimation_rmse_g}, where the autocorrelation (acf) and KL ratios show opposite trends, the PL significantly outperforms the GMM estimator. Similar results are displayed for the mean absolute error (MAE) and median absolute error (MedAE) in Figures \ref{fig:inference_results_gamma_levy_seed_mae} and \ref{fig:inference_results_gamma_levy_seed_medae} and in Tables \ref{table_a}-\ref{table_i} in the Supplementary material. To understand the results better, we delve into the details of the numerical optimization routine.

\textbf{Practical considerations in the optimization procedure}

We carried out both the GMM and PL estimation procedures with the BFGS implementation from Python's Scipy library, which requires starting points. For the GMM procedure, $\alpha$ and $\beta$ were initialized at the method of the moments estimator, using the first two moments; $\lambda$, $H$ and $\delta$ were initialized at $1$ and finally, the PL procedure was initialized at the GMM estimator. For the long memory regime displayed in Figures \ref{fig:estimation_rmse_d}-\ref{fig:estimation_rmse_i}, the GMM procedure often diverged for $n=250$, i.e.~when the discretely observed path of the trawl process was too short; thus we display results starting at $n=500$ in these figures. Besides long memory, the optimization landscape is more complicated for trawl functions with more than one parameter, e.g.~Figures \ref{fig:estimation_rmse_g}-\ref{fig:estimation_rmse_i}, where multiple local maxima exist. We found that in some of these simulations, e.g. when the GMM estimator is far away from the true value, the PL routine does not move much from the starting point. 
A task for further research is to see if running the optimization with multiple starting points or using a gradient descent procedure specifically designed for stochastically estimated gradients improves the result.

Both procedures need a set of lags $K$. We found that both the PL and GMM procedures are robust in the sense that the choice of the hyperparameter $K$ does not influence the results considerably.
For Figure \ref{fig:inference_results_gamma_levy_seed_rMSE}, we used $K = (1,3,5,10,15,20,30,40)$ in \ref{fig:estimation_rmse_d} and \ref{fig:estimation_rmse_g}, $K = (1, 3, 5, 10, 15)$ in \ref{fig:estimation_rmse_a} and \ref{fig:estimation_rmse_e}, $K = (1, 3 ,5 ,10)$ in \ref{fig:estimation_rmse_a} and \ref{fig:estimation_rmse_f}, $K= (1, 3 ,5 ,7, 10 ,12)$ in \ref{fig:estimation_rmse_h},  $K=(1, 3, 5 ,7)$ in \ref{fig:estimation_rmse_i} and $K= (1,3,5)$ in \ref{fig:estimation_rmse_c}. 

\begin{figure}[h]
 \centering
\subfloat{\includegraphics[width=.33\linewidth]{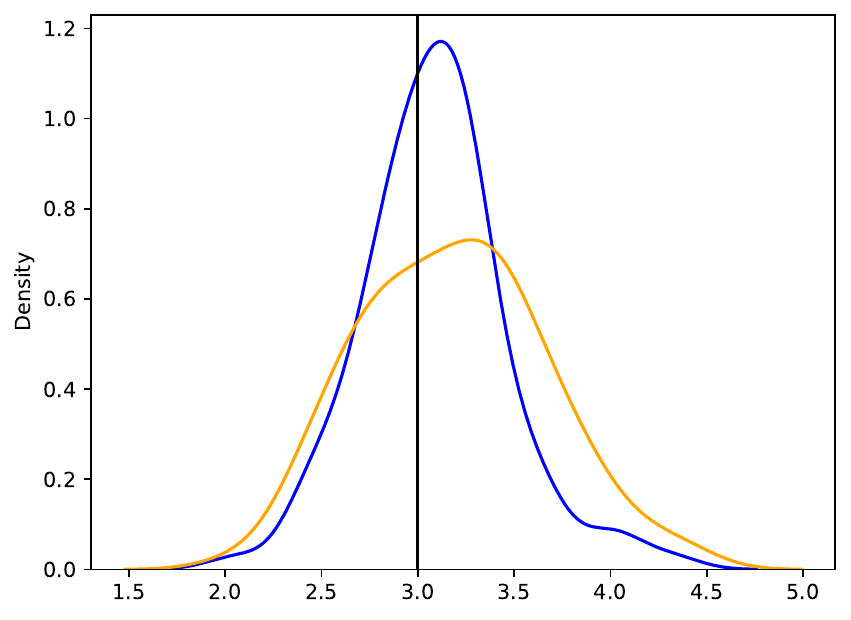}}\hfill
\subfloat{\includegraphics[width=.33\linewidth]{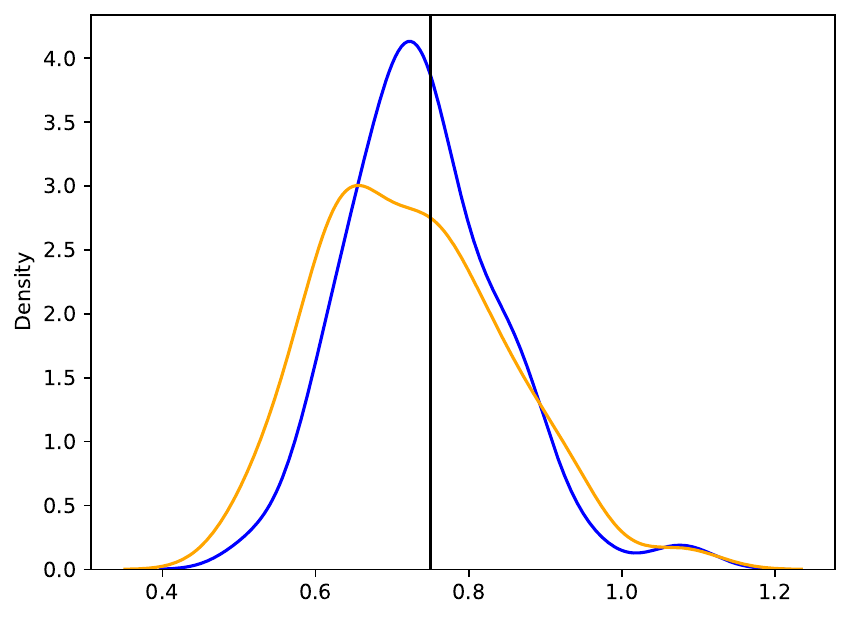}}\hfill
\subfloat{\includegraphics[width=.33\linewidth]{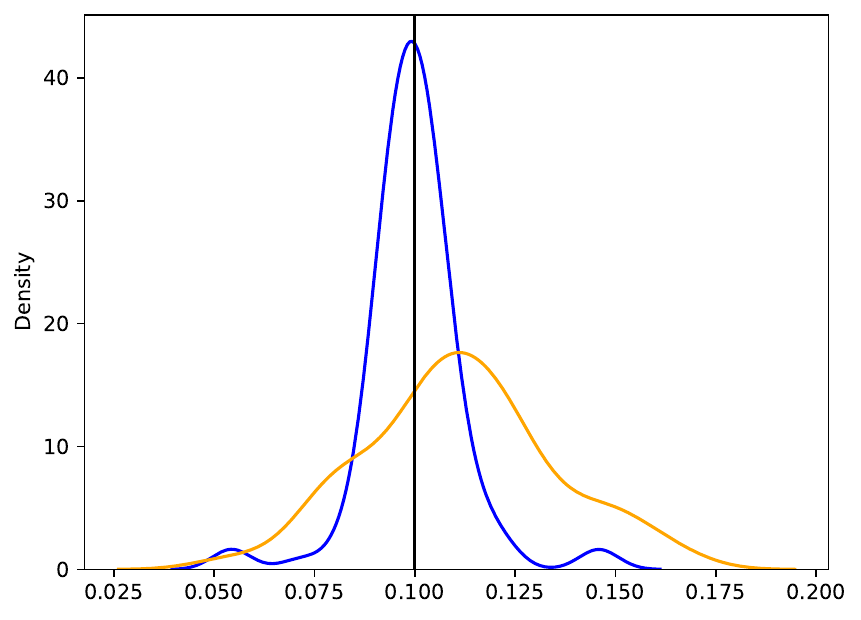}}
\caption{Kernel density estimates of the PL and GMM estimators in blue and yellow, respectively for $\alpha,\beta$ and $\lambda$, from left to right. We use the same simulation data as for Figure \ref{fig:estimation_rmse_a} with $n=1000$ and note that the PL is more concentrated around the true value, which is displayed by a vertical line.}\label{fig:empriical_distribution of the estimates}
\end{figure}

Finally, in Figure \ref{fig:empriical_distribution of the estimates} we display kernel density estimates of the parameters $\alpha,\, \beta $ and $\lambda $ inferred with the PL and GMM methodologies for the simulation study from Figure \ref{fig:estimation_rmse_a} and $n=1000$. Similar patterns in which the PL performs better can be observed for a wide range of simulation parameters.

\section{Applications to the forecasting of trawl processes}
\label{section:forecasting}
Let $\mathcal{F}_t = \sigma\left((X_s)_{s \le t}\right)$ be the $\sigma$-algebra generated by the trawl process $X$ up to time $t$ and $h>0$. In general, $X$ is not Markovian and $\ev[X_{t+h}|\mathcal{F}_t]$ is intractable, but we nevertheless approximate the distribution of $X_{t+h}|\mathcal{F}_t$ by that of $X_{t+h}| X_t$. Based on this idea, we introduce the first methodology for deterministic and probabilistic forecasting of real-valued trawl processes. For deterministic forecasting, we derive a novel conditional mean formula, which is optimal in the rMSE sense; we show that when used for forecasting, the parameters inferred by the PL methodology incur a smaller forecasting error than the ones inferred by the GMM methodology. For probabilistic forecasting, we discuss several methods to sample from the conditional distribution of $X_{t+h}|X_t$. These samples can be also used for other types of deterministic forecasting, such as conditional median forecast, which is optimal in the sense of Mean Absolute Error (MAE). Finally, we discuss the integer-valued case at the end of the section. 

We start with the conditional mean formula for deterministic forecasting.
\begin{theorem}\label{thm:non_markovian_forecast}
Assume that the L\'evy seed $L^{'}$ is integrable. Then 
\begin{equation*}
    \ev\left[X_{t+h} | X_t \right] = \frac{Leb(A_h\cap A)}{Leb(A)} X_t + \frac{Leb(A_h\backslash A)}{Leb(A)} \ev[X_t].
\end{equation*}
\end{theorem}
The above expression is a weighted average $\rho(h)X_t + (1-\rho(h))\ev[X_t]$ of the last observed value and of the mean, with $\rho(h) = \Corr{(X_t,X_{t+h})} = \frac{\textrm{Leb}(A_h \cap A)}{\textrm{Leb}(A)}$. We carry out a simulation study in the setting of the Gamma trawl process with exponential trawl function from Figure \ref{fig:estimation_rmse_a}. We use the PL and GMM methodologies to infer the parameter $\alpha,\, \beta$ and $\lambda$ from discretely observed paths of the trawl process of length $n \in \{250,750,1000,2000\}$. We further estimate $\ev[X_t] = \alpha \beta$ and $\rho(h) = e^{-\lambda h}$ and then use the forecasting formula from Theorem \ref{thm:non_markovian_forecast}. We approximate the PL and GMM out-of-sample forecasting errors (rMSE, MAE and MedAE) from $100$ simulated paths of the trawl process, each of length $2500$. We display in Table \ref{table:forecasting_error_ratio} the percentage change, i.e. $100 \cdot(\text{PL error} - \text{GMM error})/ \text{GMM error}$, as a function of $n$. We stress that the forecasting error is approximated on newly simulated paths of the trawl process, hence it is an out-of-sample error. We note that regardless of the lag at which we forecast, PL performs better than GMM, and that the improvement is bigger at smaller lags. 
\begin{table}[t]
\begin{adjustbox}{width=\textwidth}
\begin{tabular}{@{\extracolsep{3.5pt}}lrrrrrrrrrrrr@{}}
 & \multicolumn{3}{c}{$250$} & \multicolumn{3}{c}{$750$} & \multicolumn{3}{c}{$1000$} & \multicolumn{3}{c}{$2000$} \\\cline{2-4} \cline{5-7}  \cline{8-10}  \cline{11-13} 
 & MAE & MedAE & rMSE & MAE & MedAE & rMSE & MAE & MedAE & rMSE & MAE & MedAE & rMSE \\\hline
1 & -4.38 & -13.48 & -0.95 & -1.44 & -5.26 & -0.19 & -1.21 & -4.46 & -0.15 & -0.47 & -2.08 & -0.04 \\
2 & -3.80 & -9.13 & -1.51 & -1.19 & -3.43 & -0.34 & -1.00 & -2.77 & -0.26 & -0.37 & -1.15 & -0.06 \\
3 & -3.49 & -7.39 & -1.83 & -1.09 & -2.77 & -0.45 & -0.91 & -2.29 & -0.36 & -0.32 & -1.01 & -0.09 \\
4 & -3.23 & -6.32 & -1.98 & -1.01 & -2.40 & -0.53 & -0.82 & -1.97 & -0.42 & -0.29 & -0.86 & -0.10 \\
5 & -2.95 & -5.39 & -2.02 & -0.94 & -2.10 & -0.58 & -0.77 & -1.77 & -0.46 & -0.26 & -0.57 & -0.11 \\
6 & -2.68 & -4.79 & -1.98 & -0.88 & -1.84 & -0.61 & -0.70 & -1.54 & -0.49 & -0.24 & -0.52 & -0.12 \\
8 & -2.30 & -3.75 & -1.81 & -0.77 & -1.33 & -0.62 & -0.61 & -0.93 & -0.49 & -0.21 & -0.49 & -0.12 \\
10 & -1.93 & -3.02 & -1.56 & -0.65 & -1.06 & -0.56 & -0.51 & -0.75 & -0.44 & -0.18 & -0.34 & -0.10 \\
12 & -1.63 & -2.44 & -1.32 & -0.54 & -0.92 & -0.49 & -0.40 & -0.78 & -0.37 & -0.13 & -0.44 & -0.08 \\
15 & -1.20 & -1.80 & -1.01 & -0.39 & -0.68 & -0.38 & -0.29 & -0.48 & -0.28 & -0.08 & -0.18 & -0.05 \\\hline
\end{tabular}
\end{adjustbox}
\caption{Percentage change in the forecasting errors of the PL and GMM methodologies as a function of the length $n \in \{250,750,1000,2000\}$ of the path of the trawl process used to fit the parameters, displayed horizontally, and number of lags to forecast in the future $h \in \{1,2,3,4,5,6,8,10,12,15\}$, displayed vertically. We note that the parameters estimated by PL incur a smaller forecasting error.}
    \label{table:forecasting_error_ratio}
\end{table}

For probabilistic forecasting, note that $X_{t+h} = L\left(A_{t+h}\backslash A_t\right) + L(A_{t+h} \cap A_t)$ and that $L\left(A_{t+h}\backslash A_t\right)$ is independent of $X_{t} = L(A_t)$. The distribution of $X_{t+h}|X_t$ is given by the convolution of the distributions of $L\left(A_{t+h}\cap A_t\right)|X_t$ and $L(A_{t+h}\backslash A_t)$ and, to sample from $X_{t+h}|X_t$, it is enough to sample from $L\left(A_{t+h}\cap A_t\right)|X_t$ and from $L(A_{t+h}\backslash A_t)$. Assume we have a sampler for $L(A_{t+h}\backslash A_t)$, which is the case for all the examples considered in this paper. If $L' $ is Gaussian or Gamma distributed, then $L(A_{t+h}\cap A_t |X_t)$ is Gaussian, respectively Beta distributed. In general, the conditional distribution of $L(A_{t+h} \cap A_t)|X_t$ does not belong to named families, and we discuss two exact sampling techniques which generate independent samples. Note that the conditional density can be computed explicitly, as a product of known densities and a known normalizing constant.
\begin{equation*}
    p_{L(A_{t+h}\cap A_t)|X_t}(l|x_t) = \frac{p_{L(A_t \cap A_{t+h})}(l)\,  p_{L(A_t \backslash A_{t+h})}(x_t-l)}{p_{X_t}(x_t)}. 
\end{equation*}
\cite{a_r_s_f_p_o_d} provides an exact sampling algorithm for the case where the density which requires sampling is proportional to a product of densities for which exact sampling methods are available. The authors perform rejection sampling and automate the choice of both the envelope function and upper bound constant. The algorithm is efficient even when dealing with a product of two heavy-tailed distributions, such as Cauchy, and does not require any extra information on the densities. The algorithm is inefficient and gives small acceptance rates when the two densities are peaked around two different values, e.g.~$\mathcal{N}(0,0.01)$ and $\mathcal{N}(1,0.01)$. In this case, we require further information on the densities. If for example we have a concave or log-concave density, we can perform efficient rejection sampling with a piecewise-linear, respectively piecewise-exponential envelope
\citep{ccarj}. The method can also be applied if we have access to the decomposition of the density or of its logarithm into convex and concave components. The above sampling techniques can further be combined with MCMC methods, such as in Adaptive Rejection Metropolis-Hastings, although these extensions only produce dependent, asymptotic samples from the target density.

For Integer-valued trawl processes (IVT), the conditional mean forecast from Theorem \ref{thm:non_markovian_forecast} is not integer-valued, hence it is not data-coherent. Instead,  the conditional median or mode can be used, and, the conditional distributions are often part of named distributions (see Section 4 of \cite{cl_integer_trawl}). Even when this is not the case, the conditional probabilities are given by finite or countable sums, which can be calculated exactly or approximated by truncation. In both the real-valued and integer-valued cases and for both deterministic and probabilistic forecasting, the forecasting error only shows minimal improvements in spite of the significantly improved parameter fit. One potential explanation is the simplicity of our forecasting formula, which only allows conditioning on the last lag. A task for further research is to derive a forecasting formula conditional on more lags and investigate the associated forecasting error. We expect that with more lags involved, the improved PL parameter fit, together with the non-Markovianity of trawl processes will result in further improvements in the forecasting error. 
\section{Conclusion}\label{section:conclusion}
This paper develops the first likelihood-based methodology for the inference of continuous-time, real-valued trawl processes and introduces a novel deterministic forecasting formula for these processes. The contributions are threefold. 

Firstly, we reduce the variance of the estimators for the pairwise likelihood (PL) and its gradients by several orders of magnitude and show that PL inference for trawl processes is accurate and computationally efficient. 
We provide Python implementations at \cite{Leonte_Ambit_Stochastics_2022} which integrate our methodology with automatic differentiation engines, eliminating the need for manual calculations and enabling easy adaptation to fit trawl processes with other marginal distributions and trawl functions.

Secondly, we demonstrate the excellent finite sample properties of the PL estimator in a simulation study, showing a large reduction in estimation error compared to the generalized method of moments (GMM) estimator. The PL estimator consistently and significantly outperforms the GMM estimator, regardless of the number of lags used, length of the discretely observed path of the trawl process, or metric employed to evaluate the parameter fit: mean squared or absolute error, median absolute error or KL divergence.

Thirdly, we derive a novel conditional mean forecasting formula and present the first methodology for the probabilistic forecasting of continuous-time, real-valued trawl processes. In a simulation study, we demonstrate that the PL estimator outperforms the GMM estimator in out-of-sample forecasting errors.

Finally, our work begins to bridge the gap between theoretical studies of trawl processes and ambit fields and the use of deep learning techniques to fit the model parameters. We hope that this article will contribute towards both the methodological development of the field of Ambit Stochastics and its integration into routine statistical modelling toolkits.


\textbf{Acknowledgements}

We would like to thank Dan Crisan and Tony Wang for constructive discussions and comments on earlier versions of the manuscript. Dan Leonte acknowledges support from the EPSRC Centre for Doctoral Training in Mathematics of Random Systems: Analysis, Modelling and Simulation (EP/S023925/1).

\bibliographystyle{agsm} 
\bibliography{main_bibliography}
\newpage 
\renewcommand{\thesection}{}
\section{\hspace{-21pt} Appendix}
\renewcommand{\thesection}{A}
\label{appendix:parameterizations}
%
%
%
%
%
%
%

The appendix contains a list of the parameterizations of all the probability distributions used in this paper: Poisson, Negative binomial (NB), Skellam, Uniform, Beta, Gamma, inverse Gaussian (IG), Gaussian ($\mathcal{N}$), double-sided Maxwell ($\mathcal{M}$), Normal-inverse Gaussian (NIG), and Variance-gamma (VG).
%
%
 We denote $\{0,1,\ldots\}$ by $\Z^{+}$ and $(0,\infty)$ by $\R^{+}$. Let $I_{\alpha}$ and $K_\alpha$ be the modified Bessel function of the first, respectively second kind.
 
\begin{tabular}{ |c||c|c|c|  }
 \hline
 \multicolumn{4}{|c|}{Discrete Distributions} \\
 \hline \hline
 Distribution& Range & Parameters & PMF \\
 \hline
$\textrm{Poisson}(\lambda)$ & $ \Z^{+} $   & $\alpha < \beta \in \R$ &   $\frac{\lambda^x e^{-\lambda}}{x!}$  \\  \hline
 $\textrm{NB}(m,p)$& $\mathbb{Z}^{+}$ &   \begin{tabular}{c} $m \in \R^{+}$ \\$p \in [0,1]$\end{tabular} & $\frac{\Gamma(m+x)}{\Gamma(m) x!}(1-p)^m p^x$ \\ \hline
 $\textrm{Skellam}(\mu_1,\mu_2)$& $\mathbb{Z}$ & $\mu_1, \mu_2 > 0$  & $e^{-(\mu_1+\mu_2)} \left(\frac{\mu_1}{\mu_2}\right)^{x/2} I_{|x|}(2 \sqrt{\mu_1 \mu_2})$  \\ \hline
%
 \hline
 \multicolumn{4}{|c|}{Continuous Distributions} \\
 \hline \hline
 Distribution& Range & Parameters & PDF  \\[5pt]  \hline
$\mathcal{U}(\alpha,\beta)  $ & $ (\alpha,\beta) $   & $\alpha < \beta \in \R$ &   $\frac{1}{\beta - \alpha}$  \\[5pt]  \hline
 $\textrm{Beta}(\alpha,\beta)$ & $(0,1)$ & $\alpha,\beta>0$&  $\frac{x^{\alpha-1}(1-x)^{\beta-1}}{\textrm{B}(\alpha,\beta)}$ \\[5pt] \hline
  $\textrm{Gamma}(\alpha,\beta)$& $\R^{+}$    & $\alpha, \beta \in \R^{+}$   & $\frac{\beta^\alpha}{\Gamma(\alpha)} x^{\alpha-1} e^{-\beta x}$   \\ \hline
 $\textrm{IG}(\mu,\lambda)$& $\R^{+}$     & $\mu, \lambda \in \R^{+}$ & $\sqrt{\frac{\lambda}{2 \pi x^3}}  e^{-\frac{\lambda(x-\mu)^2}{2 \mu^2 x}}$
 
 \\ \hline
$\mathcal{N}(\mu,\sigma^2)$ & $\R$ & \begin{tabular}{c} $\mu \in \R$ \\$\sigma^2 \in \R^{+}$\end{tabular}
  &  $\frac{1}{\sqrt{2 \pi \sigma^2}}e^{-\frac{(x-\mu)^2}{2\sigma^2}}$ \\ \hline
   $\mathcal{M}(\mu,\sigma^2)$ & $\R$& \begin{tabular}{c} $\mu \in \R$ \\$\sigma^2 \in \R^{+}$\end{tabular} & $\frac{1}{\sqrt{2 \pi}} \frac{(x-\mu)^2}{\sigma^3} e^{-\frac{(x-\mu)^2}{2\sigma^2}}$\\ \hline
   $\textrm{NIG}(\alpha,\beta,\delta,\mu)$& $\R$  & \begin{tabular}{c} $\alpha, \beta, \mu, \delta \in \R$ \\ $ \gamma \defeq \sqrt{\alpha^2 - \beta^2} \in \R $\end{tabular}   & $\frac{\alpha  \delta  K_1\left(\alpha   \sqrt{\delta^2 + (x - \mu)^2}\right)}{\pi  \sqrt{\delta^2 + (x - \mu)^2}}                   e^{\delta \gamma + \beta  (x - \mu)}$ \\ \hline 
\end{tabular}
\begin{definition}[Normal variance-mean mixture]\label{def:normal_variance_mean_mixture}
We say $X$ follows a normal variance-mean mixture distribution with mixing density $g$ if 
\begin{equation*}
    X = c_1 + c_2 Z + c_3 \sqrt{Y} W,
\end{equation*}
where $Y$ has pdf $g$, $W \sim \mathcal{N}(0,1)$ and $Z$ and $W$ are independent. Then $X|Y \sim (c_1 + c_2 Y,c_3^2 Y)$. We discuss two particular cases.
\end{definition}
If the constants are $c_1 = \mu, \, c_2 = \beta, \, c_3 = \delta$ and the mixing density is given by $Y \sim  \textrm{IG}(1/\gamma,1)$, where $\gamma = \sqrt{\alpha^2-\beta^2}$, then $X| Y \sim \mathcal{N}(\mu+ \beta Y, \delta^2 Y)$. We obtain the the Normal-inverse Gaussian distribution, and $X \sim \textrm{NIG}(\alpha,\beta,\delta,\mu)$.



%
%
%
%



\renewcommand{\thefootnote}{\fnsymbol{footnote}}
\newpage
	\pagenumbering{arabic}
	\renewcommand*{\thepage}{S\arabic{page}}
	\renewcommand{\appendixtocname}{Supplementary material}
\renewcommand{\thesection}{}
\section{\hspace{-21pt} Supplementary material}
\renewcommand{\thesection}{S}
\begin{center}
 \large \bf{Likelihood-based inference and forecasting for trawl processes: a stochastic optimization approach}
\end{center}
\vskip 2em
\begin{center}
  \large \lineskip .75em%
  \begin{tabular}[t]{c}
    Dan Leonte and Almut E. D. Veraart \\
	Department of Mathematics, Imperial College London
  \end{tabular}\par 
\end{center}

\vspace{2\baselineskip}

The supplementary material is structured as follows.  
\begin{itemize}
    \item Section \ref{supplementary_section:proofs} contains technical derivations and proofs. Subsection \ref{supplementary_subsection:asymptotic_log_ratio_estimators} discusses the bias, variance and probabilistic bounds for the estimators of the pairwise density and its gradient. We derive expressions for the asymptotic bias and variance of these estimators and establish the corresponding convergence rates. Subsection \ref{supplementary_subsection:proofs} gives the proof of the novel deterministic forecasting formula.
\item Section \ref{supplementary_section:measure_valued_grad} derives a novel chain rule for stochastic transformations and introduces hybrid estimators, which combine PG and MVG and greatly extends the applicability of our methodology; in particular, it allows for PL inference of integer-valued trawl processes.    
\item Section \ref{supplementary_section:extendend_simulation_studies} expands on the simulation studies from Section \ref{section:variance_reduction_methods}.
\end{itemize}
\subsection{Proofs}
\label{supplementary_section:proofs}
In this section we present the proofs which were omitted from the main body of the paper.
\subsubsection{Asymptotic bias, variance and probabilistic bounds}\label{supplementary_subsection:asymptotic_log_ratio_estimators}
Fix $x_s,x_t \in \R$ and let $\xi(\bt) =  p(x_s,x_t;\bt)$ for $\bt \in \R^{d}$. We have seen in Subsection \ref{subsection:trawl_inference_as_a_stoch_optim_problem} that both the pairwise density and its gradient can be written as Monte Carlo (MC) estimators, i.e. $\xi(\bt) = \ev_{q(z;\bt)}[f(z,\bt)]$ and $\nabla_{\bt}\xi(\bt) = \ev_{q(z;\bt)}[r(z,\bt)]$ for some functions $f\colon \R \times \R^d \to \R$, $r \colon \R \times \R^d \to \R^d $ and density $q(;\bt)$ parameterized by $\bt$. The function $f$ is the product of marginal densities of the trawl process (see \eqref{eq:gradient_pre_tricks}), hence strictly positive; $h$ is given by $f(z) \nabla_{\bt} \log{q(z;\bt)} + \nabla_{\bt} f(z,\bt)$ in the SF estimator and by $\pdv{f}{z} \nabla_{\bt}z + \nabla_{\bt} f(z,\bt)$ in the PG estimator. Finally, $q$ is either the density of $L(A)$ for some set $A$, if $L^{'}$ is real-valued, or the truncation of such a density to $[0,\min{(x_s,x_t)}]$ if $L^{'}$ is positive real-valued. We show next that the properties of $q$ determine the rate of convergence of the MC estimators.

\textbf{Likelihood estimation and the Jensen gap}

Let $Z_1,\ldots,Z_N$ be a sequence of iid random variables with density $q(\cdot;\bt)$. Then the random variables $U_1, \ldots, U_N$ given by $U_i = f(Z_i)$ for $1 \le i \le n$ are strictly positive, iid, unbiased and consistent estimators for $\xi(\bt)$. Let $\overline{U} = \sum_{i=1}^N U_i/N$. By Jensen's inequality, the estimator $\log{\overline{U}}$ overestimates $\log{\xi(\bt)}$. To quantify the properties of this estimator, we derive its asymptotic bias and variance. By applying the Taylor expansion $\log{(1+x)} = \sum_{n=1}^\infty \frac{(-1)^{n+1}}{n}
x^n$, we have that
\begin{equation*}
\log{\overline{U}} = \log{\xi(\bt)} + \log{\left(1 + \left(\frac{\overline{U}}{\xi(\bt)} -1\right)\right)} = \log{\xi(\bt)} +  \left(\frac{\overline{U}}{\xi(\bt)} - 1\right) - \frac{1}{2}\left(\frac{\overline{U}}{\xi(\bt)} - 1\right)^2 + \ldots
\end{equation*} 
By taking the expectation and variance of both sides and neglecting higher-order terms, we obtain that
 \begin{align*}
   \ev[\log{\overline{U}}] &= 
    \log{\xi(\bt)} -\frac{1}{2} \ev\left[\left(\frac{\overline{U}}{\xi(\bt)} - 1\right)^2 \right] + \ldots = \log{\xi(\bt)} -  \frac{\Var{(U_1)}}{2 N \xi(\bt)^2} + \mathcal{O}(1/N^2), \\
    \Var{\left(\log{\overline{U}}\right)}    &= \Var{\left(\frac{\overline{U}}{ \xi(\bt)} -1\right)} + \ldots =  \frac{\Var{(U_1)}}{N\xi(\bt)^2} + \mathcal{O}(1/N^2),
\end{align*}
hence the asymptotic bias and variance for $\log{\overline{U}}$ are given by \begin{equation*}-\frac{1}{N} \frac{\Var{(U_1)}}{2{\xi(\bt)^2}} \text{ and }  \frac{1}{N}\frac{\Var{(U_1)}}{{\xi(\bt)^2}}, \text{ respectively}.
\end{equation*}
This highlights the importance of having low variance estimators, not only to reduce the variability, but also the bias of $\log{\overline{U}}$. Further, the estimated value of $\frac{\Var{(U_1)}}{N\xi(\bt)^2}$ can be used for an adaptive procedure, to calibrate the number of required samples for each pair $x_s,x_t$. We can also provide non-asymptotic, probabilistic bounds. By the mean value theorem, we have that $\log{\overline{U}} - \log \xi(\bt) = \frac{1}{U^{*}}(\overline{U} - \xi(\bt))$ for some random variable $U^{*}$ in between $\overline{U}$ and $\xi(\bt)$ a.s. Further
\begin{equation*}
\mathbb{P}\left(\left|\log{\overline{U}} - \log{\xi(\bt)}\right| > \epsilon \right) \le \mathbb{P}\left(\left|\log{\overline{U}} - \log{\xi(\bt)}\right| > \epsilon,\, \left|\overline{U} - \xi(\bt)\right| < \epsilon  \right)  + \mathbb{P}\left(\left|\overline{U} - \xi(\bt)\right| > \epsilon \right). 
\end{equation*} 
For $0 < \epsilon < \xi(\bt)/2$, we have that $\left|\overline{U}-\xi(\bt)\right| < \epsilon$ implies that $\frac{1}{U^{*}} < \frac{2}{\xi(\bt)}$, hence \begin{align*}
& \left\{\left|\log{\overline{U}} - \log{\xi(\bt)}\right| > \epsilon,\, \left|\overline{U} - \xi(\bt)\right| < \epsilon  \right\} \subset \left\{\frac{2}{ \xi(\bt) }\left|\overline{U} - \xi(\bt)\right| > \epsilon,\, \left|\overline{U} - \xi(\bt)\right| < \epsilon  \right\} \\ =& \left\{\frac{ \xi(\bt) }{ 2} \epsilon <\left|\overline{U} - \xi(\bt) \right| < \epsilon   \right\} \subset \left\{ \left|\overline{U} - \xi(\bt)\right| > \frac{\xi(\bt)}{2}\epsilon  \right\}
\end{align*}
and 
\begin{align*}
\mathbb{P}\left(\left|\log{\overline{U}} - \log{\xi(\bt)}\right| > \epsilon \right) &\le \mathbb{P}\left(\left|\overline{U} - \xi(\bt)\right| > \frac{\xi(\bt)}{2}\epsilon \right) + \mathbb{P}\left(\left|\overline{U} - \xi(\bt)\right| > \epsilon \right) \le 
 2 \mathbb{P}\left(\left|\overline{U} - \xi(\bt)\right| > \epsilon'\right),
\end{align*}
where $\epsilon' \defeq \max{\left(\epsilon, \frac{\xi(\bt)}{2}\right)}$. The rate of decay in the above inequality depends on the distribution of $U_i = f(Z_i)$, where $Z_i$ is distributed according to $q$. If for example $U_i$ is compactly supported, as in Example \ref{ex:pl_as_exp_for_gamma_levy_basis}, then Lemma \ref{lemma:Hoeffding's_inequality} applies and the convergence of $\overline{U} = \frac{U_1 + \ldots + U_n}{n}$ to $\xi(\bt) =\ev_{q(z;\bt)}[f(z;\bt)] = p(x_s,x_t;\bt)$ is exponential in the number of samples $n$, with an exponential rate depending on the pair $x_s,\, x_t$. The exponential convergence also holds if $U_i$ is sub-Gaussian, as in the case of the Gaussian L\'evy seed, or sub-exponential, as in the case of the IG and NIG L\'evy seeds. The exponential rates of convergence depend again on $x_s,\, x_t$ and can be deduced from Lemma \ref{lemma:chernoff}. If $U$ has a heavy tailed distribution, as is the case for Cauchy L\'evy seeds, the convergence is polynomial and not exponential. A detailed analysis of concentration inequalities can be found in \cite{MR3837109}.

\underline{Concentration inequalities}
\begin{lemma}[Chernoff bounds]\label{lemma:chernoff}
Let $X$ be a real-valued random variable. Then
\begin{equation*}
\mathbb{P}\left(X - \ev[X] > \epsilon\right) < \inf_{\lambda \ge 0} \frac{\ev\left[e^{\lambda X}\right]}{e^{\lambda \epsilon }}.
\end{equation*}
\end{lemma}
A similar bound can be obtained for $\mathbb{P}\left(X - \ev[X] < \epsilon\right)$ by replacing $X$ with $-X$ in the above inequality. We specialize the above to sub-Gaussian and sub-exponential distributions. In the following, $X_i$ are iid with mean $\mu$ and $S_n = \frac{\sum_{i=1}^n X_i}{n}$.

We say $X$ is $\sigma^2$ sub-Gaussian if for any $\lambda >0, \, \ev\left[e^{\lambda \left(X-\ev[X]\right) }\right] \le e^{\lambda^2 \sigma^2/2}$. If $X_i$ are iid, $\sigma^2$ sub-Gaussian random variables with mean $\mu$, then $S_n $ is $n\sigma^2$ sub-Gaussian and
\begin{equation*}
\mathbb{P}\left(\left|S_n- \mu \right| > \epsilon\right) < 2 \exp{\left(-\frac{n \epsilon^2}{2 \sigma^2}\right)}.
\end{equation*}
Alternatively, we say $X$ is $(\tau^2,b)$ sub-Exponential if $\ev[e^{\lambda \left(X - \ev[X]\right)} ] \le e^{\frac{\lambda^2 \tau^2}{2}}$ for $|\lambda| \le \frac{1}{b}$. Then
\begin{equation*}
\mathbb{P}\left(\left|S_n- \mu \right| > \epsilon\right) < 2 \exp{\left(-\min{\left(\frac{n \epsilon^2}{2 \tau^2},\frac{n \epsilon}{b}\right)}\right)}.
\end{equation*}
If $X_i$ are compactly supported, we can further refine the above Chernoff bounds. 
\begin{lemma}[Hoeffding’s inequality]\label{lemma:Hoeffding's_inequality} Let $X_1,\ldots,X_n$ be iid random variables with mean $\mu$ such that $a\leq X_{i}\leq b$ a.s.~for $1 \le i \le n$. Let $S_{n} = \frac{X_1 + \ldots + X_n}{n}, \, \mu = \ev[X_i] $ . Then 
\begin{equation*}
\mathbb{P}\left(\left|S_n -\mu\right| \ge  \epsilon\right)  \le  2\exp{\left( -\frac{2n\epsilon^2}{  (b-a)^2}\right)}.
\end{equation*}
\end{lemma}
\textbf{Gradient estimation}

As before, let $Z_1,\ldots,Z_N$ be a sequence of iid random variables with density $q(\cdot;\bt)$, $U_i = f(Z_i)$ and further $V_i = r(Z_i)$. Although the sequences $U_i$ and $V_i$ have independent terms, $U_i$ and $V_i$ are dependent. By taking the Taylor expansion of $u,v \to \frac{v}{u}$ 
\begin{equation*}
    \frac{v}{u}  = \frac{v_0}{u_0} + \frac{1}{u_0} \left(-\frac{v_0}{u_0} \Delta u + \Delta v \right) + \frac{2}{u_0^2}\left(\frac{v_0}{u_0}(\Delta u)^2 - \Delta u \Delta v\right) + \ldots
\end{equation*}
at $u_0 = \xi(\bt), \, v_0 = \nabla_{\bt}\xi(\bt)$ and where $\Delta u = \overline{U} - \xi(\bt), \, \Delta v = \overline{V} - \nabla_{\bt} l (\bt)$, we obtain that
\begin{align*}
    \overline{V} / \overline{U}&= \nabla_{\bt} \xi(\bt) / \xi(\bt) +  \frac{1}{\xi(\bt)}\left[-\nabla_{\bt}\log{\xi(\bt)}\left(\overline{U}-\xi(\bt)\right) + \left(\overline{V} - \nabla_{\bt} \xi(\bt)\right)\right] \\ &+ \frac{2}{\xi(\bt)^2}\left[\nabla_{\bt}\log{\xi(\bt)} \left(\overline{U}-\xi(\bt)\right)^2 -\left(\overline{U}-\xi(\bt)\right) \left(\overline{V}-\nabla_{\bt}\xi(\bt)\right) \right] + \ldots
\end{align*}
By taking the expectation and variance of both sides and neglecting higher-order terms, we obtain that
\begin{align}
    \ev[\overline{V} / \overline{U}]&= \nabla_{\bt}\log{\xi(\bt)} + \frac{2}{\xi(\bt)^2}\ev\left[\nabla_{\bt}\log{\xi(\bt)} \left(\overline{U}-\xi(\bt)\right)^2 -\left(\overline{U}-\xi(\bt)\right) \left(\overline{V}-\nabla_{\bt}\xi(\bt)\right) \right] + \ldots \notag \\
    &= \nabla_{\bt}\log{\xi(\bt)} + \frac{2}{ \xi(\bt)^2}\left[\nabla_{\bt}\log{\xi(\bt)}\Var{\left(\overline{U}\right)} -\Cov{(\overline{U},\overline{V})}\right] \notag \\
    &=\nabla_{\bt}\log{\xi(\bt)} + \frac{2}{N \xi(\bt)^2}\left[\nabla_{\bt}\log{\xi(\bt)} \Var{(U_1)} - \Cov{(U_1,V_1)}\right], \label{eq:asymp_bias_grad} \\
    \Var{(\overline{V} / \overline{U})}&=\frac{1}{\xi(\bt)^2}\Var{\left(-\nabla_{\bt}\log{\xi(\bt)}\left(\overline{U}-\xi(\bt)\right) + \left(\overline{V} - \nabla_{\bt} \xi(\bt)\right)\right)} + \ldots \notag\\
    &= \frac{2}{\xi(\bt)^2} \Var{\left(-\nabla_{\bt}\log{\xi(\bt)}\overline{U} + \overline{V}\right)} + O(1/N^2) \notag \\ &= \frac{1}{N \xi(\bt)^2} \Var{\left(\nabla_{\bt}\log{\xi(\bt)}U_1 - V_1\right)} + O(1/N^2) \label{eq:asymp_variance_grad},
\end{align}
where to obtain \eqref{eq:asymp_bias_grad} and \eqref{eq:asymp_variance_grad}, we expand the variance and covariance terms and use that $U_i$ and $V_j$ are independent when $i \neq j$. Note that Jensen's inequality does not apply to the map $\overline{U},\overline{V} \to \overline{V}/\overline{U}$, as  $\overline{U}$ and $\overline{V}$ are dependant, and the ratio estimator can have both positive and negative asymptotic bias. The asymptotic bias and variance of $\Var{(\overline{V} / \overline{U})}$ are then given by
\begin{equation*}
    \frac{1}{N} \frac{\nabla_{\bt}\log{\xi(\bt)} \Var{(U_1)} - \Cov{(U_1,V_1)}}{ \xi(\bt)^2} \text{ and } \frac{1}{N} \frac{\Var{\left(\nabla_{\bt}\log{\xi(\bt)}U_1 - V_1\right)}}{ \xi(\bt)^2},
\end{equation*}
respectively. Non-asymptotic bounds on the deviation of the gradient estimators from the gradient of the log-likelihood can be produced individually for each family of distributions $q$, by analyzing the asymptotic growth or decay of $\nabla_{\bt}z$ as $z \to \pm \infty$.

\subsubsection{Forecasting formula proof}\label{supplementary_subsection:proofs}
We now prove Theorem \ref{thm:non_markovian_forecast}, for which we need the following two results.
\begin{lemma}
\label{lemma:exchangeable}
Let $Y_1,\ldots,Y_n $ be integrable and exchangeable real-valued random variables. Then 
\begin{equation*}
    \ev[Y_1|Y_1 + \ldots + Y_n] = \frac{Y_1 + \ldots + Y_n}{n}. 
\end{equation*}
\end{lemma}
\begin{lemma}\label{lemma:unif_inetgrability}Let $X_1,X_2,\ldots$ be a sequence of weakly convergent infinitely divisible random variables, with L\'evy-Khintchine triplets $(\xi_n,a_n,l_n)$. Then for $\alpha>0$:
\begin{equation*}
\lim_{a \to \infty} \sup_{n} \int_{\R \backslash [-a,a]}|y|^\alpha \, l_n(\mathrm{d}y) = 0 \iff  \{|X_n|^\alpha \, | \, n \ge 1\} \text{ is uniformly integrable.}   
\end{equation*}
\end{lemma}Lemma \ref{lemma:exchangeable} is well known; the proof for Lemma \ref{lemma:unif_inetgrability} can be found in \cite{norvang2011local}.
\begin{proof}[Proof of Theorem \ref{thm:non_markovian_forecast}]
By the independence of $L(A_t)$ and $L(A_{t+h}\backslash A_t)$, we have that
\begin{align*}
\ev[X_{t+h} | X_t] =& \ev[L(A_{t+h}\cap A_t) | L(A_t)] + \ev[L( A_{t+h} \backslash A_t)| L(A_t)]\\ =& \ev[L(A_{t+h}\cap A_t) | L(A_t)] + \ev[L( A_{t+h} \backslash A_t)] \\
=& \ev[L(A_{t+h} \cap A_t) | L(A_t)] + \frac{Leb(A_h\backslash A)}{Leb(A)} \ev[X_t].
\end{align*}
hence it is enough to prove that
\begin{equation}\label{eq:conditional_it_is_enough}
    \ev[L( A_{t+h} \cap A_t) | L(A_t)] = \frac{Leb(A_h\cap A)}{Leb(A)} X_t.
\end{equation}
We prove \eqref{eq:conditional_it_is_enough} by analysing two cases.

\underline{Case I:} If $\frac{Leb(A \backslash A_h)}{Leb(A_h \cap A)}= \frac{m}{n} \in \mathbb{Q}$ for some positive integers $m$ and $n$, let
\begin{equation*}
   q \defeq \frac{Leb(A \backslash A_h)}{m} = \frac{Leb(A_h \cap A)}{n}.
\end{equation*}
and partition $A \backslash A_h$ into $m$ disjoint subsets $S_1,\ldots,S_m$ of area $q$ and $A \cap A_h$ into $n$ disjoint subsets $S_{m+1},\ldots,S_{m+n}$ of area $q.$ Then $L(S_1),\ldots,L(S_{m+n})$ are iid and integrable and by Lemma \ref{lemma:exchangeable}, we obtain that
\begin{align*}
    \ev\left[L(A_{t+h}\cap A_t)|L(A_t)\right] =& \sum_{j=1}^n \ev\left[ L(S_j) \bigg\rvert \sum_{i=1}^{n+m} L(S_i)\right]\\ =& \frac{n}{n+m} \sum_{i=1}^{n+m} L(S_i) = \frac{Leb(A_h \cap A)}{Leb(A)} L(A_t).
\end{align*} 
\underline{Case II:} If $\frac{Leb(A \backslash A_h)}{Leb(A_h \cap A)}= \alpha \not\in \mathbb{Q}$ let $\alpha_n \in \mathbb{Q}$ with $\alpha_n \uparrow \alpha.$ Consider a nested sequences of sets $A_t \cap A_{t+h} \subset   \ldots   \subset S_{n+1} \subset S_n \ldots \subset S_1 \subset A_t$ such that
\begin{equation*}
\bigcap_{n} S_n = A_t \cap A_{t+h} \ \textrm{  and  } \ \frac{Leb(A\backslash S_n)}{Leb(S_n)} =  \alpha_n.
\end{equation*}
By the proof of Case I, we have that $    \ev[L(S_n)|L(A_t)] = \frac{Leb(S_n)}{Leb(A)} L(A_t)$.
Since $\frac{Leb(S_n)}{Leb(A)} \to \frac{Leb(A\cap A_h)}{Leb(A)},$ we have that  $\ev[L(S_n)|L(A_t)] \to \frac{Leb(A\cap A_h)}{Leb(A)} L(A_t)$ a.s. By the uniqueness of the limit, it is enough to prove that $L(S_n) \to L(A \cap A_h)$ in $\mathcal{L}^1$, or equivalently $L\left(S_n \backslash (A_h \cap A)\right) \to 0$ in $\mathcal{L}^1$. 
We show that the sequence $L\left(S_n \backslash (A_h \cap A)\right)$ convergence in probability to $0$ and is uniformly integrable.   
We have that \begin{equation*}\ev\left[e^{i t L\left(S_n \backslash (A_h \cap A)\right)} \right] =  \mathrm{Leb}\left(S_n \backslash (A_h \cap A)\right) \, \ev\left[e^{i t L^{'}} \right] \to 0  \text { as } n \to \infty. \end{equation*} Thus $L\left(S_n \backslash (A_h \cap A)\right) \to 0$ in distribution and in probability. It remains to prove the uniform integrability property. Let $(\xi,a,l)$ be the L\'evy-Khintchine triplet of the L\'evy seed $L^{'}.$ Then the L\'evy-Khintchine triplets of $L\left(S_n \backslash (A_h \cap A)\right)$ are given by $(\xi_n, a_n, l_n) = (c_n \xi, c_n a, c_n l)$, where $c_n \defeq \textrm{Leb}\left(L\left(S_n \backslash (A_h \cap A)\right)\right) \downarrow 0$ as $n \to \infty$. We obtain that
\begin{equation*}
    \sup_{n} \int_{\R \backslash [-a,a]}|y| \, l_n(\mathrm{d}y) = \int_{\R \backslash [-a,a]}|y| \, l_1(\mathrm{d}y) =  c_1 \int_{\R \backslash [-a,a]}|y| \, l(\mathrm{d}y).
\end{equation*}
Since $L^{'}$ is integrable, it follows that $\int_{\R \backslash (-1,1)} |y|\, l(\mathrm{d}y)$ is finite  \citep[cf.][Theorem 25.3]{ken1999levy}, hence by the dominated convergence theorem 
\begin{equation*}
\lim_{a \to \infty} \sup_{n} \int_{\R \backslash [-a,a]}|y| \, l_n(\mathrm{d}y) = 0. 
\end{equation*}
By Lemma \ref{lemma:unif_inetgrability}, the sequence $L\left(S_n \backslash (A_h \cap A)\right)$ is uniformly integrable, which finishes the proof.\end{proof}
\subsection{Measure-valued gradients and hybrid estimators}
\label{supplementary_section:measure_valued_grad}
In the main part of the paper, we formulated composite likelihood inference for trawl processes as a stochastic optimization problem and expressed both the pairwise density $p(x_s,x_t;\bt) = \ev_{q(z;\bt)}[f(z,\bt)]$ and its gradient $\nabla_{\bt}p(x_s,x_t;\bt) = \nabla_{\bt} \ev_{q(z;\bt)}[f(z,\bt)]$ as Monte Carlo (MC) estimators, where $f$ depends implicitly on $x_s$ and $x_t$. We found that the conventional finite difference and score function (SF) methods yield impractical estimators for the gradient of the log density, as they suffer from large biases and variances. However, we demonstrated the feasibility of composite likelihood inference by deriving lower variance estimators with the method of pathwise gradients (PG), which can be easily applied to a wide range of continuous distributions. To utilize the PG methodology, there are three requirements: the pathwise gradient $\nabla_{\bt}z$ must be available numerically for samples $z$ with density $q(\cdot;\bt)$, the function $f$ must be differentiable in $z$ and the interchange between differentiation and integration must be valid. Under these conditions, the MC  estimator for the gradient is then given by
\begin{equation*}
    \nabla_{\bt}p(x_s,x_t;\bt) = \nabla_{\bt}\ev_{q(z;\bt)}[f(z,\bt)] = \ev_{q(z;\bt)}\left[ \pdv{f}{z}(z,\bt) \nabla_{\bt} z + \nabla_{\bt}f(z,\bt)  \right].
\end{equation*}
We greatly extend the gradient estimation methodology by relaxing the first two requirements while retaining the last one. 
We start by explaining why such extensions are important. Firstly, PG are generally not available for discrete distributions \citep[cf.][p. 595 - 597]{FU2006575}; secondly, even if $f$ is smooth, the partial derivative $\pdv{f}{z}$ might be difficult to compute. Illustrating both issues is the Skellam L\'evy basis from Example \ref{ex:distr_skellam}, for which the pairwise mass function $p(x_s,x_t;\bt)$ is given by
\begin{align*}
     &\sum_{k=-\infty}^{\infty} \mathrm{Skellam}(k; \nu_1 s_{21}, \nu_2 s_{21}) \, \mathrm{Skellam}(x_t-k;\nu_1 s_{12}, \nu_2 s_{12}) \, \mathrm{Skellam}(x_s -k; \nu_1 s_{11}, \nu_2 s_{11})\\
     &= e^{-(\nu_1+\nu_2)(s_{21}+s_{12} + s_{11})} \sum_{k=-\infty}^{\infty}  \left(\frac{\nu_1}{\nu_2}\right)^\frac{x_s+x_t-k}{2} I_{|k|}\left(2\sqrt{\nu_1 \nu_2} s_{21}\right) I_{|x_t-k|}\left(2\sqrt{\nu_1 \nu_2} s_{12}\right) I_{|x_s-k|}\left(2\sqrt{\nu_1 \nu_2} s_{11}\right),
\end{align*}
where $s_{11},\, s_{21}, \, s_{12}$ are the Lebesgue measures of the slices $A_s \backslash A_t, \, A_t \backslash A_s,\, A_s \cap A_t$ (see Figure \ref{fig:slice_partition_and_2_trawls} and \eqref{eq:discrete_ivt_skellam}).
The above can be estimated via Monte Carlo samples by taking the expectation over any of $L(A_s\backslash A_t), \, L(A_t \backslash A_s), \, L(A_s \cap A_t)$; working with $L(A_s \backslash A_t)$, we obtain that
\begin{equation*}
p(x_s,x_t;\bt) = e^{-(\nu_1+\nu_2)\textrm{Leb}(A) }\left(\frac{\nu_1}{\nu_2}\right)^{\frac{x_t}{2}}\ev\left[I_{|Z|}\left(2\sqrt{\nu_1 \nu_2} s_{21}\right) I_{|x_t-Z|}\left(2\sqrt{\nu_1 \nu_2} s_{12}\right) \right],
\end{equation*}
where $Z \sim \textrm{Skellam}(\nu_1 s_{11}, \nu_2 s_{11})$ and $I$ is the modified Bessel function of the firs kind. Although $I_k(x)$ is differentiable in both arguments, we are not aware of computer implementations for $\pdv{I}{k}$. Further, PG are not available for the Skellam distribution. Motivated by these two issues, we give an overview of MVG and explain how MVG can be used to obtain low-variance MC estimators for the gradient of the pairwise mass function for Integer-valued trawl (IVT) processes; we illustrate with the Poisson and Skellam L\'evy bases. For the cases when MVG are not available (e.g.~the Negative Binomial distribution), we derive hybrid estimators, which combine PG and MVG and provide a unified inference methodology, applicable to both integer-valued and real-valued trawl processes. In doing so, we formalize a chain rule for the class of stochastic transformations given by conditional sampling, which can be applied outside the trawl processes framework to derive low-variance gradient estimators in a wide range of stochastic optimization tasks, e.g.~in deep learning.

Given a probability mass function $q(\cdot;\bt)$ parameterized by $\bt \in \R^d$, the MVG method uses the decomposition of the signed measure induced by the unnormalized probability mass functions $\nabla_{\bt_i} q(\cdot;\bt)$ into $c^{+}_i q^{+}_i(\cdot; \bt) - c^{-}_i q^{-}_i(\cdot;\bt)$, where $c^{+}_i, \, c^{-}_i$ are positive constants and $q^{+}_i(\cdot;\bt), \, q^{-}_i(\cdot;\bt)$ are 
probability measures parameterized by $\bt$ for $i=1,\ldots,d$. In shorthand, we have 
\begin{equation*}
\nabla_{\bt}q(\cdot;\bt) = \bm{c}^{+} \bm{q}^{+}(\cdot;\bt) -  \bm{c}^{-} \bm{q}^{-}(\cdot;\bt),
\end{equation*}
where $\bm{c}^{+}, \, \bm{c}^{-}$ are $d$ dimensional vectors with positive entries and $\bm{q}^{+}(\cdot;\bt),\,  \bm{q}^{-}(\cdot;\bt)$ are $d$ dimensional vectors of probability measure. 
We discuss the practical implementation details, non-uniqueness of the pair $(\bm{q}^{+},\bm{q}^{-})$ and variance reduction by coupling in the Poisson and Skellam cases.
\begin{example}[Poisson distribution]\label{ex:MVG_poisson}
 Let $Z \sim \textrm{Poisson}(\lambda)$ with $q(\cdot;\lambda) = \sum_{j=0}^{\infty} \frac{\lambda^j}{j!} \delta_j$, where $\delta_j$ is the Dirac delta function at $j$. Then the unnormalized probability mass function  
\begin{equation*}
\pdv{q}{\lambda}(\cdot;\lambda) = e^{-\lambda} \sum_{j=0}^{\infty} \frac{\lambda^j}{j!} \delta_{j+1}-e^{-\lambda} \sum_{j=0}^{\infty} \frac{\lambda^j}{j !} \delta_j
\end{equation*}
can be decomposed in at least two ways. Firstly, we can set
\begin{equation*}
q^{+} =\frac{1}{c} e^{-\lambda} \sum_{j=\lceil \lambda \rceil}^{\infty}\left(\frac{j-\lambda}{\lambda}\right) \frac{\lambda^j}{j !} \delta_j  
  \text{  and } q^{-}=\frac{1}{c} e^{-\lambda}\left[\delta_0+\sum_{j=1}^{\lfloor \lambda\rfloor}\left(\frac{\lambda-j}{\lambda}\right) \frac{\lambda^j}{j !} \delta_j\right],
\end{equation*} 
where 
\begin{equation*}
c \defeq c^{+} = c^{-} = \sum_{j=\lceil \lambda\rceil}^{\infty}\left(\frac{j-\lambda}{\lambda}\right) \frac{\lambda^j}{j !},
\end{equation*}
and where $\lfloor \cdot \rfloor$ and $\lceil \cdot \rceil$ are the floor, respectively ceiling functions. This is the Hahn-Jordan decomposition of $q$ into measures with disjoint support. A simpler decomposition is given by 
\begin{equation*}q^{+} = e^{-\lambda} \sum_{j=0}^{\infty} \frac{\lambda^j}{j!} \delta_{j+1}  \text{ and } q^{-} = e^{-\lambda} \sum_{j=0}^{\infty} \frac{\lambda^j}{j !} \delta_j,
\end{equation*}
which are the probability mass functions of a Poisson distribution shifted by one unit, respectively of a Poisson distribution, and $c^{+}= c^{-} = 1$. The latter decomposition is much easier to implement on a computer. Using the latter decomposition, the MVG estimator is given by
\begin{equation}
\nabla_{\bt}\ev\left[f(Z,\bt)\right] = \ev\left[\nabla_{\bt}f(Z,\bt)\right] 
 + \ev\left[f(Z^{+},\bt)\right] - \ev\left[f(Z^{-},\bt)\right],\label{eq:poisson_nice_decomposition}
\end{equation} 
where $Z,\, Z^{+}-1,\, Z^{-} \sim \textrm{Poisson}(\lambda)$. The samples from $Z^{+}$ can be generated by adding $1$ to the samples from $Z^{-}$, halving the simulation computational cost. Using dependent samples for $Z^{+}$ and $Z^{-}$ instead of independent ones generally reduces the variance of the resulting estimator, as observed in Example 4.12 of \citep{pflug2012optimization} and the simulation study from \citep{reparam_tutorial}. 
\end{example}
\begin{example}[Skellam distribution]
Note that $ Z_1 - Z_2 \sim \textrm{Skellam}(\lambda_1,\lambda_2)$, where $Z_1,Z_2 \sim \textrm{Poisson}(\lambda_i)$ are independent. Thus we can obtain MVG for the Skellam distribution from the MVG of the Poisson distribution.
\end{example}
Next, we discuss hybrid estimators for distributions $q$ for which neither PG nor MVG is directly available, generalizing Lemma \ref{lemma:chain_rule_pathwise_grads_with_cal_S}. These estimators can be implemented in practice when samples from $q$ can be generated by sequential conditional sampling from distributions for which either PG or MVG are available. An illustrative example is the Negative Binomial (NB) distribution: if $Y \sim \textrm{Gamma}\left(m,\frac{1-p}{p}\right)$ and $Z|Y \sim \textrm{Poisson}(Y),$ then $Z \sim \textrm{NB}(m,p)$. To this end, we formalize the chain rule applied to the composition of stochastic transformations $\bt \mapsto Y(\bt) \mapsto Z\left(Y(\bt)\right)$, where each of the sampling procedures $\bt \mapsto Y(\bt)$ and $Y(\bt) \mapsto Z\left(Y(\bt)\right)$ have either PG or MVG. In the following, $q_Y$ is a density and $q_c$ can be either a density or a probability mass function.
\begin{lemma}[Chain rule for stochastic transformations]\label{lemma:hybrid_estimators}
Let $Y,\,Z$ be random variables with distributions $q_{Y}(\cdot;\bt),\, q(\cdot;\bt)$ where the conditional distribution of $Z|Y$ is given by $q_c(\cdot;Y)$ and such that we can sample from $q_{Y}$ and $q_c(\cdot;Y)$ for any value of $Y$. Suppose that we can compute the PG $\nabla_{\bt}Y$ of $q_{Y}$. 

i) Suppose further that we can compute the PG $\nabla_{Y}Z$ of $q_c(\cdot;Y)$ for any value of $Y$. Then 
\begin{equation}
    \nabla_{\bt}\ev\left[f(Z,\bt)\right] 
   = \ev\left[\pdv{f}{z}(Z,\bt) \nabla_{Y}Z  \, \nabla_{\bt}Y\right] + \ev\left[\nabla_{\bt}f(Z,\bt)\right].\label{eq:stochastic_chain_rule_pg_pg}
\end{equation}
ii) Suppose further that an MVG decomposition $\bm{c}^{+}(Z_1),\, \bm{c}^{-}(Y),\,\bm{q}^{+}_c(\cdot;Y),\,  \bm{q}^{-}_c(\cdot;Y)$ of $q_c(\cdot;Y)$ is available and that conditional on any value of $Y$, we can sample random variables $Z^{+}|Y, \, Z^{-}|Y$ distributed according to $\bm{q}_c^{+}(\cdot;Y)$ and $\bm{q}_c^{-}(\cdot;Y)$. Then 
\begin{equation}
    \nabla_{\bt}\ev\left[f(Z,\bt)\right] =  \ev\left[\left(\bm{c}^{+}f(Z^{+},\bt) - \bm{c}^{-}f(Z^{-},\bt) \right)\nabla_{\bt}Y\right]  + \ev[\nabla_{\bt}f(Z,\bt)].\label{eq:stochastic_chain_rule_pg_mvg}
\end{equation}
\end{lemma}
\begin{proof}[Proof of Lemma \ref{lemma:hybrid_estimators}]\label{proof:hybrid_lemma} 
By the law of total expectation with $g(y,\bt) \defeq \ev_{q_c(z;y)}\left[f(z,\bt)\right]$, we have that 
\begin{align}
\nabla_{\bt}\ev_{q(z;\bt)}\left[f(z,\bt)\right] &= \nabla_{\bt}\ev_{q_{Y}(y;\bt)}\left[ \ev_{q_c(z;y)}\left[f(z,\bt)\right]\right] = \nabla_{\bt}\ev_{q_{Y}(y;\bt)}\left[g(y,\bt)\right] \notag \\ &= \ev_{q_{Y}(y;\bt)}\left[ \pdv{g}{y}(y,\bt) \nabla_{\bt}y + \nabla_{\bt} g(y,\bt)\right].\label{eq:hybrid_mvg_master_eq}
\end{align}
We analyze separately the two terms from \eqref{eq:hybrid_mvg_master_eq}. Since $\nabla_{\bt} g(y,\bt) = \nabla_{\bt} \ev_{q_c(z;y)}\left[f(z,\bt)\right] = \ev_{q_c(z;y)}\left[\nabla_{\bt} f(z,\bt)\right]$, we have that
\begin{equation}
\ev_{q_{Y}(y;\bt)}\left[\nabla_{\bt} g(y,\bt)\right] = \ev_{q_{Y}(y;\bt)}\left[\ev_{q_c(z;y)}\left[\nabla_{\bt} f(z,\bt)\right]\right] = \ev_{q(z;\bt)}\left[\nabla_{\bt}f(z,\bt)\right].\label{eq:easy_term_for_both_1_i_and_1_ii}
\end{equation}
For part i), we have that
\begin{equation}
\pdv{g}{y}(y,\bt) = \pdv{}{y}\ev_{q_c(z;y)}\left[f(z,\bt)\right] =\ev_{q_c(z;y)}\left[\pdv{f}{z}(z,\bt)\nabla_{\bt}z\right].\label{eq:part_1_i}
\end{equation}
Plugging \eqref{eq:part_1_i} and \eqref{eq:easy_term_for_both_1_i_and_1_ii} back in \eqref{eq:hybrid_mvg_master_eq}, we obtain the result from \eqref{eq:stochastic_chain_rule_pg_pg}.

For part ii), we have that
\begin{equation}
\pdv{g}{y}(y,\bt) = \pdv{}{y}\ev_{q_c(z;y)}\left[f(z,\bt)\right] =c^{+}(y)\ev_{q^{+}_c(z;y)}\left[f(z^{+},\bt)\right] - c^{-}(y) \ev_{q^{-}_c(z^{-};y)}\left[f(z^{-},\bt)\right]\label{eq:part_1_ii}
\end{equation}
Plugging \eqref{eq:part_1_ii} and \eqref{eq:easy_term_for_both_1_i_and_1_ii} back in \eqref{eq:hybrid_mvg_master_eq}, we obtain the result from \eqref{eq:stochastic_chain_rule_pg_mvg}. 
\end{proof}
Part i) is similar to Lemma \ref{lemma:chain_rule_pathwise_grads_with_cal_S} and can be applied to derive the pathwise gradients for the NIG distribution, see Example \ref{ex:nig} in the appendix. Part ii) is illustrated in the next example.
\begin{example}[Negative binomial distribution with 1.ii]\label{example:nb_hydrib_estimator} Note that if $Z \sim \textrm{Poisson}(Y)$, where $Y \sim \textrm{Gamma}\left(m,\frac{1-p}{p}\right)$, then $Z \sim \textrm{NB}(m,p)$. Let $\bt = (m,p)$ and $q(z,\bt) = \textrm{NB}(z;m,p)$. By Lemma \ref{lemma:hybrid_estimators} and by using the MVG for Poisson distributions from \eqref{eq:poisson_nice_decomposition}, in which $c^{+} = c^{-} = 1$ and $Z|Y,\,Z^{+}-1|Y,\, Z^{-}|Y$ are all $\textrm{Poisson}(Y)$ distributed, we obtain that
\begin{equation*}
\nabla_{\bt}\ev\left[f(Z,\bt)\right] =   \ev\left[\left(f(Z+1,\bt) - f(Z,\bt)\right)\nabla_{\bt}Y\right] + \ev[\nabla_{\bt}f(Z,\bt)],
\end{equation*}
where the PG $\nabla_{\bt} Y$ for the Gamma distribution can be computed as in Example \ref{ex:gamma}.
\end{example}

\subsection{Extended simulation study for trawl processes}\label{supplementary_section:extendend_simulation_studies}
We present further simulation study results for the variance reduction properties of our gradient estimation methodology and for the improved finite sample performance of the PL estimator over the GMM estimator.

\textbf{Variance reduction}

In Table \ref{table:variance_reduction_sim_study_main_body} of Subsection \ref{subsection:variance_reduction_sim_study} we displayed the bias and standard deviation of the gradient estimators. In practice, we are more interested in the absolute deviations from the true value of the gradient, as displayed in Table \ref{table:extended_sim_study_table}, rather than the bias. We note that the PG estimator always has a smaller deviation from the true value, regardless of the metric: mean absolute error (MAE), median absolute error (MedAE) or root-mean-square error (rMSE). 
\begin{table}[htb]
\begin{adjustbox}{width=\textwidth}
\begin{tabular}{@{\extracolsep{3.5pt}}llrrrrrrrrrrrr@{}}
   &   & \multicolumn{3}{c}{$m=0$} & \multicolumn{3}{c}{$m=1$} & \multicolumn{3}{c}{$m=2$} & \multicolumn{3}{c}{$m=3$}   \\ \cline{3-5} \cline{6-8}  \cline{9-11}  \cline{12-14} 
&  & MAE & MedAE & rMSE & MAE & MedAE & rMSE & MAE & medAE & rMSE & MAE & MedAE & rMSE \\
\hline
SF & $\alpha$ & 5.44 & 4.72 & 7.56 & 1.51 & 1.23 & 2.44 & 1.13 & 0.92 & 1.83 & 0.95 & 0.78 & 1.53 \\
   & $\beta$ & 4.02 & 3.82 & 5.11 & 1.51 & 1.28 & 2.35 & 1.16 & 0.99 & 1.85 & 0.91 & 0.77 & 1.47 \\
   & $H$ & 9.89 & 8.67 & 13.44 & 1.72 & 1.40 & 2.72 & 1.15 & 0.95 & 1.84 & 0.91 & 0.74 & 1.46 \\
   & $\delta$  & 6.07 & 5.27 & 8.08 & 1.17 & 0.99 & 1.86 & 0.81 & 0.67 & 1.30 & 0.64 & 0.52 & 1.04 \\
      \hline   
PG & $\alpha$  & 3.45 & 3.26 & 4.44 & 1.20 & 1.02 & 1.86 & 0.93 & 0.82 & 1.47 & 0.74 & 0.62 & 1.19 \\
   & $\beta$ & 4.02 & 3.82 & 5.11 & 1.51 & 1.28 & 2.35 & 1.16 & 0.99 & 1.85 & 0.91 & 0.77 & 1.47 \\
   & $H$ & 1.14 & 0.95 & 1.69 & 0.93 & 0.77 & 1.50 & 0.68 & 0.60 & 1.08 & 0.52 & 0.44 & 0.85 \\
   & $\delta$ & 0.67 & 0.58 & 1.01 & 0.65 & 0.55 & 1.05 & 0.51 & 0.45 & 0.81 & 0.40 & 0.33 & 0.64 \\
\hline 
\end{tabular}
\end{adjustbox}
\caption{MAE, MedAE and RMSE of the SF and PG gradient estimators for each value $m$ of the degree of the Taylor polynomial used as control variate. Note that PG estimator improves significantly over the SF estimator. The true values of the gradient rounded to the nearest integer are $(-77,  17,  70, -56)$. The results are obtained from the same simulation study as for Table \ref{table:variance_reduction_sim_study_main_body}.}
\label{table:extended_sim_study_table}
\end{table}

\textbf{Parameter inference}

\begin{figure}[p]
 \centering
\subfloat[$(\alpha, \beta, \lambda) = (3,0.75,0.1)$]{\label{}\includegraphics[width=.33\linewidth]{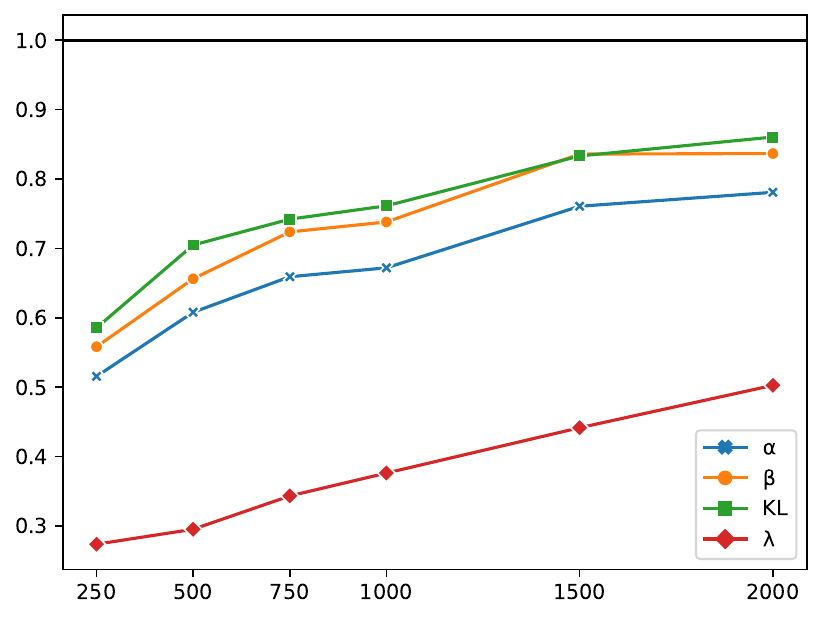}}\hfill
\subfloat[$(\alpha, \beta, \lambda) = (3,0.75,0.25)$]{\label{}\includegraphics[width=.33\linewidth]{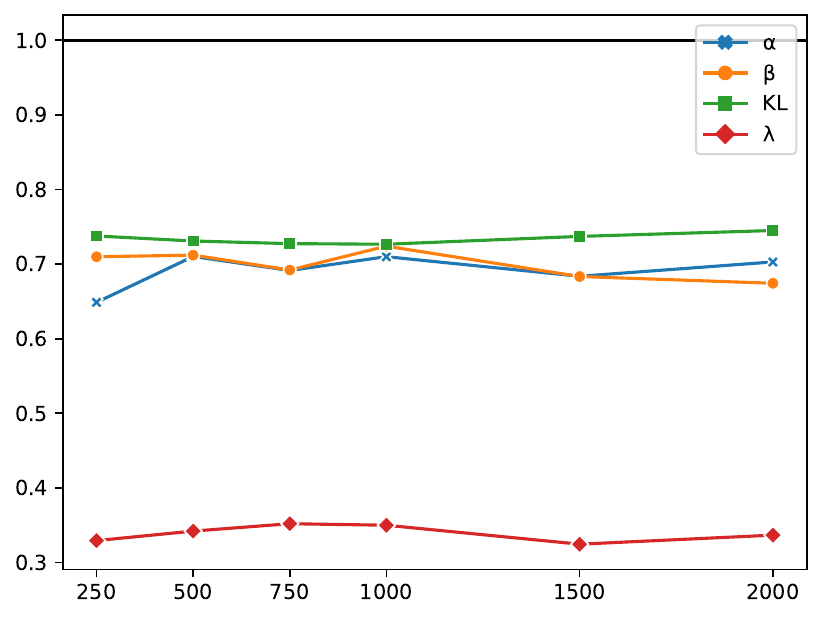}}\hfill
\subfloat[$(\alpha, \beta, \lambda) = (3,0.75,0.4)$]{\label{}\includegraphics[width=.33\linewidth]{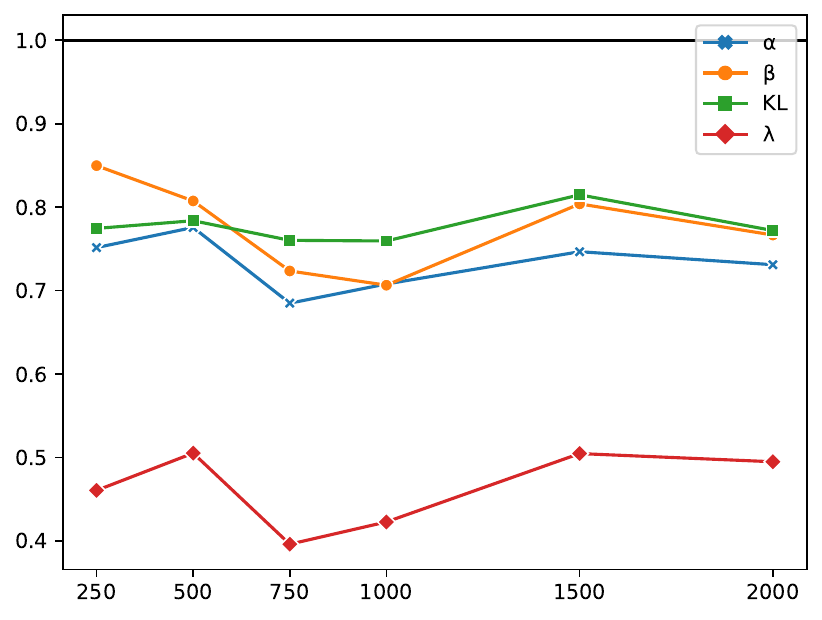}}\par 
\subfloat[$(\alpha, \beta, H, \delta) = (3,0.75,0.5,1)$]{\label{}\includegraphics[width=.33\linewidth]{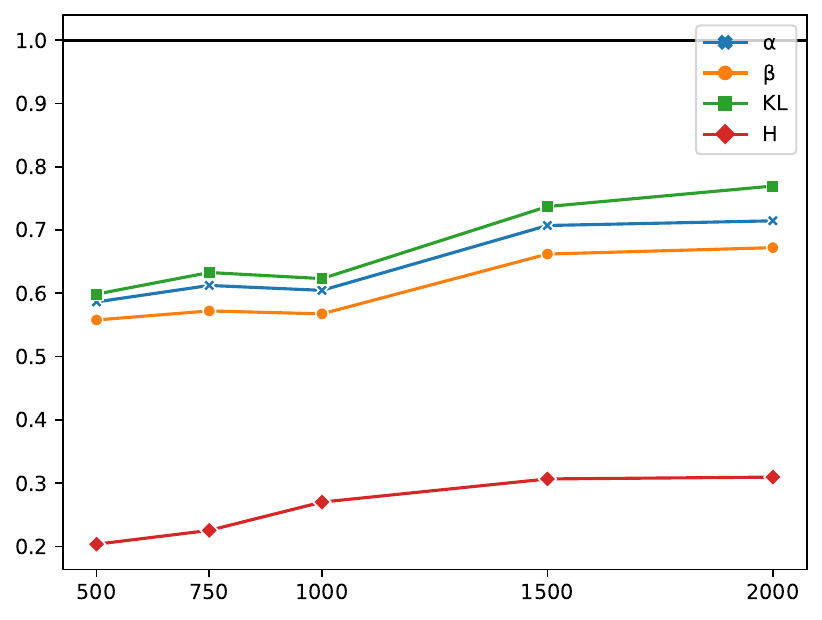}}\hfill
\subfloat[$(\alpha, \beta, H, \delta) = (3,0.75,1.5,1)$]{\label{}\includegraphics[width=.33\linewidth]{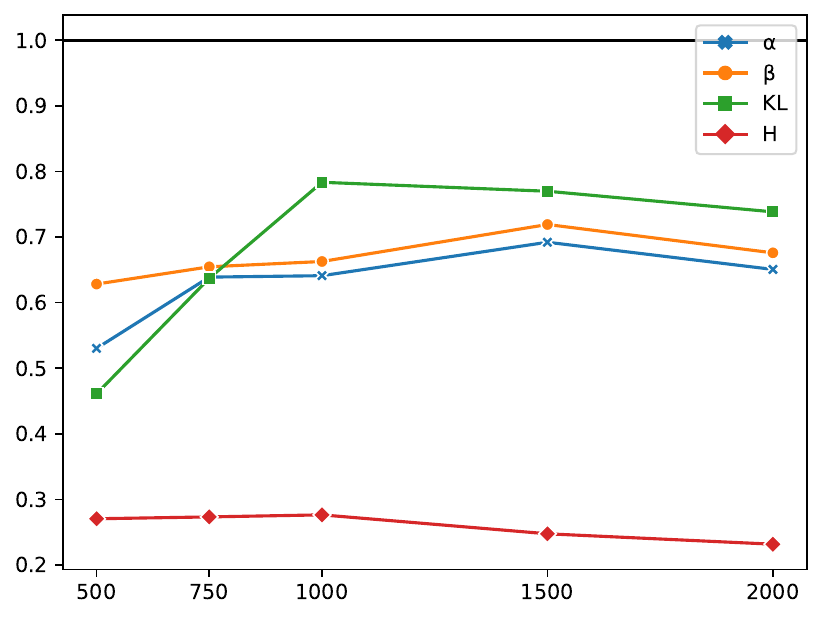}}\hfill
\subfloat[$(\alpha, \beta, H, \delta) = (3,0.75,2.5,1)$]{\label{}\includegraphics[width=.33\linewidth]{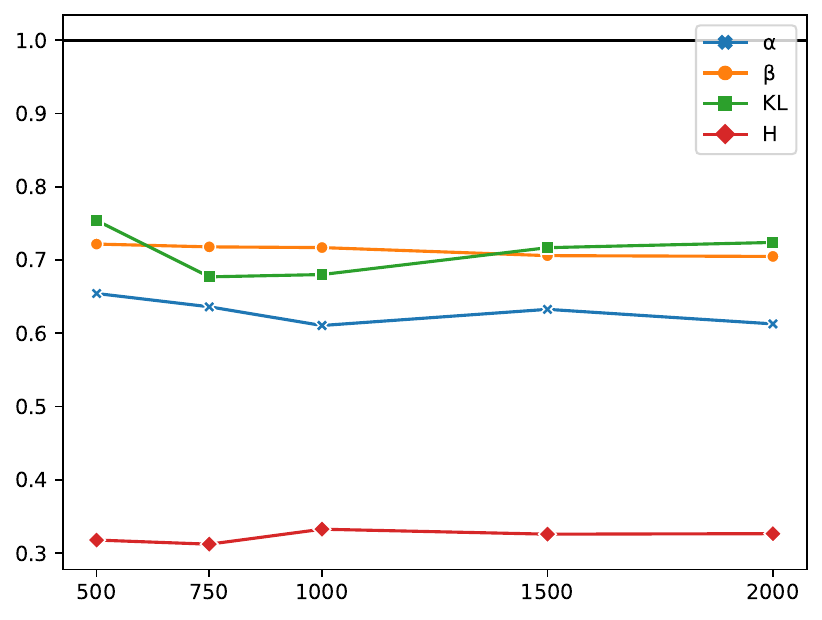}}\par 
\subfloat[$(\alpha, \beta, H, \delta) = (4,3,0.5,0.75)$]{\label{}\includegraphics[width=.33\linewidth]{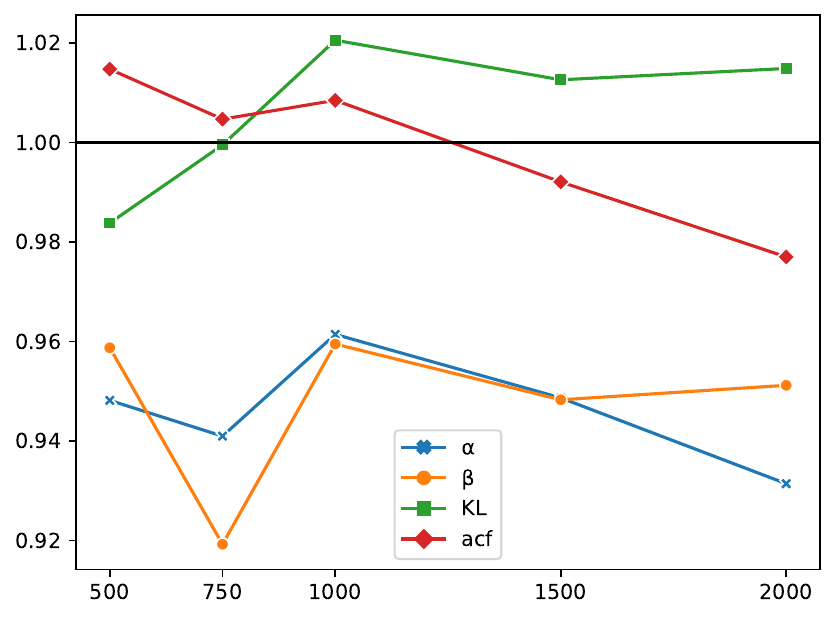}}\hfill
\subfloat[$(\alpha, \beta, H, \delta) = (4,3,1,1)$]{\label{}\includegraphics[width=.33\linewidth]{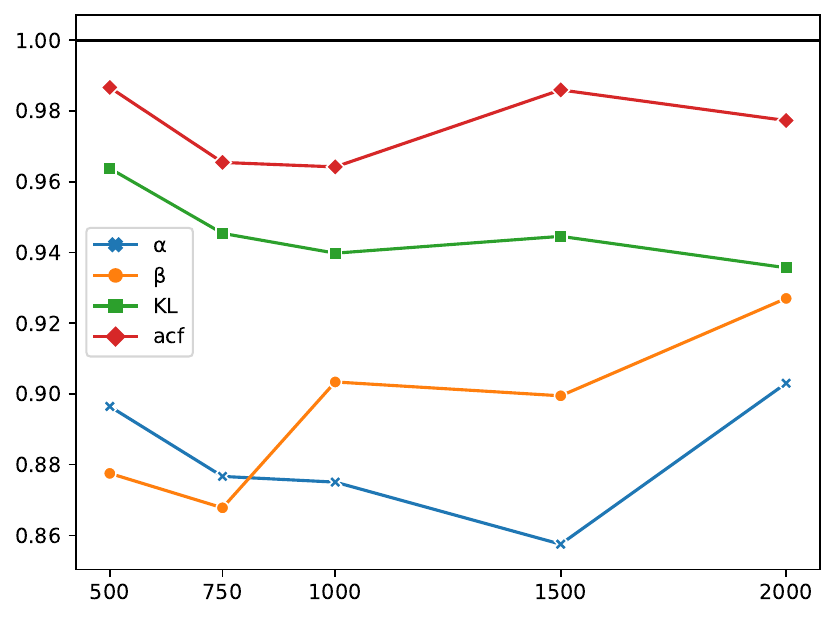}}\hfill
\subfloat[$(\alpha, \beta, H, \delta) = (4,3,2,3) $]{\label{}\includegraphics[width=.33\linewidth]{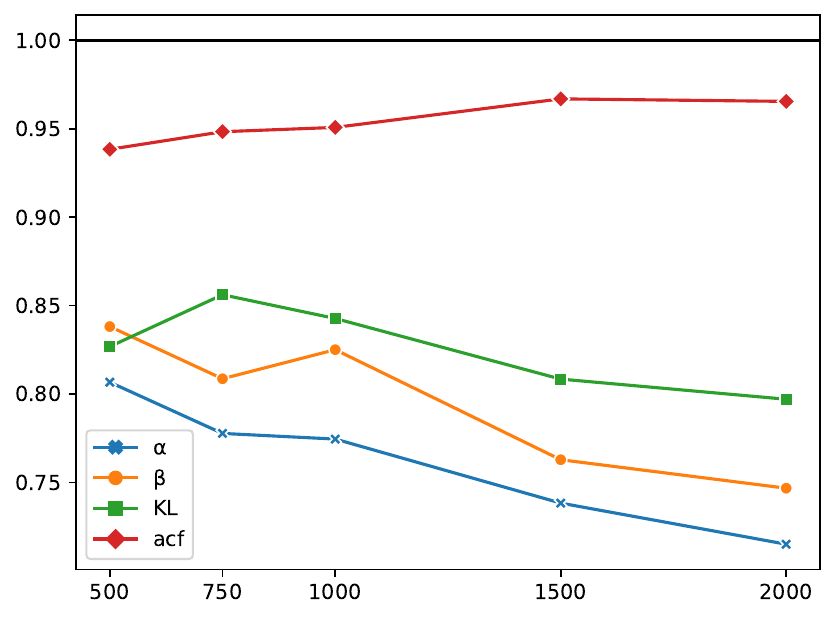}}
\caption{Ratio between estimation errors (MAE, mean KL divergence and mean weighted $L^1$ distance) of the PL and GMM estimators for the parameters of the trawl process $X$ with Gamma L\'evy basis. The results are obtained from the same simulation study as for Figure \ref{fig:inference_results_gamma_levy_seed_rMSE}.}
\label{fig:inference_results_gamma_levy_seed_mae}
\end{figure}

\begin{figure}[p]
 \centering
\subfloat[$(\alpha, \beta, \lambda) = (3,0.75,0.1)$]{\label{}\includegraphics[width=.33\linewidth]{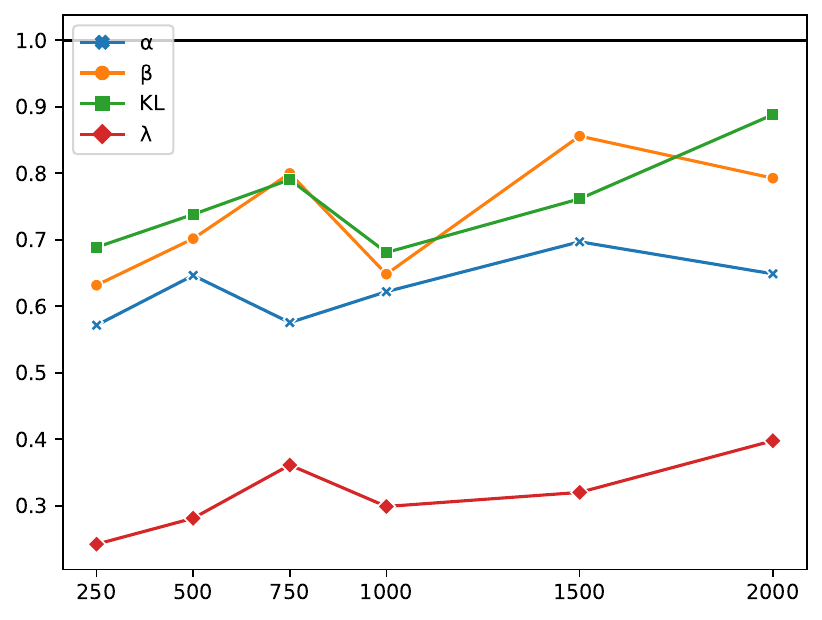}}\hfill
\subfloat[$(\alpha, \beta, \lambda) = (3,0.75,0.25)$]{\label{}\includegraphics[width=.33\linewidth]{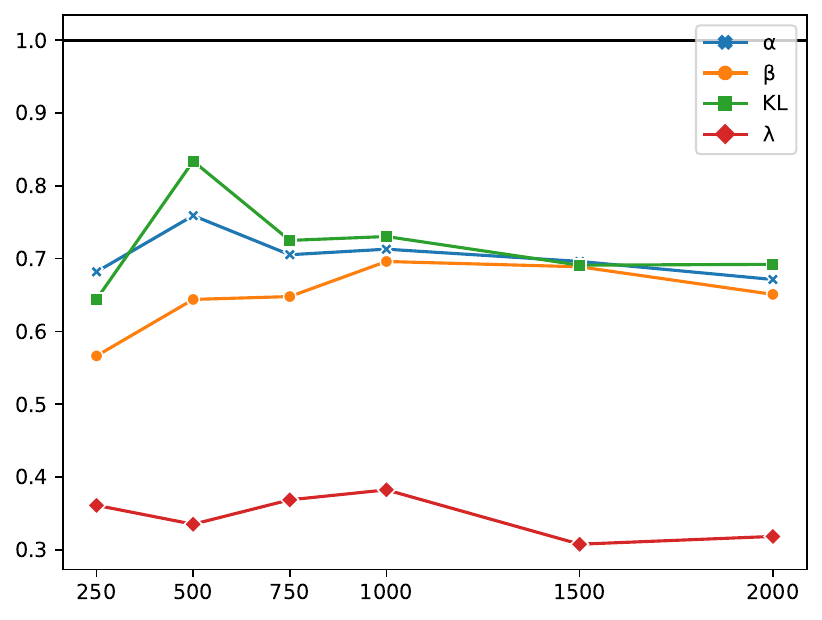}}\hfill
\subfloat[$(\alpha, \beta, \lambda) = (3,0.75,0.4)$]{\label{}\includegraphics[width=.33\linewidth]{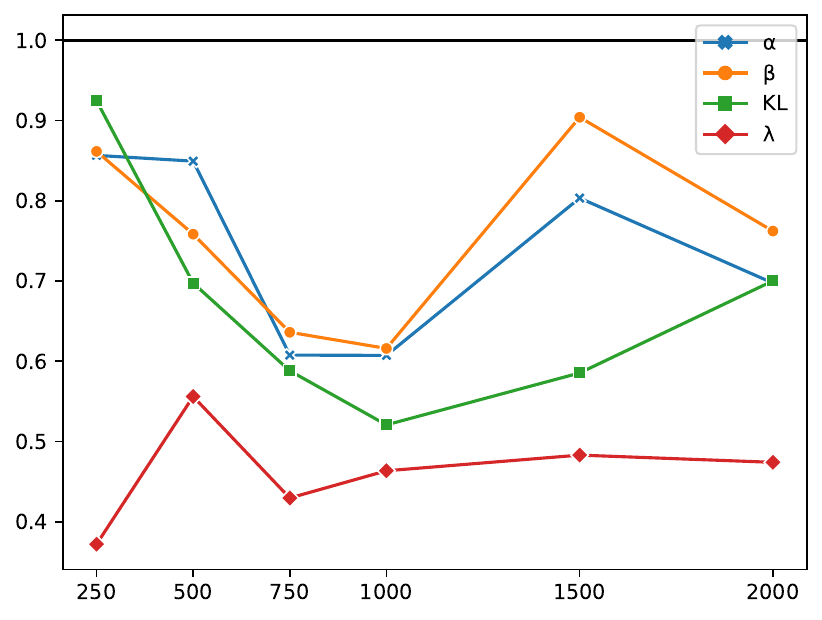}}\par 
\subfloat[$(\alpha, \beta, H, \delta) = (3,0.75,0.5,1)$]{\label{}\includegraphics[width=.33\linewidth]{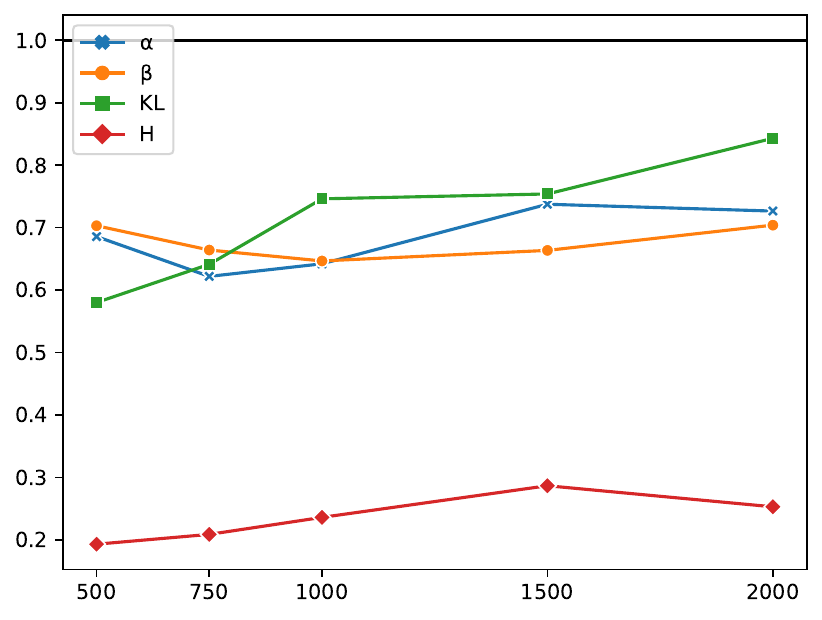}}\hfill
\subfloat[$(\alpha, \beta, H, \delta) = (3,0.75,1.5,1)$]{\label{}\includegraphics[width=.33\linewidth]{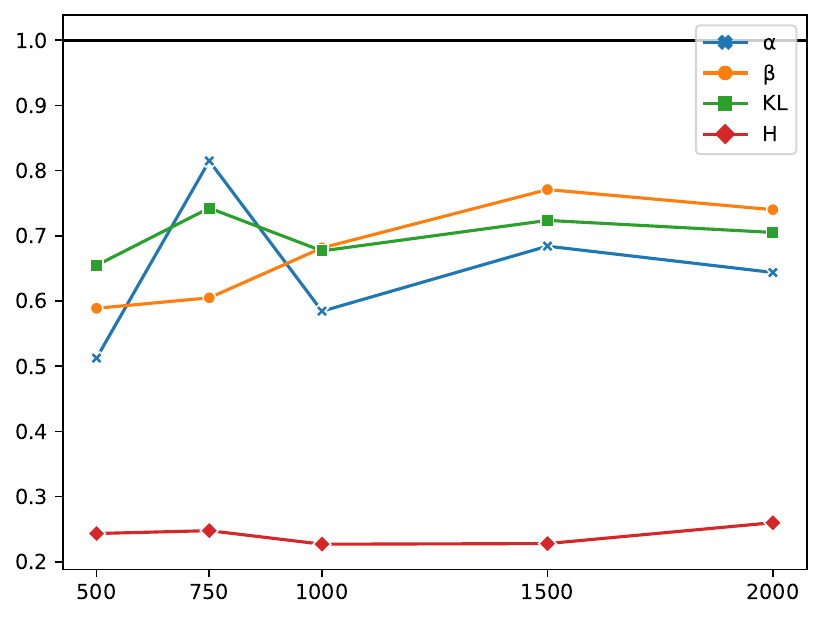}}\hfill
\subfloat[$(\alpha, \beta, H, \delta) = (3,0.75,2.5,1)$]{\label{}\includegraphics[width=.33\linewidth]{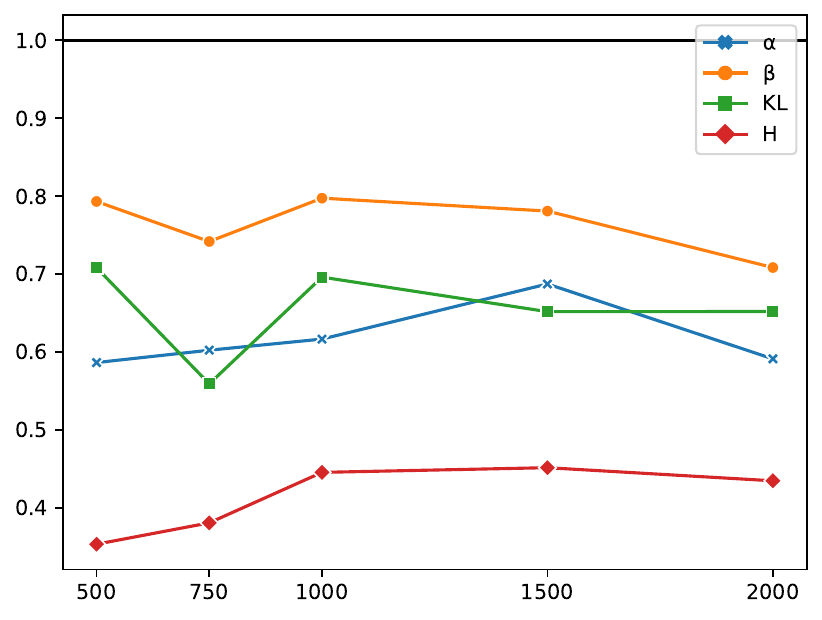}}\par 
\subfloat[$(\alpha, \beta, H, \delta) = (4,3,0.5,0.75)$]{\label{}\includegraphics[width=.33\linewidth]{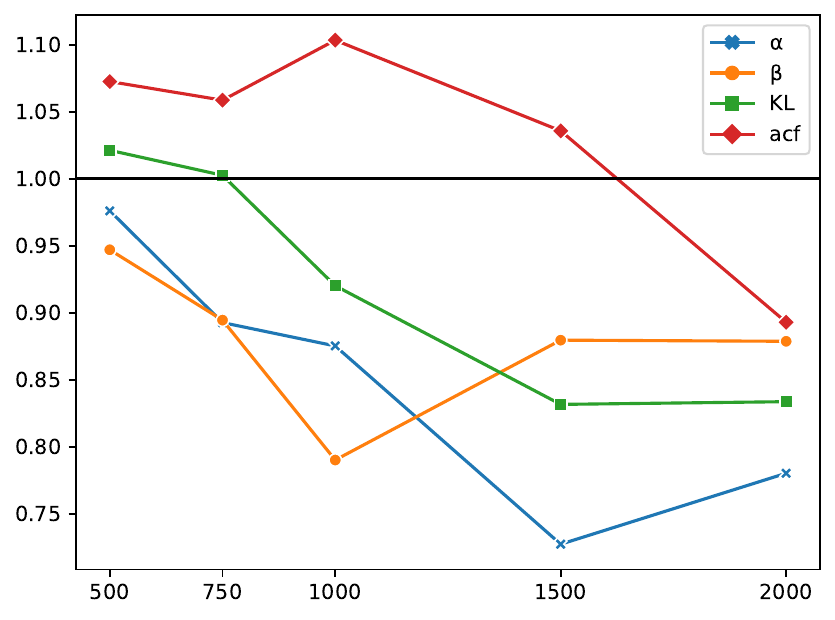}}\hfill
\subfloat[$(\alpha, \beta, H, \delta) = (4,3,1,1)$]{\label{}\includegraphics[width=.33\linewidth]{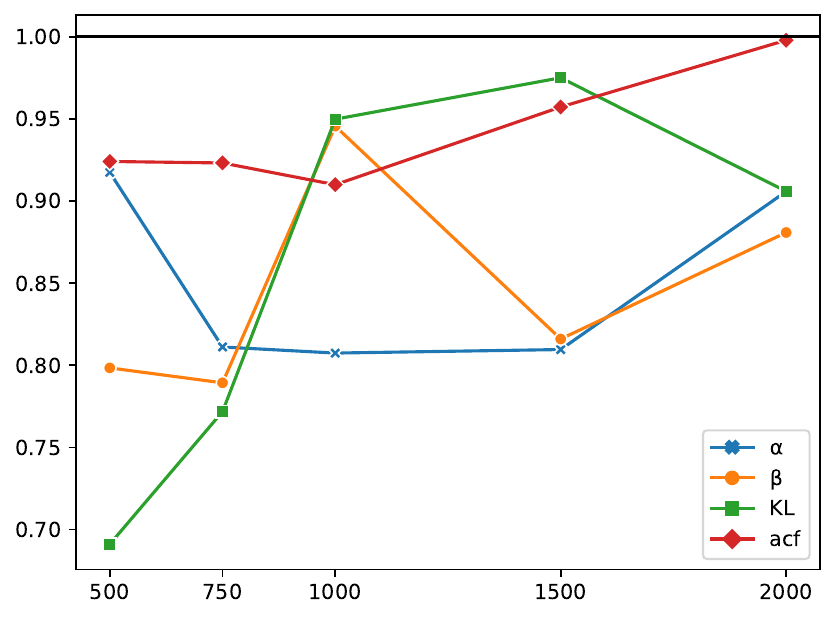}}\hfill
\subfloat[$(\alpha, \beta, H, \delta) = (4,3,2,3) $]{\label{}\includegraphics[width=.33\linewidth]{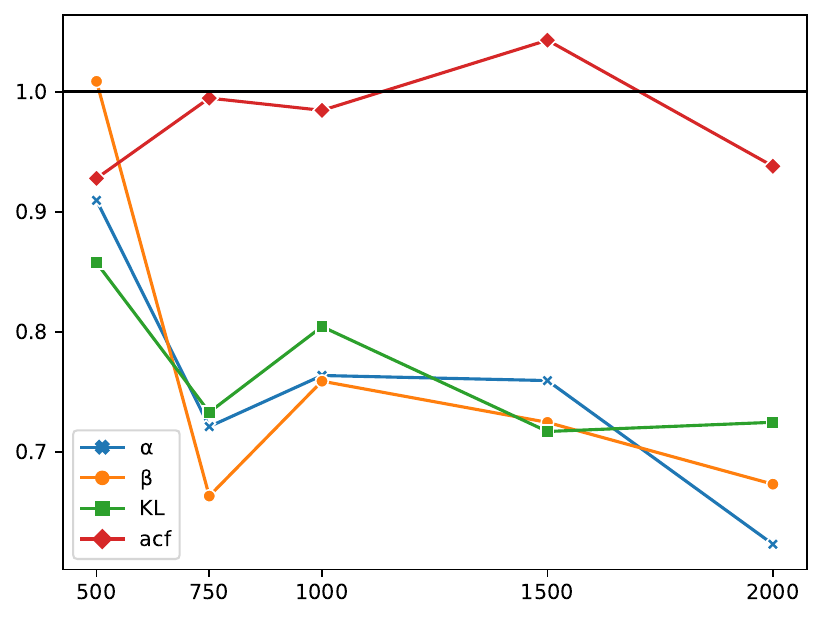}}
\caption{Ratio between estimation errors (MedAE, median KL divergence and median weighted $L^1$ distance) of the PL and GMM estimators for the parameters of the trawl process $X$ with Gamma L\'evy basis. The results are obtained from the same simulation study as for Figure \ref{fig:inference_results_gamma_levy_seed_rMSE}.}
\label{fig:inference_results_gamma_levy_seed_medae}
\end{figure}

In Figure \ref{fig:inference_results_gamma_levy_seed_rMSE} of Subsection \ref{subsection:parameter_inference_results} we showed that the PL estimator achieves a lower rMSE than the GMM estimator. We show in Figures \ref{fig:inference_results_gamma_levy_seed_mae} and \ref{fig:inference_results_gamma_levy_seed_medae} that the improvement is maintained if instead of rMSE we use MAE or MedAE. We also provide Tables \ref{table_a}-\ref{table_i} with all the metrics from Figures \ref{fig:inference_results_gamma_levy_seed_rMSE}, \ref{fig:inference_results_gamma_levy_seed_mae} and \ref{fig:inference_results_gamma_levy_seed_medae} together. 
\newpage
\begin{table}
\renewcommand\thetable{4a} 
\begin{adjustbox}{width=\textwidth}
\begin{tabular}{@{\extracolsep{3.5pt}}lrrrrrrrrrrr@{}}
 & \multicolumn{3}{c}{$\alpha$} & \multicolumn{3}{c}{$\beta$} & \multicolumn{2}{c}{KL} & \multicolumn{3}{c}{$\lambda$} \\ \cline{2-4} \cline{5-7}  \cline{8-9} \cline{10-12}  
 & rMSE & MAE & MedAE & rMSE & MAE & MedAE & MKL & MedKL & rMSE & MAE & MedAE \\\hline
2000 & 0.84 & 0.80 & 0.65 & 0.89 & 0.85 & 0.79 & 0.87 & 0.89 & 0.72 & 0.52 & 0.40 \\
1500 & 0.79 & 0.76 & 0.70 & 0.84 & 0.84 & 0.86 & 0.83 & 0.76 & 0.65 & 0.42 & 0.32 \\
1000 & 0.74 & 0.68 & 0.62 & 0.77 & 0.75 & 0.65 & 0.77 & 0.68 & 0.62 & 0.39 & 0.30 \\
750 & 0.69 & 0.66 & 0.58 & 0.72 & 0.72 & 0.80 & 0.74 & 0.79 & 0.60 & 0.36 & 0.36 \\
500 & 0.60 & 0.61 & 0.65 & 0.65 & 0.66 & 0.70 & 0.71 & 0.74 & 0.54 & 0.29 & 0.28 \\
250 & 0.45 & 0.52 & 0.57 & 0.49 & 0.57 & 0.63 & 0.59 & 0.69 & 0.52 & 0.27 & 0.24 \\ \hline
\end{tabular}
\end{adjustbox}
\caption{$(\alpha, \beta, \lambda) = (3,0.75,0.1)$.}
\label{table_a}
\end{table}

\begin{table}.
\renewcommand\thetable{4b} 
\begin{adjustbox}{width=\textwidth}
\begin{tabular}{@{\extracolsep{3.5pt}}lrrrrrrrrrrr@{}}
 & \multicolumn{3}{c}{$\alpha$} & \multicolumn{3}{c}{$\beta$} & \multicolumn{2}{c}{KL} & \multicolumn{3}{c}{$\lambda$} \\ \cline{2-4} \cline{5-7}  \cline{8-9} \cline{10-12}  
 & rMSE & MAE & MedAE & rMSE & MAE & MedAE & MKL & MedKL & rMSE & MAE & MedAE \\\hline
2000 & 0.69 & 0.70 & 0.67 & 0.70 & 0.67 & 0.65 & 0.74 & 0.69 & 0.58 & 0.34 & 0.32 \\
1500 & 0.68 & 0.68 & 0.70 & 0.70 & 0.68 & 0.69 & 0.74 & 0.69 & 0.57 & 0.32 & 0.31 \\
1000 & 0.72 & 0.71 & 0.71 & 0.73 & 0.72 & 0.70 & 0.73 & 0.73 & 0.59 & 0.35 & 0.38 \\
750 & 0.70 & 0.69 & 0.71 & 0.72 & 0.69 & 0.65 & 0.73 & 0.72 & 0.59 & 0.35 & 0.37 \\
500 & 0.69 & 0.71 & 0.76 & 0.70 & 0.71 & 0.64 & 0.73 & 0.83 & 0.58 & 0.34 & 0.34 \\
250 & 0.66 & 0.65 & 0.68 & 0.72 & 0.71 & 0.57 & 0.74 & 0.64 & 0.57 & 0.33 & 0.36 \\ \hline
\end{tabular}
\end{adjustbox}
\caption{$(\alpha, \beta, \lambda) = (3,0.75,0.25)$.}
\label{table_b}
\end{table}

\begin{table}
\renewcommand\thetable{4c} 
\begin{adjustbox}{width=\textwidth}
\begin{tabular}{@{\extracolsep{3.5pt}}lrrrrrrrrrrr@{}}
 & \multicolumn{3}{c}{$\alpha$} & \multicolumn{3}{c}{$\beta$} & \multicolumn{2}{c}{KL} & \multicolumn{3}{c}{$\lambda$} \\ \cline{2-4} \cline{5-7}  \cline{8-9} \cline{10-12}  
 & rMSE & MAE & MedAE & rMSE & MAE & MedAE & MKL & MedKL & rMSE & MAE & MedAE \\\hline
2000 & 0.74 & 0.73 & 0.70 & 0.78 & 0.77 & 0.76 & 0.77 & 0.70 & 0.70 & 0.50 & 0.47 \\
1500 & 0.76 & 0.75 & 0.80 & 0.79 & 0.80 & 0.90 & 0.81 & 0.59 & 0.71 & 0.50 & 0.48 \\
1000 & 0.72 & 0.71 & 0.61 & 0.74 & 0.71 & 0.62 & 0.76 & 0.52 & 0.65 & 0.42 & 0.46 \\
750 & 0.72 & 0.68 & 0.61 & 0.76 & 0.72 & 0.64 & 0.76 & 0.59 & 0.63 & 0.40 & 0.43 \\
500 & 0.75 & 0.78 & 0.85 & 0.80 & 0.81 & 0.76 & 0.78 & 0.70 & 0.71 & 0.51 & 0.56 \\
250 & 0.72 & 0.75 & 0.86 & 0.81 & 0.85 & 0.86 & 0.77 & 0.92 & 0.68 & 0.46 & 0.37 \\\hline
\end{tabular}
\end{adjustbox}
\caption{$(\alpha, \beta, \lambda) = (3,0.75,0.4)$.}
\label{table_c}
\end{table}

\begin{table}
\renewcommand\thetable{4d} 
\begin{adjustbox}{width=\textwidth}
\begin{tabular}{@{\extracolsep{3.5pt}}lrrrrrrrrrrr@{}}
  & \multicolumn{3}{c}{$\alpha$} & \multicolumn{3}{c}{$\beta$} & \multicolumn{2}{c}{KL} & \multicolumn{3}{c}{$H$} \\\cline{2-4} \cline{5-7}  \cline{8-9}  \cline{10-12}  
 & rMSE & MAE & MedAE & rMSE & MAE & MedAE & MKL & MedKL & rMSE & MAE & MedAE \\\hline
2000 & 0.72 & 0.71 & 0.73 & 0.66 & 0.67 & 0.70 & 0.77 & 0.84 & 0.56 & 0.31 & 0.25 \\
1500 & 0.69 & 0.71 & 0.74 & 0.63 & 0.66 & 0.66 & 0.74 & 0.75 & 0.55 & 0.31 & 0.29 \\
1000 & 0.52 & 0.60 & 0.64 & 0.48 & 0.57 & 0.65 & 0.62 & 0.75 & 0.52 & 0.27 & 0.24 \\
750 & 0.59 & 0.61 & 0.62 & 0.54 & 0.57 & 0.66 & 0.63 & 0.64 & 0.47 & 0.23 & 0.21 \\
500 & 0.53 & 0.59 & 0.69 & 0.51 & 0.56 & 0.70 & 0.60 & 0.58 & 0.45 & 0.20 & 0.19 \\\hline
\end{tabular}
\end{adjustbox}
\caption{$(\alpha, \beta, H, \delta) = (3,0.75,0.5,1)$.}
\label{table_d}
\end{table}

\begin{table}
\renewcommand\thetable{4e} 
\begin{adjustbox}{width=\textwidth}
\begin{tabular}{@{\extracolsep{3.5pt}}lrrrrrrrrrrr@{}}
  & \multicolumn{3}{c}{$\alpha$} & \multicolumn{3}{c}{$\beta$} & \multicolumn{2}{c}{KL} & \multicolumn{3}{c}{$H$} \\\cline{2-4} \cline{5-7}  \cline{8-9}  \cline{10-12}  
 & rMSE & MAE & MedAE & rMSE & MAE & MedAE & MKL & MedKL & rMSE & MAE & MedAE \\\hline
2000 & 0.66 & 0.65 & 0.64 & 0.66 & 0.68 & 0.74 & 0.74 & 0.71 & 0.48 & 0.23 & 0.26 \\
1500 & 0.70 & 0.69 & 0.68 & 0.68 & 0.72 & 0.77 & 0.77 & 0.72 & 0.50 & 0.25 & 0.23 \\
1000 & 0.66 & 0.64 & 0.58 & 0.68 & 0.66 & 0.68 & 0.78 & 0.68 & 0.53 & 0.28 & 0.23 \\
750 & 0.59 & 0.64 & 0.81 & 0.67 & 0.65 & 0.60 & 0.64 & 0.74 & 0.52 & 0.27 & 0.25 \\
500 & 0.42 & 0.53 & 0.51 & 0.60 & 0.63 & 0.59 & 0.46 & 0.65 & 0.52 & 0.27 & 0.24 \\\hline
\end{tabular}
\end{adjustbox}
\caption{$(\alpha, \beta, H, \delta) = (3,0.75,0.75,1)$.}
\label{table_e}
\end{table}

\begin{table}
\renewcommand\thetable{4f} 
\begin{adjustbox}{width=\textwidth}
\begin{tabular}{@{\extracolsep{3.5pt}}lrrrrrrrrrrr@{}}
  & \multicolumn{3}{c}{$\alpha$} & \multicolumn{3}{c}{$\beta$} & \multicolumn{2}{c}{KL} & \multicolumn{3}{c}{$H$} \\\cline{2-4} \cline{5-7}  \cline{8-9}  \cline{10-12}  
 & rMSE & MAE & MedAE & rMSE & MAE & MedAE & MKL & MedKL & rMSE & MAE & MedAE \\\hline
2000 & 0.64 & 0.61 & 0.59 & 0.69 & 0.70 & 0.71 & 0.72 & 0.65 & 0.57 & 0.33 & 0.43 \\
1500 & 0.64 & 0.63 & 0.69 & 0.68 & 0.71 & 0.78 & 0.72 & 0.65 & 0.57 & 0.33 & 0.45 \\
1000 & 0.62 & 0.61 & 0.62 & 0.67 & 0.72 & 0.80 & 0.68 & 0.70 & 0.58 & 0.33 & 0.45 \\
750 & 0.62 & 0.64 & 0.60 & 0.68 & 0.72 & 0.74 & 0.68 & 0.56 & 0.56 & 0.31 & 0.38 \\
500 & 0.65 & 0.65 & 0.59 & 0.71 & 0.72 & 0.79 & 0.75 & 0.71 & 0.56 & 0.32 & 0.35 \\\hline
\end{tabular}
\end{adjustbox}
\caption{$(\alpha, \beta, H, \delta) = (3,0.75,2.5,1)$.}
\label{table_f}
\end{table}

\begin{table}
\renewcommand\thetable{4g} 
\begin{adjustbox}{width=\textwidth}
\begin{tabular}{@{\extracolsep{3.5pt}}lrrrrrrrrrrr@{}}
 & \multicolumn{3}{c}{$\alpha$} & \multicolumn{3}{c}{$\beta$} & \multicolumn{2}{c}{KL} & \multicolumn{3}{c}{acf} \\\cline{2-4} \cline{5-7}  \cline{8-9}  \cline{10-12}  
 & rMSE & MAE & MedAE & rMSE & MAE & MedAE & MKL & MedKL & rWMSE & WMAE & WMedAE \\\hline
2000 & 1.03 & 0.93 & 0.78 & 0.99 & 0.95 & 0.88 & 1.01 & 0.83 & 0.98 & 0.96 & 0.89 \\
1500 & 1.03 & 0.95 & 0.73 & 0.99 & 0.95 & 0.88 & 1.01 & 0.83 & 0.99 & 0.99 & 1.04 \\
1000 & 1.04 & 0.96 & 0.88 & 1.03 & 0.96 & 0.79 & 1.02 & 0.92 & 1.01 & 1.02 & 1.10 \\
750 & 1.01 & 0.94 & 0.89 & 0.99 & 0.92 & 0.89 & 1.00 & 1.00 & 1.00 & 1.01 & 1.06 \\
500 & 0.99 & 0.95 & 0.98 & 0.97 & 0.96 & 0.95 & 0.98 & 1.02 & 1.01 & 1.03 & 1.07 \\\hline
\end{tabular}
\end{adjustbox}
\caption{$(\alpha, \beta, H, \delta) = (4,3,0.5,0.75)$.}
\label{table_g}
\end{table}
\begin{table}
\renewcommand\thetable{4h} 
\begin{adjustbox}{width=\textwidth}
\begin{tabular}{@{\extracolsep{3.5pt}}lrrrrrrrrrrr@{}}
 & \multicolumn{3}{c}{$\alpha$} & \multicolumn{3}{c}{$\beta$} & \multicolumn{2}{c}{KL} & \multicolumn{3}{c}{acf} \\\cline{2-4} \cline{5-7}  \cline{8-9}  \cline{10-12}  
 & rMSE & MAE & MedAE & rMSE & MAE & MedAE & MKL & MedKL & rWMSE & WMAE & WMedAE \\\hline
2000 & 0.91 & 0.90 & 0.91 & 0.92 & 0.93 & 0.88 & 0.94 & 0.91 & 0.98 & 0.97 & 1.00 \\
1500 & 0.92 & 0.86 & 0.81 & 0.91 & 0.90 & 0.82 & 0.94 & 0.98 & 0.99 & 0.98 & 0.96 \\
1000 & 0.90 & 0.88 & 0.81 & 0.90 & 0.90 & 0.95 & 0.94 & 0.95 & 0.96 & 0.94 & 0.91 \\
750 & 0.91 & 0.88 & 0.81 & 0.91 & 0.87 & 0.79 & 0.95 & 0.77 & 0.97 & 0.94 & 0.92 \\
500 & 0.94 & 0.90 & 0.92 & 0.93 & 0.88 & 0.80 & 0.96 & 0.69 & 0.99 & 0.99 & 0.92 \\\hline
\end{tabular}
\end{adjustbox}
\caption{$(\alpha, \beta, H, \delta) = (4,3,1,1)$.}
\label{table_h}
\end{table}
\begin{table}
\renewcommand\thetable{4i} 
\begin{adjustbox}{width=\textwidth}
\begin{tabular}{@{\extracolsep{3.5pt}}lrrrrrrrrrrr@{}}
 & \multicolumn{3}{c}{$\alpha$} & \multicolumn{3}{c}{$\beta$} & \multicolumn{2}{c}{KL} & \multicolumn{3}{c}{acf} \\\cline{2-4} \cline{5-7}  \cline{8-9}  \cline{10-12}  
 & rMSE & MAE & MedAE & rMSE & MAE & MedAE & MKL & MedKL & rWMSE & WMAE & WMedAE \\\hline
2000 & 0.75 & 0.72 & 0.62 & 0.79 & 0.75 & 0.67 & 0.80 & 0.72 & 0.97 & 0.95 & 0.94 \\
1500 & 0.75 & 0.74 & 0.76 & 0.80 & 0.76 & 0.72 & 0.81 & 0.72 & 0.97 & 0.95 & 1.04 \\
1000 & 0.79 & 0.77 & 0.76 & 0.85 & 0.83 & 0.76 & 0.84 & 0.80 & 0.95 & 0.92 & 0.98 \\
750 & 0.80 & 0.78 & 0.72 & 0.83 & 0.81 & 0.66 & 0.86 & 0.73 & 0.95 & 0.91 & 0.99 \\
500 & 0.80 & 0.81 & 0.91 & 0.85 & 0.84 & 1.01 & 0.83 & 0.86 & 0.94 & 0.90 & 0.93 \\\hline
\end{tabular}
\end{adjustbox}
\caption{$(\alpha, \beta, H, \delta) = (4,3,2,3).$}
\label{table_i}
\end{table}
\newpage
%

\end{document}